\documentclass[%
 aip,
 amsmath,amssymb,
 reprint,%
]{revtex4-1}

\usepackage{graphicx}
\usepackage{dcolumn}
\usepackage{bm,amsmath,amssymb,color}
\usepackage{subcaption}

\usepackage[utf8]{inputenc}
\usepackage[T1]{fontenc}
\usepackage{mathptmx}

\usepackage{framed}
\DeclareGraphicsExtensions{.eps,.pdf,.png,.jpg,.jpeg,.xcf}
\graphicspath{{./figures/}} 

\begin{document}
	
\preprint{AIP/123-QED}

\title[a freely-vibrating circular cylinder in mixed convection flow]{Hydrodynamic and thermal characteristics of a freely-vibrating circular cylinder in mixed convection flow}

\author{B. Liu}\email{a0098961@u.nus.edu}
\affiliation{Department of Mechanical Engineering, National University of Singapore, Singapore 119077}%
\date{\today}

\begin{abstract}
The hydrodynamic and thermal characteristics of a freely-vibrating circular cylinder in mixed convection flow are numerically investigated at low Reynolds numbers. The numerical investigations are conducted for a range of parameters, $2.0 \leq Ur \leq 10.0$, $ 0.7 \leq Pr \leq 10.0$ and $0.5 \leq Ri \leq 2.0$. On the other hand, the Reynolds number and structural coefficients are fixed at $Re=100$, $m^*=10.0$ and $\zeta=0.01$. The structural responses and onset of vortex-induced vibration (VIV) are documented for various environments, e.g., different reduced velocity, Prandtl numbers and Richardson numbers. On the other hand, the influence of structural dynamics on the heat transfer over a heated circular cylinder is recorded and discussed as well. A secondary VIV lock-in region is found in the cases of high Richardson number $Ri=2.0$ for high $Ur$ values, in which the buoyancy-driven flow is non-trivial. A wide VIV lock-in region is formed with tremendous energy transfer between fluid and structure, which is extremely meaningful for hydropower harvesting. The influences of Prandtl and Richardson numbers on the hydrodynamics, structural dynamics and heat transfer are discussed in detail. The temperature contours are found concentrating around cylinder's surface in the cases of high Prandtl numbers, which are also associated with high $\overline{Nu}$ values. The influence on heat convection over cylinder's surface is quantified via the calculation of mean Nusselt number and its fluctuation for different circumstances. The energy transfer coefficient is employed to quantify the energy transfer between fluid and structure in mixed convection flow. The phase angle difference between the transverse displacement and lift force is used to support the discussions of energy transfer. A stabilized finite element formulation in Arbitrary Lagrangian-Eulerian description is derived to simulate the fluid and structural dynamics in mixed convection flow. The obtained numerical results match well with literature and the established empirical formula. To the knowledge of the author, this is the first time that the hydrodynamic, structural and thermal characteristics of a freely-vibrating circular cylinder in mixed convection flow subjecting to transverse buoyancy force are reported. 
\end{abstract}

\maketitle

\begin{table*}[!htp]
	\begin{framed}
	\centering \textbf{Nomenclature}
  \begin{center}
	\begin{tabular}{p{1.0cm}p{6.0cm}p{2cm}p{1.0cm}p{6.0cm}} 
		$\bm{g}$ & gravitational acceleration vector &  & $Nu$ & Nusselt number \\
		$Pr$ & Prandtl number ($=\frac{\nu}{\alpha}$) &  & $Gr$ & Grashof number ($=\frac{g\beta (T_H-T_C) D^3}{\nu^2}$)\\
		$Ra$ &  Rayleigh number ($\frac{g \beta (T_H-T_C) D^3}{\alpha \nu} = Gr Pr$) &  & $Re$ &  Reynolds number ($=\frac{U_{\infty} D}{\nu}$) \\
		$\nu$ & kinematic viscosity &  & $\alpha$ & thermal diffusivity\\
		$\beta$ & volumetric thermal expansion coefficient &  & $\rho_0$ & reference fluid density\\
		$c_p$ & constant pressure specific heat &  &  $q$ & surface heat flux \\
		$Ri$ &  Richardson number ($=\frac{Gr}{Re^2} = \frac{Ra}{Re}$) &  & $\bm{h}$ & surface traction vector\\
		$\bm{u}$ & fluid velocity vector &  & $p$ & pressure\\
		$\bm{u}^*$ &  dimensionless fluid velocity vector &  & $p^*$ & dimensionless pressure\\
		$T$ & temperature &  & $\theta$ & dimensionless temperature\\
		$\bm{x}^*$ &  dimensionless coordinates &  & $\tau$ & dimensionless time\\
		$m^*$ &  mass ratio &  & $\zeta$ & damping ratio\\
		$\bm{f}_n$ & structural frequency vector &  & $Ur$ & reduced velocity\\
		$f_{vs}$ & vortex shedding frequency &  & $St$ & Strouhal number\\
		$A_x$ & dimensionless streamwise vibration amplitude &  & $A_y$ & dimensionless transverse vibration amplitude\\
		$D$ & characteristic length (cylinder's diameter) &  & $U_{\infty}$ & freestream velocity\\
		$\bm{\Phi}^*$ &  dimensionless location of structure &  & $\bm{\eta}^*$ & dimensionless displacement of structure\\
		$\bm{u}^*_m$ & dimensionless mesh velocity vector &  & $\omega$ & spanwise vorticity\\
		$C_d$ & drag coefficient  &  & $C_l$ & lift coefficient\\
		$C_e$ &  energy transfer coefficient &  & $H$ & dimensionless height of computational domain \\	
		$L_u$ &  dimensionless upstream distance &  & $L_d$ & dimensionless downstream distance\\
	\end{tabular}
\end{center}
\end{framed}
\end{table*}

\section{Introduction} \label{sec:intro}
Thermo-physical property in fluid convection arises in many fields of science and engineering, e.g., nuclear reactors, turbine engines, fuel cells, solar energy collectors, thermal power plants and life science research. The conservation equations describing the energy transportation in fluid were well developed~\cite{de1951thermodynamics,hirschfelder1964molecular,chapman1970mathematical} a prior. The energy flux imposes temperature gradients in the fluid system and influences the neighborhood fluid dynamics. Although a number of analytical solutions for conduction heat transfer problems are available~\cite{carslaw1959conduction,ozisik2002boundary}, in many practical circumstances, the geometry and boundary conditions are so complicated that an analytical solution is impossible. Hence, the numerical solutions were sought-after by the research scientists to explore the subtle physics in thermo-fluids~\cite{ibanez2002advanced,ozicsik2017finite,patankar2018numerical,lewis2004fundamentals,reddy2010finite}.  

In many engineering applications, the flow over a circular cylinder is a canonical form used to acquire the fundamental understanding of fluid dynamics and heat transfer. In contrast to the iso-thermal fluid flow, the natural convection component could perturb the neighborhood flow field and induces complicated flow regimes in wake. In 1990, Biswas et. al. (1990)~\cite{biswas1990numerical} documented unsteady mixed convection heat transfer over a square obstacle in a horizontal channel. Their results shown that the mixed convection can initiate periodicity and asymmetry in the wake at low Reynolds numbers, in contrast to the forced convection flow. Sanitjai \& Goldstein (2004)~\cite{sanitjai2004heat} investigated the thermal characteristics over a rigid cylinder in forced convection flow for $2 \times 10^3 < Re < 9 \times 10^4$ and $7 < Pr < 176$. More recently, Juncu (2007)~\cite{Juncu2007} numerically investigated the heat transfer characteristics in the forced convection flow over two tandem cylinder for $1 < Re < 30$ and $0.1 < Pr < 100$. It was reported that the heat transfer characteristics for the evolution of the system for $RePr > 1$ is completely different than that of $RePr < 1$.  For a heated cylinder, it was found by Biswas \& Sarkar (2009)~\cite{biswas2009effect} that hydrodynamic instabilities grow and flow becomes unsteady periodic if the fluid is severely influenced by thermal buoyancy. 

In the previous studies, the primary focus is the thermal characteristics in flow over a stationary cylinder. However, the intensive structural motions in fluid flow has a tremendous influence on the hydrodynamics in fluid-structure interaction (FSI) problems. In iso-thermal and incompressible flow, many well-documented numerical investigations~\cite{Sarpkaya1979JoAM,Blackburn2001JoFaS,Williamson2004ARFM,Sarpkaya2004JoFaS,Prasanth2008JoFM,Bearman2011JoFaS,zhu2019wake} had been published to assess the complex coupling of fluid and structural dynamics. Furthermore, in multi-body systems, e.g., side-by-side\cite{Williamson1985JoFM,Carini2014JoFM,liu2016interaction,liu2016effect}, tandem~\cite{Borazjani2009Jofm,Assi2010JoFM}, near-wall~\cite{Li2016JoFaS,liu2020numerical,Ju2020JoFM} or array~\cite{joshi2016flow,Tang2020PoFa} configurations, the interference could be extremely complex and subtly linked with hydrodynamic instabilities. However, the thermal effect of fluid flow is missed in these studies, which is key parameter in many engineering applications. It is well documented by scientists that the heat energy transfer in convection flow has an enormous impact to hydrodynamic stabilities~\cite{venkatasubbaiah2009mixed,sengupta2010instabilities,sengupta2011linear,sengupta2012instabilities}. Recently, Kan et. al. (2020)~\cite{Khan2020IJoTS} conducted an excellent numerical investigation for a freely-vibrating square cylinder in forced convection flow. Yang et. al. (2020)~\cite{Yang2020IJoHaMT} also documented a study of freely-vibrating circular cylinder in forced convection flow. In this article, we are very interested in the effect of buoyancy force on the momentum transportation in mixed convection flow, where the Navier-Stokes and energy equations are strongly coupled via Boussinesq approximation. Furthermore, the hydrodynamic responses of a square cylinder are significantly different from a circular cylinder, due to the fixed separation points of boundary layers and the occurrence of galloping in the post VIV lock-in. Particularly, we would like to know more about the relationship between mixed convection flow, and the fluid and structure stabilities. To this end, in this article we investigate the hydrodynamic and thermal characteristics for flow over a freely-vibrating circular cylinder subjecting to transverse buoyancy-driven flow. To the knowledge of author, it has not been reported in literature previously. Hence the primary focuses in this article are to address the following questions:
\begin{itemize}
	\item How does the structural dynamics accommodate itself in mixed convection flow, e.g., the response of VIV lock-in?
	\item What is the influence of structural dynamics on the heat transfer over a vibrating cylinder?
	\item How does the thermal effect interfere the kinetic energy transfer between fluid and structure?
\end{itemize}
In particular, the energy transfer between fluid and structure is a primary concern in hydropower harvesting. Liu \& Jaiman (2018)~\cite{liu2018dynamics} found that the maximum energy transfer occurs at about $90^{\circ}$ phase angle difference between the lift force and transverse vibration during VIV lock-in. For $90^{\circ}$ phase angle difference, the transverse motion acquires the maximum acceleration from the lift force. In contrast, the energy transfer are almost suppressed during pre and post lock-in regions for a vibrating cylinder in isothermal flow~\cite{liu2018dynamics}. In this article, we noticed a secondary VIV lock-in region and the corresponding enhanced energy transfer for a vibrating cylinder in mixed convection flow. The maximum energy transfer is again confirmed for phase angle difference at $90^{\circ}$ in the current investigation.  

It is known that buoyancy-driven flow is created by strong temperature gradients in a fluid flow field, which occurs in common flow situations. However, its significance can vary depending upon flow regimes. In mixed convection flow, the component of natural convection is controlled by Prandtl and Richardson numbers. Typically, the natural convection is negligible at $Ri \lesssim 0.1$ for forced convection flow. On the other hand, the natural convection becomes dominant at $Ri \gtrsim 10$. In a mixed convection flow ($0.1 \lesssim Ri \lesssim 10$), both forced and natural convection are significant. For Richardson numbers less than 0.15, the flow was characterized by broadening of the wake; whereas Richardson numbers greater than 0.15 revealed separation delay and attached twin vortices behind the cylinder~\cite{biswas2009effect}. The variation of Richardson number signify the influence of buoyancy-driven flow on the fluid momentum transportation.
In this investigation, we are particularly interested in thermal characteristics in mixed convection flow, but the force convection component is dominant over the natural convection. Hence the range of Richardson number is chosen relatively close to forced convection flow regime, $Ri = [0.5, 2.0]$. On the other hand, as Prandtl number quantifies the diffusivity of fluid momentum over heat energy, it plays a significant role to determine heat transfer and hydrodynamic characteristics in near wake for mixed convection flow. To investigate the influences of Prandtl number ($Pr$) and Richardson number ($Ri$) on mixed convection flow, the values of $Pr$ and $Ri$ are chosen as $0.7 \leq Pr \leq 10.0$ and $0.5 \leq Ri \leq 2.0$ respectively, considering their practical significance. 

The structure of this article is organized as follow. At first, the governing equations and the derived numerical formulations are introduced in Section~\ref{sec:govern_num}. Following that, the setup of computational domain and the validation of derived numerical formulations are presented in Section~\ref{sec:scheme}. Subsequently, the obtained numerical results of structural dynamics, hydrodynamics and energy transfer in mixed convection flow are discussed in Section~\ref{sec:results}. Finally, the conclusion remarks are drawn in Section~\ref{sec:conclusion}.

\section{Governing equation and numerical formulation} \label{sec:govern_num}
In current problem, the fluid density is assumed to be uniform over the computational domain, except for the buoyancy term, in which the density is taken as a function of temperature. This assumption leads to the Boussinesq Approximation (BA) if the temperature difference level are maintained within certain limits. The unsteady Navier-Stokes equation is strongly coupled with the energy equation to simulate mixed convection flow in the article. The unsteady Navier-Stokes equation is spatially discretized with 4-nodes quadrilateral elements in a stabilized finite element formulation, in which the fluid-structure interaction is precisely tracked with the body-conformal meshes. The fluid solver is coupled with the structural solver using a second-order stagger-partitioned weakly-coupling fluid-structure interaction scheme. The fluid and structural dynamics are coupled in Arbitrary Lagrangian-Eulerian (ALE) description. The unconditional-stable second-order accurate generalized-$\alpha$ time integration schemes are employed for both fluid and structure solvers to march in time. In this section, the relevant variables are explained at the appropriate location in the contents. For the detailed description of the other variables, please refer to the nomenclature in this article.
\subsection{governing equations and non-dimensionalization} \label{sec:govern}
The unsteady, incompressible and Newtonian Navier-Stokes equation in its conservative form is coupled with the energy equation via the Boussinesq approximation, as presented in Equation~\eqref{eq:ns}, to simulate the mixed convection flow over an elastically-mounted cylinder, where $\bm{g} = [0, -g]' = [0, -9.81]'$ is the gravitational acceleration vector. On the other hand, the energy equation is coupled with the Navier-Stokes equation through its advection term ($(\bm{u}\cdot \nabla)T$). The superscript ($'$) is a transpose operator. The $\bm{u}$, $p$ and $T$ are the fluid velocity, pressure and temperature respectively. $\alpha$ is the thermal diffusivity. Equation~\eqref{eq:ns4} and Equation~\eqref{eq:ns5} are the boundary conditions prescribed by the Dirichlet ($\Gamma_D$) and Neumann ($\Gamma_N$) domain boundaries respectively, where $\tilde{\bm{h}}$ and $\tilde{q}$ refer to the prescribed surface traction vector and heat flux. Equation~\eqref{eq:ns6} is the initial states of the flow field. 

\begin{subequations} \label{eq:ns}
	\begin{align}
	\nabla \cdot \bm{u} &= 0 \qquad \qquad \qquad \quad \;\;\; \forall \bm{x} \in \Omega^f(t) \label{eq:ns1}\\
	\rho_0 \big(\partial_t \bm{u} + (\bm{u}\cdot\nabla)\bm{u}\big) &= \nabla \cdot \bm{\sigma}\{\bm{u},p\} + \rho \bm{g} \quad\; \forall \bm{x} \in \Omega^f(t) \label{eq:ns2}\\
	\partial_t T + (\bm{u} \cdot \nabla) T &= \alpha \nabla^2 T \qquad \qquad \quad \;\;\; \forall \bm{x} \in \Omega^f(t)\label{eq:ns3}\\
	\bm{u} &= \tilde{\bm{u}};\; T = \tilde{T} \qquad \quad \;\;\;\;\;\; \forall \bm{x} \in \Gamma^f_D(t) \label{eq:ns4}\\
	\bm{\sigma} \{\bm{u},p\} \cdot \bm{n} &= \tilde{\bm{h}};\; \alpha (\nabla T)\cdot \bm{n} = \tilde{q} \quad \forall \bm{x} \in \Gamma^f_{N}(t) \label{eq:ns5}\\
	\bm{u} &= \tilde{\bm{u}}_0;\; T = \tilde{T}_0 \qquad \quad\;\;\; \forall \bm{x} \in \Omega^f(0) \label{eq:ns6}
	\end{align}
\end{subequations}
The term $\partial_t (\cdot)$ refers to the time derivative of a variable with respect to the spatial coordinates ($\bm{x}$). The Cauchy stress tensor ($\bm{\sigma}$) is defined as
\begin{subequations}
	\begin{align}
	\bm{\sigma}\{\bm{u},p\} &= -p\bm{I} + 2\mu \epsilon(\bm{u}) \\
	\epsilon(\bm{u}) &= \frac{1}{2}\Big[\nabla \bm{u} + (\nabla \bm{u})'\Big]
	\end{align}
\end{subequations}
where $p$, $\bm{I}$, $\mu$ and $\epsilon(\bm{u})$ respectively are the fluid pressure, the identity matrix, the dynamic viscosity and the strain rate tensor. Natural convection is generated by the density difference induced by the temperature differences within a fluid system. Because of the small density variations present in these type of flows, a general incompressible flow approximation is normally adopted. Assuming the density is a function of temperature, $\rho = \rho(T)$, elementary thermodynamics states that $\beta = -\frac{1}{\rho}(\partial \rho/\partial T)_p$. Hence the density of the fluid at constant pressure depends on the temperature, which can be written as
\begin{eqnarray}
\rho = \rho_0\big[1.0-\beta (T-T_{C})\big]
\end{eqnarray}
where $\rho_0$ is the reference fluid density. These considerations lead to the Boussinesq Approximation in the $y$-component fluid momentum equation.

A mixed convection state is one in which both natural and forced convection are present. The buoyancy effects become comparable to the forced flow effects at moderate Richardson numbers. Since the flow is partially dominated by forced convection, a reference velocity ($U_{\infty}$) value is normally known. Introducing the following dimensionless groups in Equation~\eqref{eq:dim}, where $U_{\infty}$ is available, the governing equations for the mixed convection flow can be non-dimensionalized into the form in Equation~\eqref{eq:ns_dim}, where $\bm{n}_g = [0, -1]'$ is the unit vector of gravitational force.
\begin{subequations} \label{eq:dim}
	\begin{align}
	\bm{x}^* &= \frac{\bm{x}}{D}; \quad \bm{u}^* = \frac{\bm{u}}{U_{\infty}}; \quad \tau = \frac{tU_{\infty}}{D} \\
	p^* &= \frac{p+\rho_0 g z}{\rho_0 U^2_{\infty}}; \quad \theta = \frac{T-T_{C}}{T_H-T_C} 
	\end{align}
\end{subequations}
\begin{figure*}[!htp] 
	\begin{subequations} \label{eq:ns_dim}
		\begin{align}
		\nabla \cdot \bm{u}^*  &= 0 \qquad \qquad \qquad \qquad \qquad \qquad \qquad \qquad \qquad \;\;\;\;\;\;\;\;\; \forall \bm{x}^* \in \Omega^f(\tau) \label{eq:ns_dim1}\\
		\partial_{\tau} \bm{u}^* + (\bm{u}^* \cdot \nabla) \bm{u}^* &= -\nabla p^* +\frac{1}{Re} \nabla \cdot \big(\nabla \bm{u}^* + (\nabla \bm{u}^*)'\big) - \big(\frac{Gr}{Re^2}\bm{n}_g\big) \theta \qquad \forall \bm{x}^* \in \Omega^f(\tau) \label{eq:ns_dim2}\\
		\partial_{\tau} \theta + (\bm{u}^* \cdot \nabla) \theta &= \frac{1}{Re Pr} {\nabla}^{2} \theta \qquad \qquad \qquad \qquad \qquad \qquad \qquad \;\;\;\;\;\;\;\;\;\; \forall \bm{x}^* \in \Omega^f(\tau) \label{eq:ns_dim3}\\
		\bm{u}^* &= \tilde{\bm{u}}^*;\quad \theta = \tilde{\theta} \qquad \qquad \qquad \qquad \qquad \qquad \;\;\;\;\;\;\;\;\;\;\;\;\;\;\; \forall \bm{x}^* \in \Gamma^f_D(\tau) \label{eq:ns_dim4}\\
		\big(-p^*\bm{I} + \frac{2}{Re} \epsilon(\bm{u}^*)\big) \cdot \bm{n} &= \tilde{\bm{h}}^*; \quad \big(\frac{1}{Re Pr} \nabla \theta\big)\cdot \bm{n} = \tilde{q}^* \qquad \qquad \qquad \quad \;\;\;\;\;\;\;\;\;\;\;\; \forall \bm{x}^* \in \Gamma^f_{N}(\tau) \label{eq:ns_dim5}\\
		\bm{u}^* &= \tilde{\bm{u}}^*_0; \quad \theta = \tilde{\theta}_0 \qquad \qquad \qquad \qquad \qquad \qquad \;\;\;\;\;\;\;\;\;\;\;\;\; \forall \bm{x}^* \in \Omega^f(0) \label{eq:ns_dim6}
		\end{align}
	\end{subequations}
\end{figure*}

In Equation~\eqref{eq:dim}, the superscript ($*$) indicates the dimensionless groups. $\tau$ and $\theta$ respectively are the dimensionless time and temperature. The $T_H$ and $T_C$ represent the highest and coolest temperature in the computational domain. The value of $z$ is the elevation height in the direction of gravitational acceleration.

\subsection{numerical formulations of fluid-structure interaction simulation} \label{sec:fem}
The numerical formulation is derived by spatially discretizing the governing equations in their primitive variables in Equation~\eqref{eq:ns_dim} using a stabilized finite element formulation at first. Similar to the Navier-Stokes equation, the energy equation in Equation~\eqref{eq:ns_dim3} involves the advection term $(\bm{u}^*\cdot \nabla) \theta$ too. which causes spurious oscillations in velocity field. Hence a residual-based stabilized finite element formulations, Galerkin Least Squares (GLS)~\cite{brooks1982streamline} and Pressure Stabilizing Petrov Galerkin (PSPG)~\cite{hughes1986new}, are employed to stabilize the spurious oscillation in velocity field with numerical diffusion and circumvent the Ladyzhenskaya-Babuska-Brezzi (LBB) condition of the velocity-pressure field. In the finite element formulations, we define appropriate sets of finite trial solution spaces ($S^h_u$, $S^h_{\theta}$ and $S^h_p$) for velocity, temperature and pressure, and their finite test function spaces ($V^h_u$, $V^h_{\theta}$ and $V^h_p$) respectively, as shown in Equation~\eqref{eq:space}.
\begin{subequations}\label{eq:space}
	\begin{align}
	S^h_u &= \{ \bm{u}^{*h} | \bm{u}^{*h} \in (H^{1h})^d, \bm{u}^{*h} \doteq \tilde{\bm{u}}^{*h} \quad \forall \bm{x}^* \in \Gamma^f_D(\tau)\}\\
	V^h_u &= \{\bm{\psi}^{h}_u | \bm{\psi}^h_u \in (H^{1h})^d, \bm{\psi}^h_u \doteq \bm{0} \quad \quad\;\;\; \forall \bm{x}^* \in \Gamma^f_D(\tau)\}\\
	S^h_{\theta} &= \{ \theta^h | \theta^{h} \in H^{1h}, \theta^{h} \doteq \tilde{\theta}^{h} \qquad \quad \;\;\;\; \forall \bm{x}^* \in \Gamma^f_D(\tau)\}\\
	V^h_{\theta} &= \{ \psi^h_{\theta} | \psi^h_{\theta} \in H^{1h}, \psi^h_{\theta} \doteq \bm{0} \qquad \quad \;\;\;\; \forall \bm{x}^* \in \Gamma^f_D(\tau)\}\\
	S^h_p &= V^h_p = \{\psi^h_p | \psi^h_p \in H^{1h}\}
	\end{align}
\end{subequations} 
where the superscript ($h$) indicates a finite function space, e.g., $S^h_u \subset S_u$. The value of $d$ refers to the number of space dimension. $H^{1h}$ is a finite dimensional space defined in Equation~\eqref{eq:H}, where $P^1$ is the piece-wise linear polynomial and $\mathcal{E}$ denotes the set of elements resulting from the spatial discretization.     
\begin{eqnarray} \label{eq:H}
H^{1h} = \{\Theta^h | \Theta^h \in C^0,\Theta^h |_{\Omega^e} \in P^1, \forall \Omega^e \in \mathcal{E} \}
\end{eqnarray}
Hence the stabilized finite element formulation of Equation~\eqref{eq:ns_dim} can be written as: for all $\bm{\psi}^h_u \in V^h_u, \psi^h_p \in V^h_p, \psi^h_{\theta} \in V^h_{\theta}$, find $\bm{u}^{*h} \in S^h_u, p^{*h} \in S^h_p, \theta^h \in S^h_{\theta}$ such that Equation~\eqref{eq:fem} is satisfied.  
\begin{figure*}[!htp]
	\begin{eqnarray} \label{eq:fem}
	& &\underbrace{\int_{\Omega^f} \Big[\bm{\psi}^h_u \cdot \Big(\partial_{\tau} {\bm{u}^{*h}} |_{\mathcal{X}} + \big((\bm{u}^{*h}-\bm{u}^{*h}_m) \cdot \nabla\big) \bm{u}^{*h}+\big(\frac{Gr}{Re^2} \bm{n}_g\big) \theta^{h}\Big) + \epsilon(\bm{\psi}^h_u):\bm{\sigma}\{\bm{u}^{*h},p^{*h}\} \Big] d\Omega - \int_{\Gamma^f_N} \bm{\psi}^h_u \cdot \tilde{\bm{h}}^{*h}\; d\Gamma  + \int_{\Omega^f} \big[ \psi^h_p \nabla \cdot \bm{u}^{*h}\big] d\Omega}_{\mathcal{B}_G([\bm{\psi}^h_u,\psi^h_p],[\bm{u}^{*h},p^{*h},\theta^h])} \nonumber\\
	& & \underbrace{- \int_{\Gamma^f_{N(out)}} \bm{\psi}^h_u \cdot \big[-\frac{1}{Re}(\nabla \bm{u}^{*h})'\cdot \bm{n}\big] d\Gamma}_{\mathcal{B}_{corr}(\bm{\psi}^h_u,\bm{u}^{*h})} + \underbrace{\sum \limits^{n_{el}}_{e=1} \int_{\Omega^f} \tau_m \Big[\big((\bm{u}^{*h}-\bm{u}^{*h}_m) \cdot \nabla\big) \bm{\psi}^h_u - \frac{1}{Re} \nabla^2 \bm{\psi}^h_u + \nabla \psi^h_p\Big]\cdot}_{\mathcal{B}_S([\bm{\psi}^h_u,\psi^h_p],[\bm{u}^{*h},p^{*h},\theta^h])}\nonumber\\
	 & &\underbrace{\Big[\partial_{\tau} \bm{u}^{*h}|_{\mathcal{X}} + \big((\bm{u}^{*h}-\bm{u}^{*h}_m) \cdot \nabla\big) \bm{u}^{*h} - \frac{1}{Re} \nabla^2 \bm{u}^{*h} + \nabla p^{*h}  + \big(\frac{Gr}{Re^2} \bm{n}_g \big) \theta^h\Big]d\Omega +\sum \limits^{n_{el}}_{e=1} \int_{\Omega^f} \tau_c \Big[(\nabla \cdot \bm{\psi}^h_u)(\nabla \cdot \bm{u}^{*h})\Big]d\Omega}_{\mathcal{B}_S([\bm{\psi}^h_u,\psi^h_p],[\bm{u}^{*h},p^{*h},\theta^h])} \nonumber\\
	& & + \underbrace{\int_{\Omega^f} \Big[\psi^h_{\theta} \cdot \big(\partial_{\tau} \theta^h |_{\mathcal{X}} + \big((\bm{u}^{*h}-\bm{u}^{*h}_m) \cdot \nabla \big) \theta^h\big) + \frac{1}{Re Pr} \nabla \psi^h_{\theta} \cdot \nabla \theta^h\Big]d\Omega}_{\mathcal{B}_G(\psi^h_{\theta},[\bm{u}^{*h},\theta^h])} + \underbrace{\sum \limits^{n_{el}}_{e=1} \int_{\Omega^f} \tau_{\theta} \Big[\big((\bm{u}^{*h}-\bm{u}^{*h}_m) \cdot \nabla\big) \psi^h_{\theta} - \frac{1}{Re Pr} \nabla^2 \psi^h_{\theta} \Big]\cdot}_{\mathcal{B}_S(\psi^h_{\theta},[\bm{u}^{*h},\theta^h])}\nonumber\\
	& & \underbrace{\Big[\partial_{\tau} \theta^h |_{\mathcal{X}} + \big((\bm{u}^{*h}-\bm{u}^{*h}_m) \cdot \nabla \big) \theta^h - \frac{1}{Re Pr} \nabla^2 \theta^h\Big]d\Omega}_{\mathcal{B}_S(\psi^h_{\theta},[\bm{u}^{*h},\theta^h])} - \underbrace{\int_{\Gamma^f_N} \big[\psi^h_{\theta} \tilde{q}^{*h}\big] d\Gamma}_{\mathcal{B}_G(\psi^h_{\theta},\theta^h)} = 0 \qquad \forall [\bm{\psi}^h_u,\psi^h_p,\psi^h_{\theta}] \in V^h_{u} \times V^h_p \times V^h_{\theta}
	\end{eqnarray}
\end{figure*}
The $\mathcal{B}_G([\bm{\psi}^h_u,\psi^h_p],[\bm{u}^{*h},p^{*h},\theta^h])$, $\mathcal{B}_G(\psi^h_{\theta},[\bm{u}^{*h},\theta^h])$ and $\mathcal{B}_G(\psi^h_{\theta},\theta^h)$ terms are derived from the unsteady and incompressible Navier-Stokes equation and the conservation of energy equation respectively, based on the standard Galerkin method in finite element framework. The boundary integral $\mathcal{B}_{corr}(\bm{\psi}^h_u,\bm{u}^{*h})$ is a correction term~\cite{heywood1996artificial} for the "do-nothing" outflow boundary condition to avoid reverse numerical flux. The $\mathcal{B}_S([\bm{\psi}^h_u,\psi^h_p],[\bm{u}^{*h},p^{*h},\theta^h])$ and $\mathcal{B}_S(\psi^h_{\theta},[\bm{u}^{*h},\theta^h])$ are the stabilization terms based on the GLS and PSPG formulations. The term $\partial_{\tau} (\cdot)|_{\mathcal{X}}$ refers to the spatial time derivative with respect to the fixed referential coordinates ($\mathcal{X}$) and the dimensionless time ($\tau$) in ALE description. The stabilization parameters ($\tau_m$, $\tau_c$ and $\tau_{\theta}$) are defined as
\begin{subequations}
	\begin{align}
	\tau_m &= \Big[\Big(\frac{2||\bm{u}^*||}{h}\Big)^2+\Big(\frac{4}{Re h^2}\Big)^2\Big]^{-1/2}\\
	\tau_c &= \frac{h}{2} ||\bm{u}^*||\gamma \quad \text{for} \; \gamma = \left\{\begin{array}{ll}
	Re_u /3 \quad 0 < Re_u \leqslant 3 \\
	1 \quad \quad \quad 3 < Re_u
	\end{array} \right.\\
	\tau_{\theta} &= \Big[\Big(\frac{2||\bm{u}^*||}{h}\Big)^2+\Big(\frac{4}{Re Pr h^2}\Big)^2\Big]^{-1/2}
	\end{align}
\end{subequations}
where $Re_u$ and $h$ respectively are the local Reynolds number and the size of element. The value of $\bm{u}_m$ is the dimensionless mesh velocity. The fluid solver is coupled with the structural solver via satisfying the kinematic and dynamic constraints along the fluid-structure interface ($\Gamma^{fs}$), as shown in Equation~\eqref{eq:fsi}.
\begin{subequations}\label{eq:fsi}
	\begin{align}
	&\bm{u}^*(\bm{\Phi}^*(\bm{\mathcal{X}},\tau),\tau) = \partial_{\tau}\bm{\Phi}^*(\bm{\mathcal{X}},\tau) \quad \;\;\; \forall \bm{\mathcal{X}} \in \Gamma^{fs}(\tau)\\
	&\bm{h}^*(\bm{\Phi}^*(\bm{\mathcal{X}},\tau),\tau) = \nonumber\\
	&- \bm{h}^{*cyl} (\bm{\Phi}^*(\bm{\mathcal{X}},\tau),\tau) \qquad \qquad \qquad \forall \bm{\mathcal{X}} \in \Gamma^{fs}(\tau)
	\end{align}
\end{subequations}
where $\bm{h}^*$ and $\bm{h}^{*cyl}=[h^{*cyl}_x,h^{*cyl}_y]$ respectively are the dimensionless fluid and structural stresses along the fluid-structure interface. The value of $\bm{\Phi}^*$ is the dimensionless location of the structure, which is defined in Equation~\eqref{eq:disp}, where $\bm{\eta}^*$ is the dimensionless displacement of structure at time $\tau$. In this investigation, the referential coordinates can be taken as the initial position of the rigid circular cylinder. The values of $\partial_{\tau} \bm{\Phi}^*$ and $\partial^2_{\tau} \bm{\Phi}^*$ are defined as the structural velocity and acceleration respectively.
\begin{eqnarray} \label{eq:disp}
	\bm{\Phi}^*(\bm{\mathcal{X}},\tau) = \bm{\eta}^*(\bm{\mathcal{X}},\tau)+\bm{\mathcal{X}} \quad \forall \bm{\mathcal{X}} \in \Omega^s(\tau)
\end{eqnarray}
The superscript ($s$) indicates the structural variables. Hence the governing equation of the structure can be formulated as follow
\begin{eqnarray}
	& &\partial^2_{\tau}\bm{\Phi}^* + \bm{c} \partial_{\tau} \bm{\Phi}^* + \bm{k} \bm{\eta}^* = \bm{h}^{*cyl} \quad \forall \bm{\mathcal{X}} \in \Omega^s(\tau) \\
	& &\bm{c} = 2 \zeta \sqrt{\bm{k} m^s}; \quad \bm{k} = (2 \pi \bm{f}_n)^2 m^s \nonumber\\
	& &Ur = U_{\infty}/(f_{ny}D); \quad m^s = m^* (0.25\pi D^2 L \rho_0) \nonumber
\end{eqnarray}
where $\bm{c}=[c_x, c_y]'$, $\bm{k}=[k_x, k_y]$ and $\bm{f}_{n} = [f_{nx}, f_{ny}]'$ respectively are the resultant damping coefficient vector, the resultant stiffness coefficient vector and the structural frequency vector. $m^*$ is the mass ratio and $\zeta$ is the damping ratio. The reduced velocity ($Ur$) is defined based on the structural frequency in the transverse direction ($f_{ny}$). In this investigation, it is assumed that the structural frequencies in transverse and streamwise directions are identical. The values of $D=1.0$, $L=1.0$ and $m^s$ are the diameter, spanwise length and mass of the cylinder. In the ALE description, the coordinates of mesh nodes are mapped using a popular harmonic model, whose strong form reads as 
\begin{eqnarray}
-\nabla \cdot (\alpha_e \nabla \bm{d}^*) &=& 0 \quad \forall \bm{x}^* \in \Omega^f (\tau)\\
\tilde{\bm{d}}^* &=& \bm{\eta}^* \;\; \forall \bm{x}^* \in \Gamma^{fs} (\tau)\nonumber\\
\tilde{\bm{d}}^* &=& \bm{0} \;\;\;\; \forall \bm{x}^* \in \Gamma^{f}(\tau)/\Gamma^{fs}(\tau)\nonumber
\end{eqnarray}
where $\bm{d}^*$ are the dimensionless grid displacement with respect to the fixed referential framework $\bm{\mathcal{X}}$. The value of $\alpha_e$ is a "stiffness parameter" of the mesh~\cite{stein2003mesh}. 

To couple the fluid and structural solvers, a second-order staggered-partitioned weakly-coupling FSI scheme is implemented. Both the solutions to the fluid and structural solvers march in time with the popular second-order unconditionally stable generalized $\alpha$ time integration schemes. For the detailed formulations of the FSI scheme and the generalized $\alpha$ time integration schemes, please refer to Dettmer \& Peri{\'c} (2013) ~\cite{Dettmer2013IJfNMiE}, Chung \& Hulbert (1993)~\cite{Chung1993Joam}, Jansen et. al. (2000)~\cite{Jansen2000Cmiamae}.

In many mixed convection flow applications, there are two important quantities of interest: the rate of heat transfer, Nusselt number ($Nu$), and the hydrodynamic responses of the submerged structure, the hydrodynamic forces ($C_d$ and $C_l$) and Strouhal number ($St$), as shown in Equation~\eqref{eq:nu}.
\begin{subequations} \label{eq:nu}
	\begin{align}
	Nu (s) &= - \nabla\theta(s) \cdot \bm{n}(s); \quad \overline{Nu} = \frac{1}{\ell}\int^{\ell}_{s=0} Nu (s)\; ds \\
	C_d &= \frac{h^{*cyl}_{x}}{0.5 \rho_0 U^2_{\infty} D L}; \quad C_l = \frac{h^{*cyl}_{y}}{0.5 \rho_0 U^2_{\infty} D L} \\
	C_e &= \int C_l v^* d\tau; \quad St=\frac{f_{vs}D}{U_{\infty}}
	\end{align}
\end{subequations}
In particular, the energy transfer coefficient ($C_e$) in transverse direction is a important parameter in the analysis of fluid-structure interaction, which indicates the total kinetic energy transferred between fluid and structure for a time interval. In this investigation, we employ these global quantities to analyze and quantify the hydrodynamic and thermal characteristics of an elastically-mounted cylinder in different flow regimes.

\section{Problem statement and validation} \label{sec:scheme}
In this section, the configuration of the computational domain and its hydrodynamic \& thermal boundary conditions are presented at first. Subsequently, the mesh and time convergence analyses are conducted to determined the optimal spatial and temporal discretizations for the simulations. The implemented numerical formulations are validated with literature. The obtained numerical results match well with the results in literature.
\subsection{problem setup and boundary conditions}
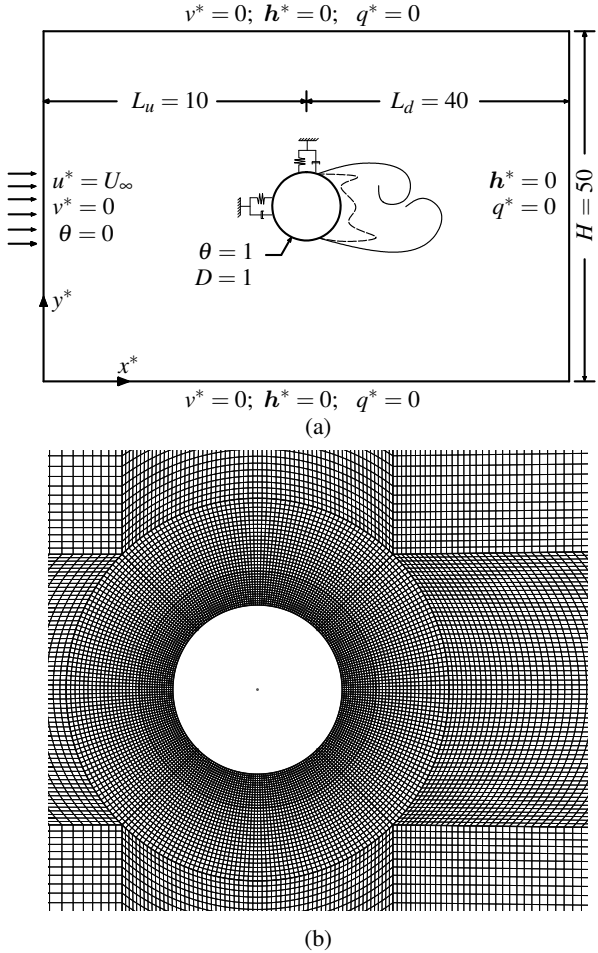
\begin{figure}[!htp]
	\begin{subfigure}{0.5\textwidth}
	\includegraphics[trim=0.1cm 7cm 0.1cm 9cm,scale=0.4,clip]{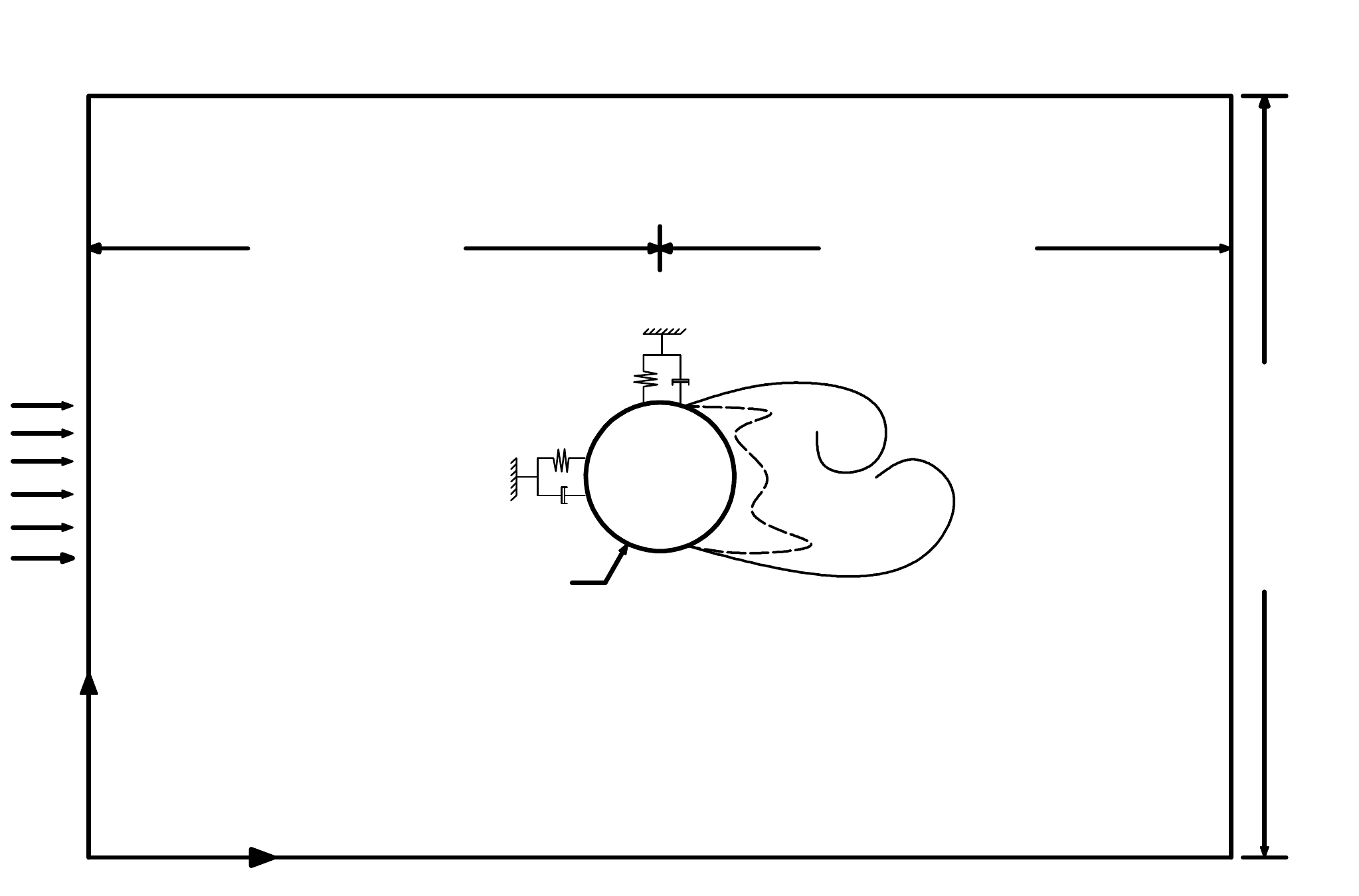}
	\begin{picture}(0,0)
	\put(-220,35){\small $y^*$}
	\put(-195,10){\small $x^*$}
	\put(-220,80){\small $u^* = U_{\infty}$}
	\put(-220,70){\small $v^* = 0$}
	\put(-218,60){\small $\theta = 0$}
	\put(-55,80){\small $\bm{h}^* = 0$}
	\put(-53.5,70){\small $q^* = 0$}
	\put(-170,143){\small $v^* = 0;$}
	\put(-140,143){\small $\bm{h}^* = 0;$}
	\put(-105,143){\small $q^* = 0$}	
	\put(-170,-3){\small $v^* = 0;$}
	\put(-140,-3){\small $\bm{h}^* = 0;$}
	\put(-105,-3){\small $q^* = 0$}		
	\put(-165,53){\small $\theta = 1$}
	\put(-166,43){\small $D = 1$}
	\put(-190,110){\small $L_u = 10$}
	\put(-92,110){\small $L_d = 40$}	
	\put(-22,60){\rotatebox{90}{\small $H = 50$}}		
	\end{picture}
	\caption{}
	\label{fig:schem1}
\end{subfigure}
\begin{subfigure}{0.5\textwidth}
	\includegraphics[trim=0.1cm 6cm 0.1cm 5cm,scale=0.27,clip]{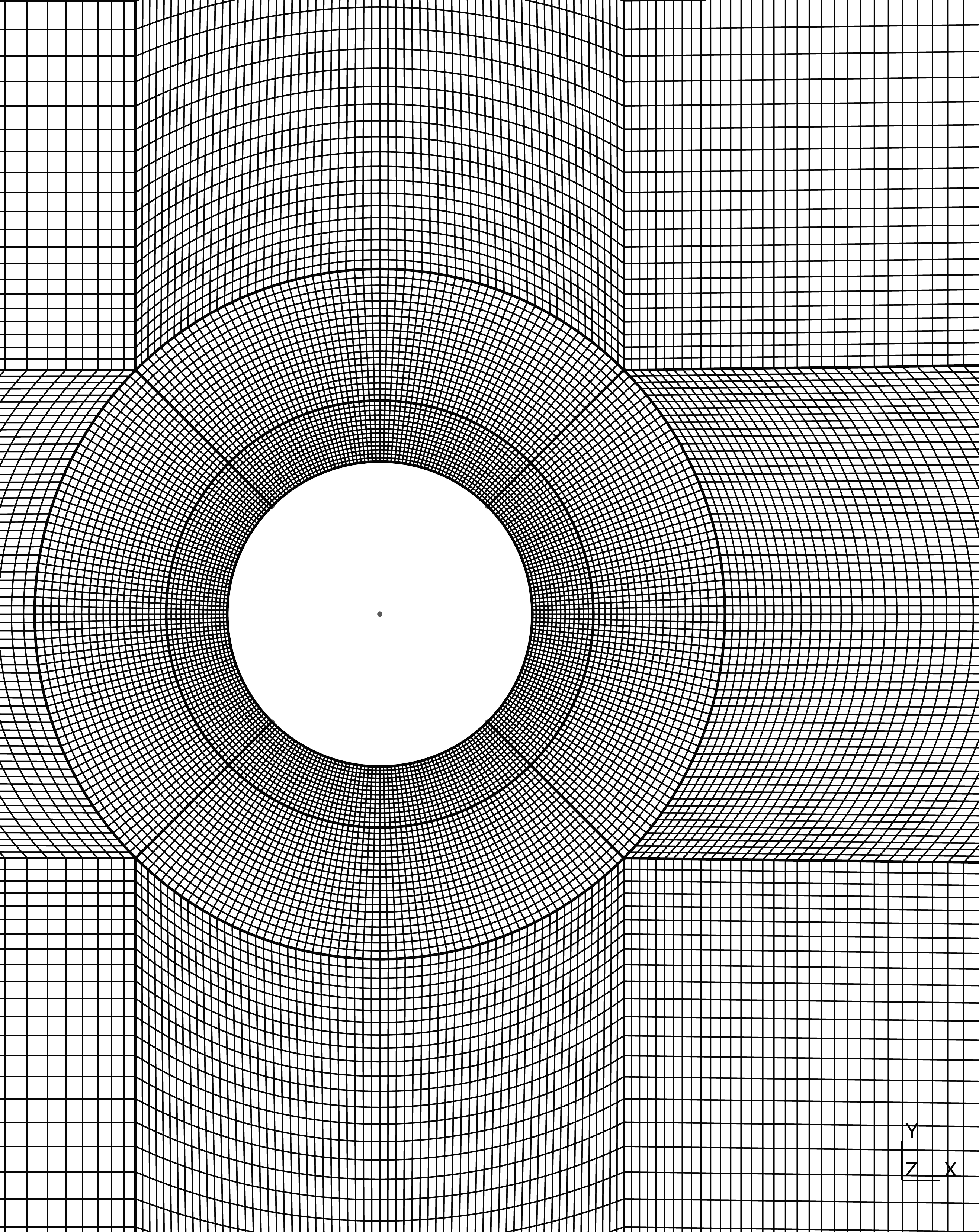}
	\caption{}
	\label{fig:schem2}
\end{subfigure}
	\caption{Schematic diagram and mesh discretization: (a) schematic diagram of an elastically-mounted cylinder in mixed convection flow and (b) mesh discretization around the wake behind the cylinder}
	\label{fig:scheme}
\end{figure}
In this investigation, the numerical investigation is conducted for the flow over an elastically-mounted circular cylinder in mixed convection flow. As illustrated in Figure~\ref{fig:schem1}, the cylinder is initially situated at the origin of the axes $\bm{x}^* = (0,0)$, where is ten-diameter downstream the inlet ($L_u = 10$), forty-diameter upstream the outlet ($L_d = 40$) and twenty-five-diameter away from the upper and lower traction-free boundaries. The blockage ratio is 2\%. The diameter of the cylinder is taken as the characteristic length, $D = 1.0$. The rigid circular cylinder is attached with spring-damper systems and allowed to vibrate along x (streamwise) and y (transverse) axes. In this article, the Reynolds number, mass and damping ratios are fixed at $Re=100.0$, $m^*=10.0$ and $\zeta^*=0.01$ respectively. The traction-free thermal and momentum boundary conditions are imposed along the boundaries of the computational domain, as shown in Figure~\ref{fig:schem1}, except the inlet and the cylinder. A uniform velocity ($\tilde{\bm{u}}^* = [U_{\infty},0.0]'$) and a homogeneous temperature ($\theta=0.0$) are imposed along the inlet. The uniform temperature ($\theta=1.0$) is prescribed over the cylinder's surface. The instantaneous velocity and location of the cylinder is determined by the dynamics of its vibrating motion. The structured boundary-layer meshed surround the cylinder's surface and radiate outward at a mesh growth rate less than $1.03$ to avoid mesh skewness, as exhibited in Figure~\ref{fig:schem2}. The height of the first boundary-layer mesh is controlled well below the linear viscus sublayer $y^+ =1.0$, where $y^+$ is the dimensionless wall distance. Unlike the isothermal incompressible fluid flow, we noticed that the height of the first boundary layer mesh should be much less than $y^+=1.0$ for the mixed convection flow, so as to achieve the mesh convergence, e.g., $y^+ \lesssim 0.3$.    

The hydrodynamic and thermal characteristics of the vibrating cylinder is investigated in different flow regimes by varying the reduced velocity, the Prandtl number and Richardson number. The characteristic phenomena of VIV usually occur within the range of  $Ur \in [2.0,10.0]$, e.g., VIV lock-in, fluttering and galloping. The flow of the gases and water is particularly of our focus in practice; hence the range of the Prandtl number is taken as $Pr \in [0.7, 2.0]$. For mixed convection flow, the values of Richardson number are chosen from 0.5 to 2.0, in which the forced convection is assumed to be relatively influential to the natural convection in this investigation. 

\subsection{validation and convergence analysis}
As illustrated in Figure~\ref{fig:schem2}, the computational domain is discretized with four-node quadrilateral elements. The fluid velocity vector, pressure and temperature are collocated at each node of the elements.
\begin{figure*}
	\centering
	\begin{subfigure}{0.5\textwidth}
	\centering
	\includegraphics[trim=0.0cm 0.1cm 0.1cm 0.1cm,scale=0.25,clip]{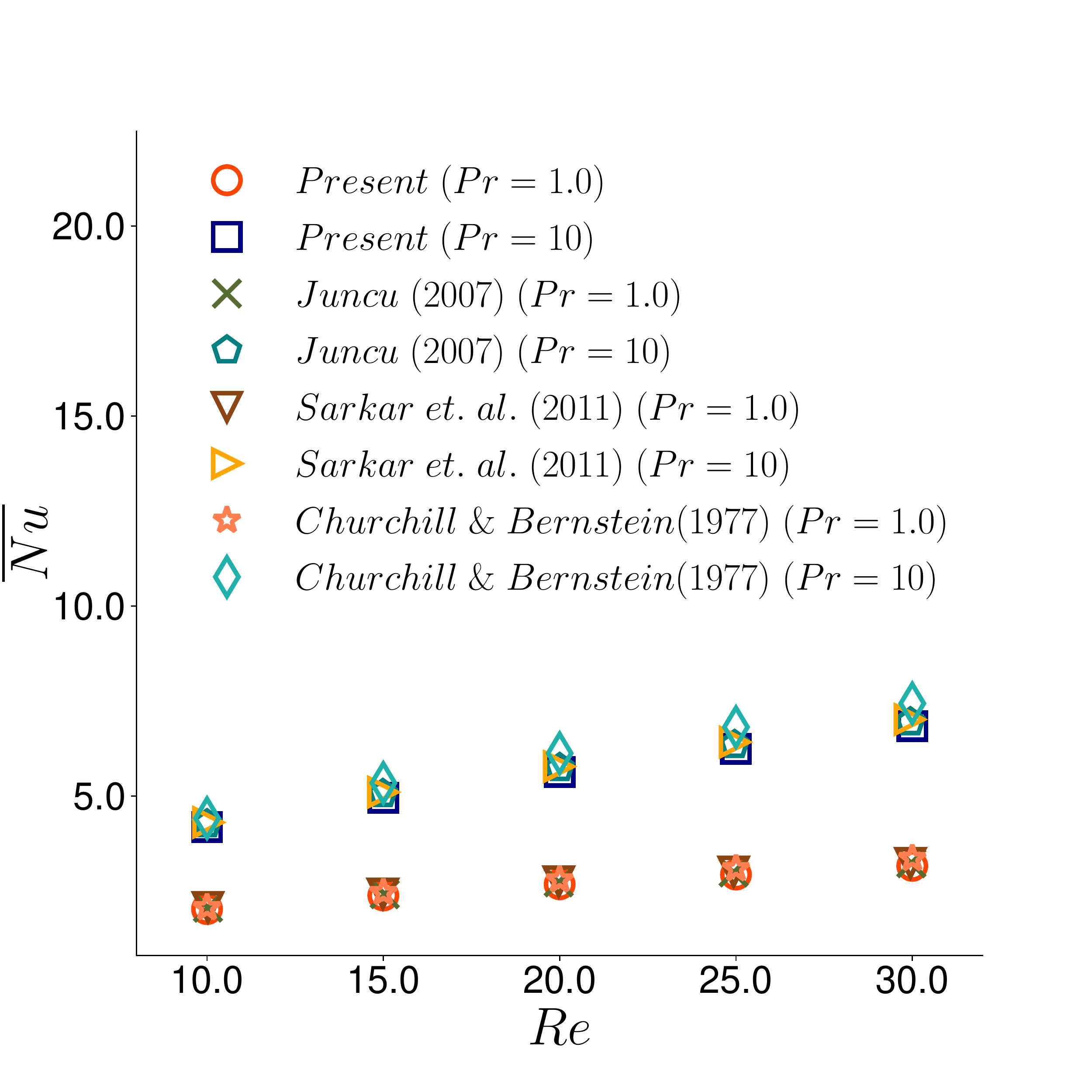}
	\caption{}
	\label{fig:val1}
\end{subfigure}%
\begin{subfigure}{0.5\textwidth}
	\centering
	\includegraphics[trim=0.0cm 0.1cm 0.1cm 0.1cm,scale=0.25,clip]{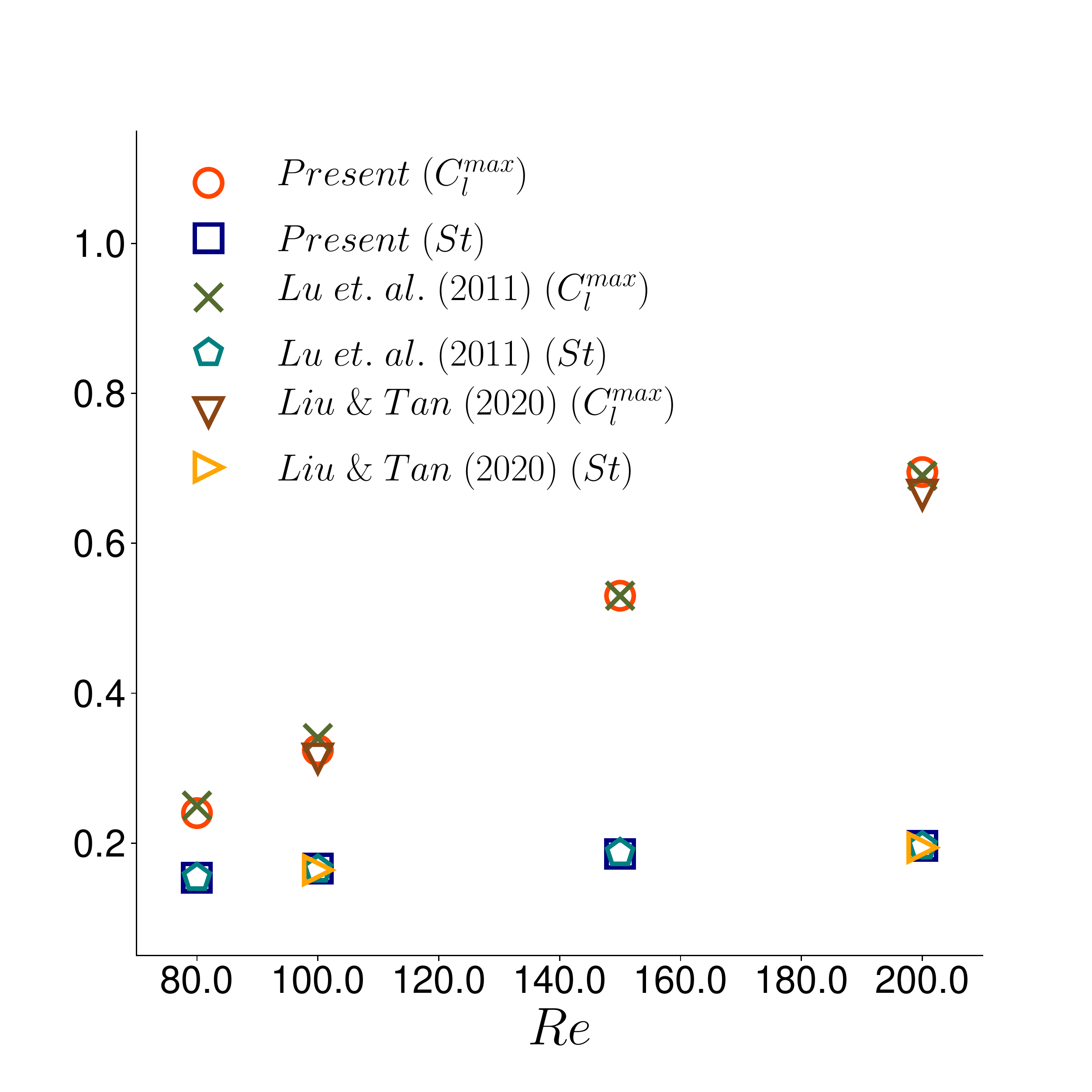}
	\caption{}
	\label{fig:val2}
\end{subfigure}
\begin{subfigure}{0.5\textwidth}
	\centering
	\includegraphics[trim=0.0cm 0.1cm 0.1cm 0.1cm,scale=0.25,clip]{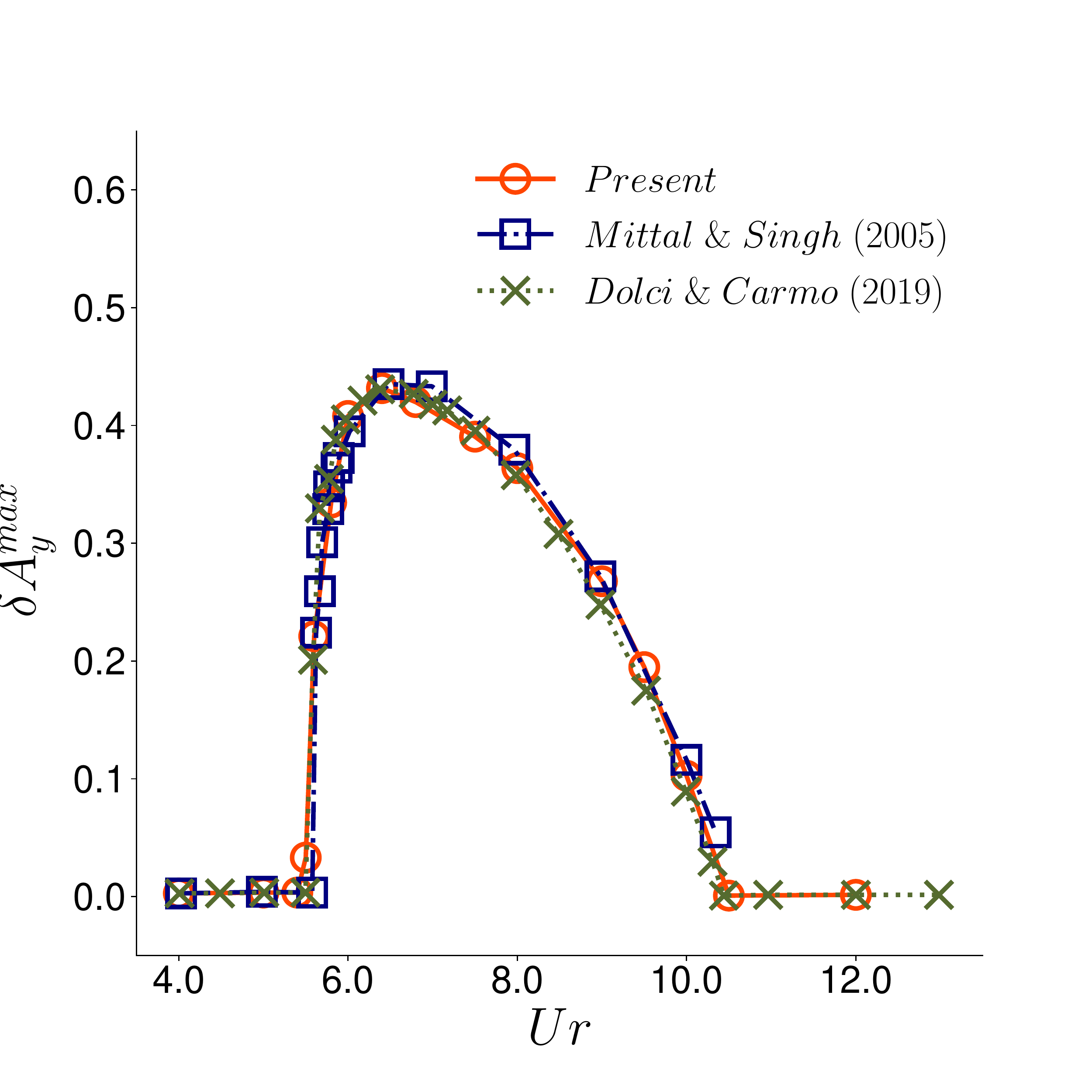}
	\caption{}
	\label{fig:val3}
\end{subfigure}%
\begin{subfigure}{0.5\textwidth}
	\centering
	\includegraphics[trim=0.0cm 0.1cm 0.1cm 0.1cm,scale=0.25,clip]{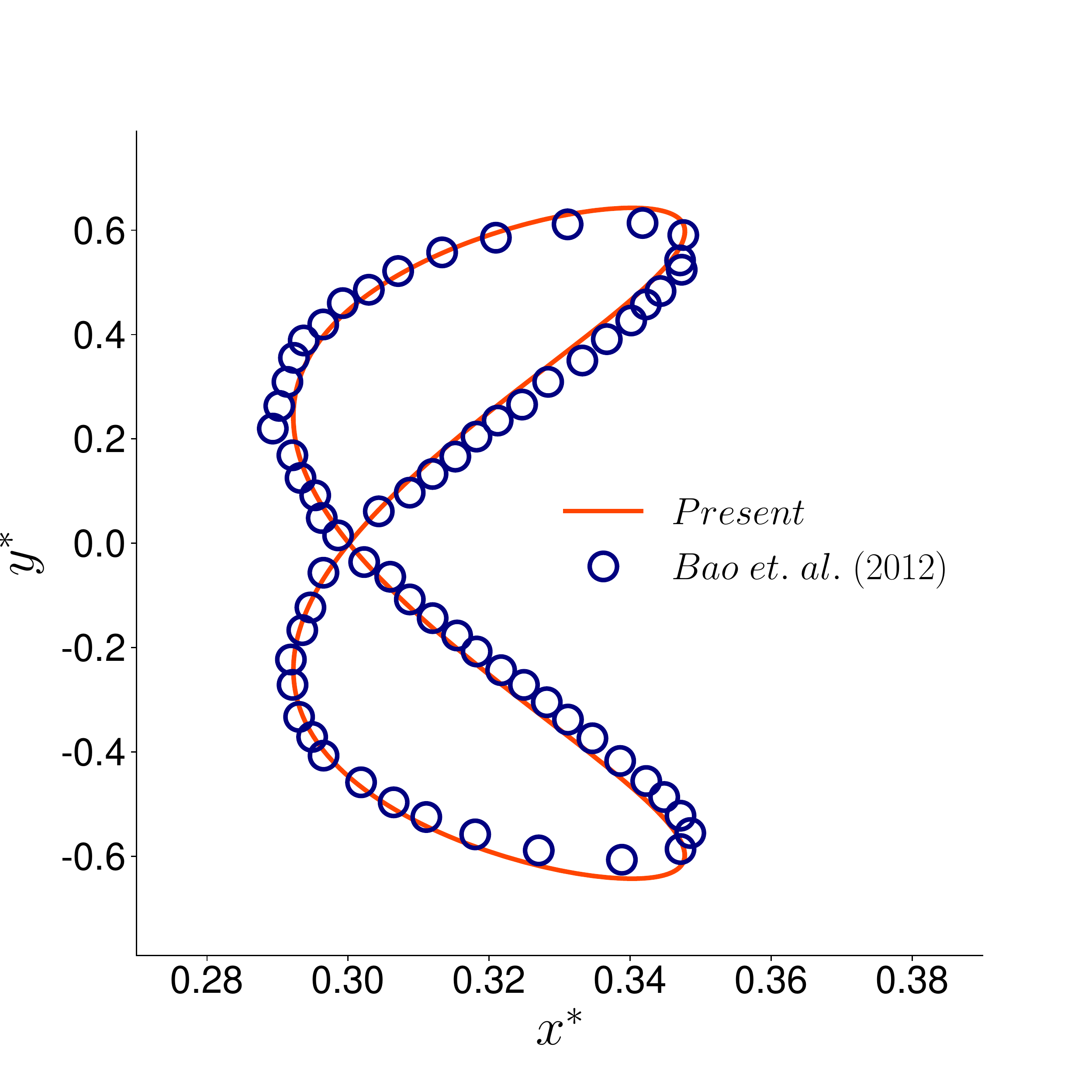}
	\caption{}
	\label{fig:val4}
\end{subfigure}
	\caption{Validation of the implemented numerical algorithm: (a) flow past a circular cylinder in forced convection flow; (b) flow past a circular cylinder in isothermal fluid flow; (c) flow past an elastically-mounted cylinder at $Re=33$, $m^*=4.73$ and $\zeta=0$ and (d) flow pas an elastically-mounted cylinder at $Re=150$, $m^*=2.55$, $\zeta=0$ and $Ur = 5.0$}
	\label{fig:val}
\end{figure*}
The mesh convergence analysis is carried out for simulation of flow over a circular cylinder in mixed convection flow at $Re=100$, $Pr=0.8$, $Ri = 1.0$ and $dt=0.01$. From Table~\ref{tab:mesh}, it is evident that the error of the hydrodynamic and thermal responses are within 1\% for the mesh resolution M2, where $\overline{C_d}$, $C^{rms}_l$, $St$ and $\overline{Nu}$ respectively are the mean drag coefficient, the rooted-mean-square lift coefficient, Strouhal number and the mean Nusselt number. Hence M2 is used for all simulations in this article. The time convergence analysis in Table~\ref{tab:time} shows that the time step $dt=0.01$ is optimal for the investigation of mixed convection flow, in which the errors of the hydrodynamic and thermal responses are within 1\%.
\begin{table}
	\caption{Flow past a circular cylinder in mixed convection flow at $Re=100$, $Pr = 0.7$, $Ri=1.0$, $dt=0.01$ and different mesh resolutions}
	\begin{ruledtabular}\label{tab:mesh}
		\begin{tabular}{lp{1.5cm}p{1cm}p{1cm}p{1cm}p{1cm}}
			MESH~\footnote{The value in the bracket refers to the number of nodes around the cylinder.} & NODES &$\overline{C_d}$&$C^{rms}_l$& $St$ & $\overline{Nu}$\\
			\hline
			M3 (128) & $2.6 \times 10^4$ & 1.300 (0.5\%) & 0.251 (1.9\%) & 0.175 (0.0\%)  & 5.106 (1.8\%)\\
			M2 (256) & $4.8 \times 10^4$ & 1.306 (0.0\%) & 0.255 (0.4\%) & 0.175 (0.0\%) & 5.191 (0.2\%)\\
			M1 (512) & $8.9 \times 10^4$ & 1.306 & 0.256 & 0.175 & 5.198 \\
		\end{tabular}
	\end{ruledtabular}
\end{table}
\begin{table}
	\caption{Flow past a circular cylinder in mixed convection flow at $Re=100$, $Pr = 0.7$, $Ri=1.0$, M2 (256) and different time steps}
	\begin{ruledtabular} \label{tab:time}
		\begin{tabular}{lp{1cm}p{1cm}p{1cm}p{1cm}}
			TIME STEP &$\overline{C_d}$&$C^{rms}_l$&$St$ & $\overline{Nu}$\\
			\hline
			$dt=0.020$ & 1.287 (1.5\%) & 0.236 (8.0\%) & 0.175 (0.0\%) & 5.096 (2.2\%)\\
			$dt=0.010$ & 1.306 (0.0\%)  & 0.255 (0.8\%) & 0.175 (0.0\%)  & 5.191 (0.4\%) \\
			$dt=0.005$ & 1.307  & 0.257 & 0.175  & 5.211 \\
		\end{tabular}
	\end{ruledtabular}
\end{table}

Subsequently, the derived numerical formulation is validated with the results in literature. It can be seen that the obtained mean Nusselt numbers from the derived formulation for heat convection flow match well with the literature~\cite{Juncu2007,sarkar2011unsteady} and the empirical formula~\cite{churchill1977correlating}, as shown in Figure~\ref{fig:val1}. The derived formulation for the unsteady, incompressible and isothermal fluid flow is validated with literature~\cite{lu2011numerical,liu2020nitsche} by comparing the hydrodynamic forces in Figure~\ref{fig:val2}. The structural dynamics obtained from the implemented ALE formulation is validated by comparing the values of the dimensionless transverse fluctuation in literature~\cite{mittal2005vortex,dolci2019bifurcation,bao2012two}. The maximum values of a dimensionless quantity are defined based on its root-mean-square and mean values. For instance,
\begin{subequations} \label{eq:max}
	\begin{align}
	A^{max}_y &= \sqrt{2}A^{rms}_y + \overline{A_y}; \quad \delta A^{max}_y = \sqrt{2}A^{rms}_y \\
	A^{rms}_y &= \sqrt{\frac{\sum \limits^n_i \big(A_y(i)-\overline{A_y}\big)^2}{n}}
	\end{align}
\end{subequations}
where $A^{max}$, $\delta A^{max}_y$, $A^{rms}_y$ and $\overline{A_y}$ are the maximum transverse displacement, the maximum transverse fluctuation, the root-mean-square transverse fluctuation and the mean transverse displacement of the cylinder respectively. The value of $n$ is the total number of sampled data. Figure~\ref{fig:val3} shows good agreement of the VIV lock-in responses with literature. Figure~\ref{fig:val4} demonstrates the typical figure-eight trajectory obtained from the derived numerical formulation for the freely-vibrating cylinder in the isothermal cross flow.

\section{Results and discussion} \label{sec:results}
As the flow past a heated cylinder, the thermal action predominantly takes place in wake. The heat energy is rapidly transferred from the cylinder to wake downstream and perturbs the shear-layer interactions, due to the presence of large temperature gradients. Consequently, the structural dynamics of the cylinder is altered, which reversely affect temperature distribution in near wake too. In mixed convection flow, the influence of heat transfer is primarily controlled by two parameters, Prandtl and Richardson numbers. By varying Prandtl number, we know how diffusive is the heat energy field with respect to fluid momentum. On the other hand, intuitively speaking, Richardson number indicates how much the fluid momentum transportation is perturbed by heat energy field. In this section, we primarily focus on the presentation and discussion of the complex interaction and regimes between fluid, structure and heat transfer for different Prandtl and Richardson numbers, e.g., $Pr \in [0.7, 10]$ and $Ri \in [0.5, 2.0]$, where the forced convection is relatively stronger than natural convection.   

\subsection{structural dynamics in mixed convection flow} \label{sec:struct}
\begin{figure*}
	\centering
	\begin{subfigure}{0.5\textwidth}
	\centering
	\includegraphics[trim=0.0cm 0.1cm 0.1cm 0.1cm,scale=0.25,clip]{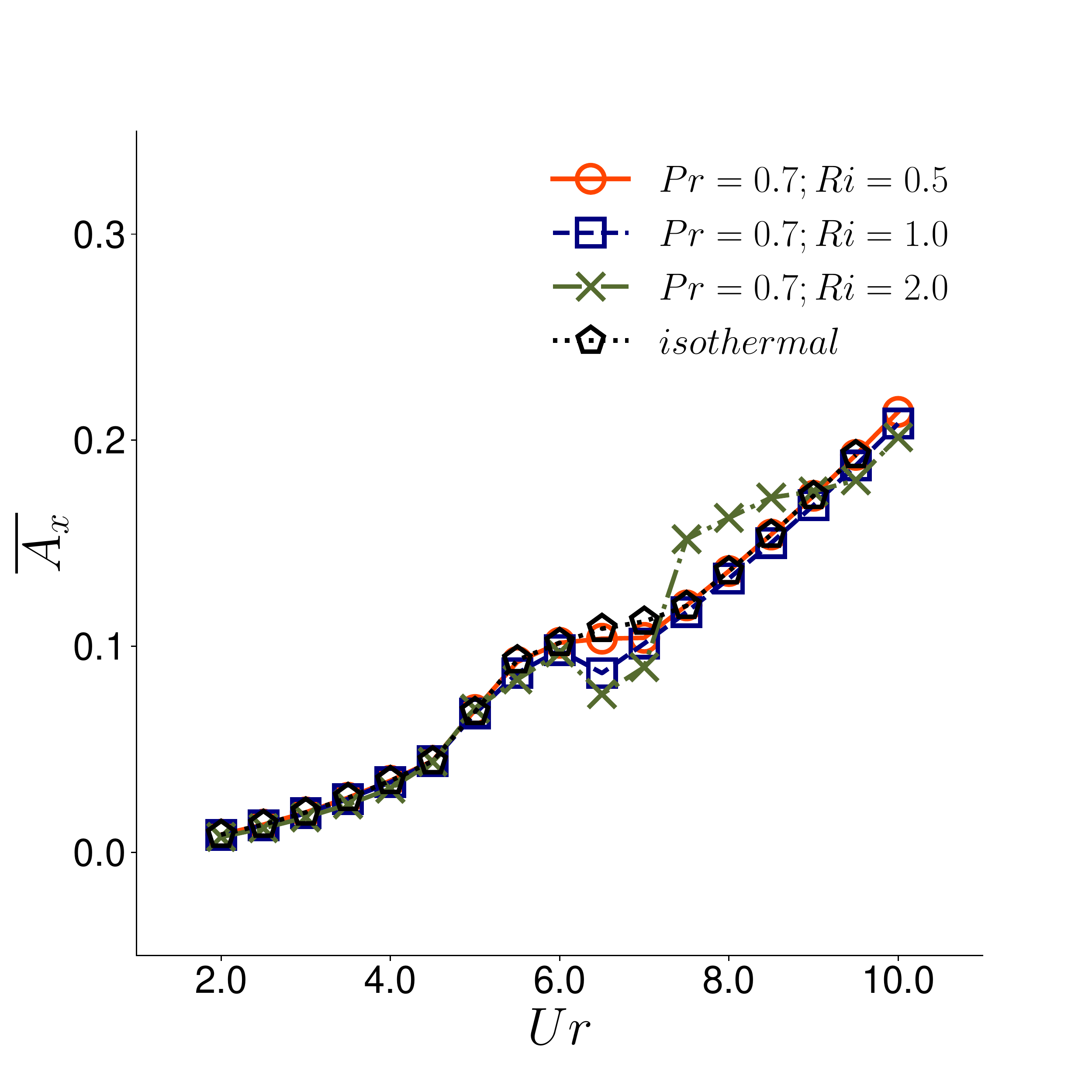}
	\caption{}
	\label{fig:Ax1}
\end{subfigure}%
\begin{subfigure}{0.5\textwidth}
	\centering
	\includegraphics[trim=0.0cm 0.1cm 0.1cm 0.1cm,scale=0.25,clip]{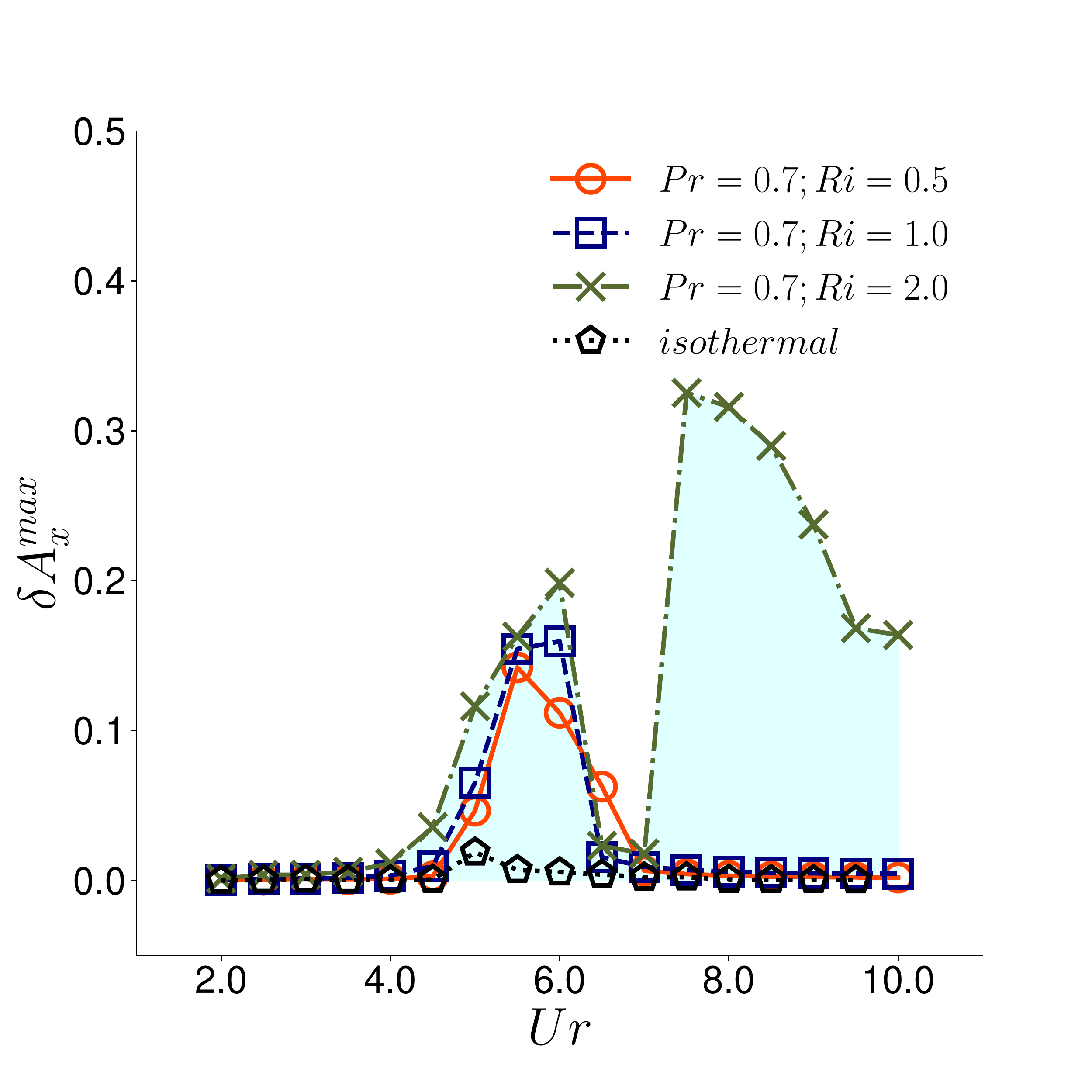}
	\caption{}
	\label{fig:Ax2}
\end{subfigure}
\begin{subfigure}{0.5\textwidth}
	\centering
	\includegraphics[trim=0.0cm 0.1cm 0.1cm 0.1cm,scale=0.25,clip]{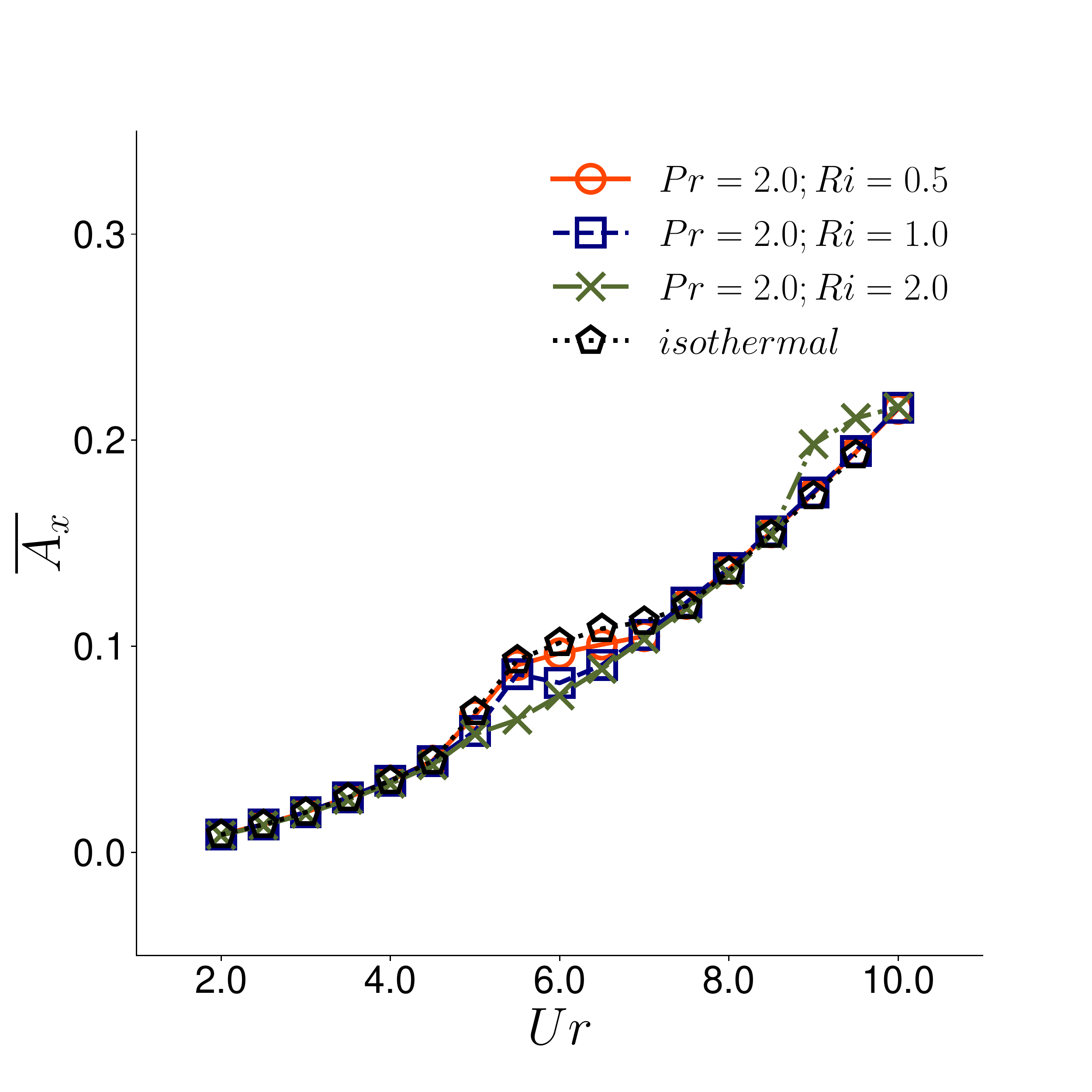}
	\caption{}
	\label{fig:Ax3}
\end{subfigure}%
\begin{subfigure}{0.5\textwidth}
	\centering
	\includegraphics[trim=0.0cm 0.1cm 0.1cm 0.1cm,scale=0.25,clip]{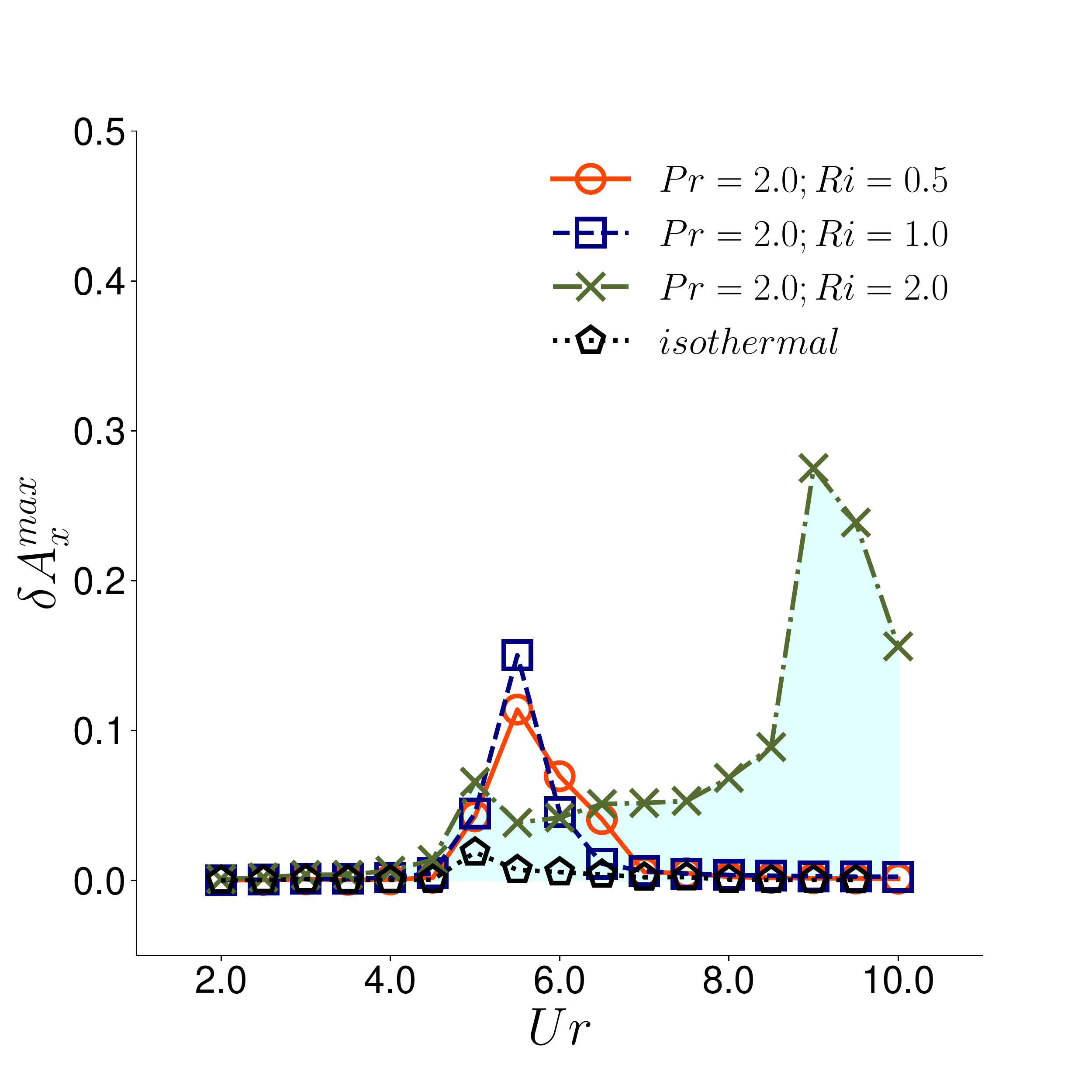}
	\caption{}
	\label{fig:Ax4}
\end{subfigure}
\begin{subfigure}{0.5\textwidth}
	\centering
	\includegraphics[trim=0.0cm 0.1cm 0.1cm 0.1cm,scale=0.25,clip]{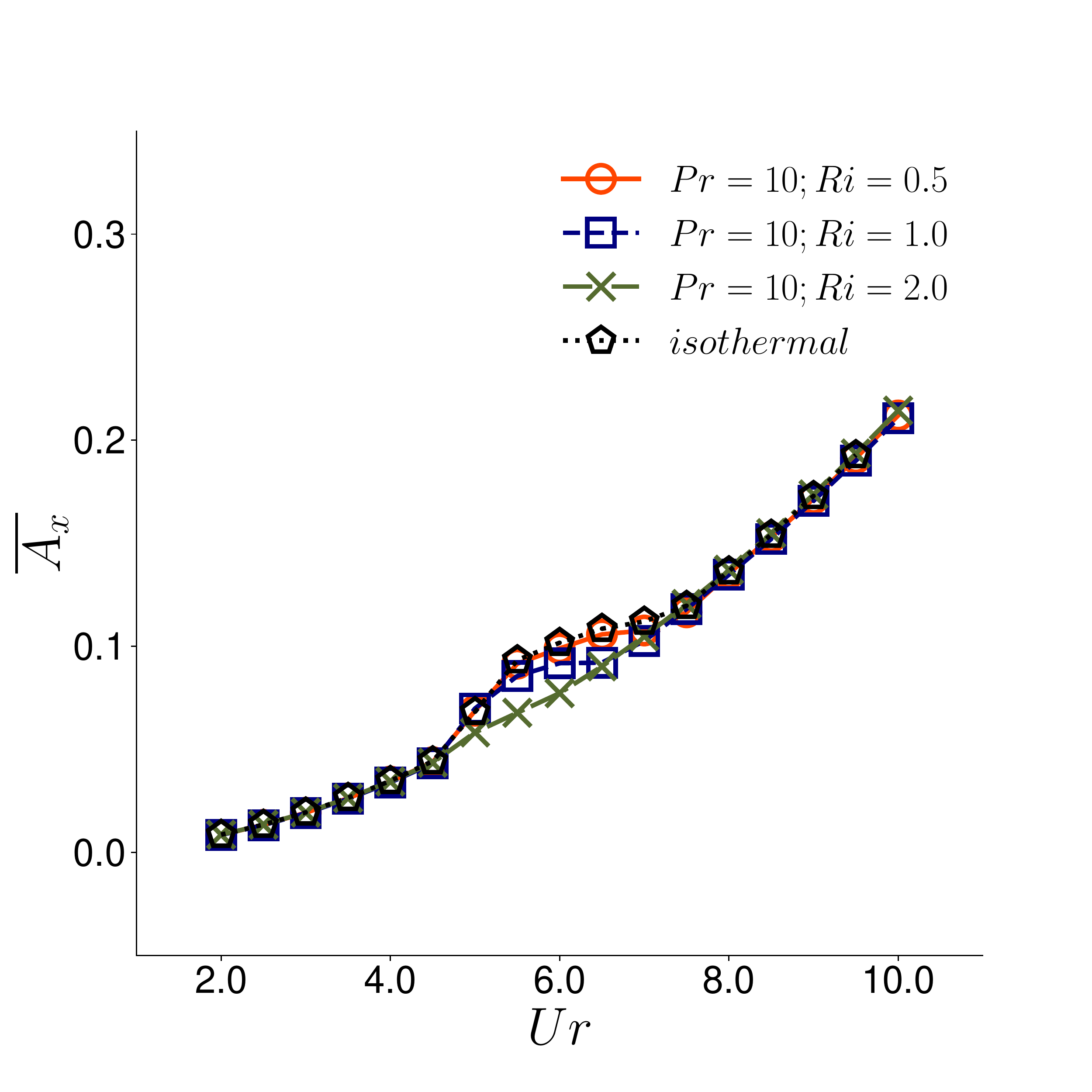}
	\caption{}
	\label{fig:Ax5}
\end{subfigure}%
\begin{subfigure}{0.5\textwidth}
	\centering
	\includegraphics[trim=0.0cm 0.1cm 0.1cm 0.1cm,scale=0.25,clip]{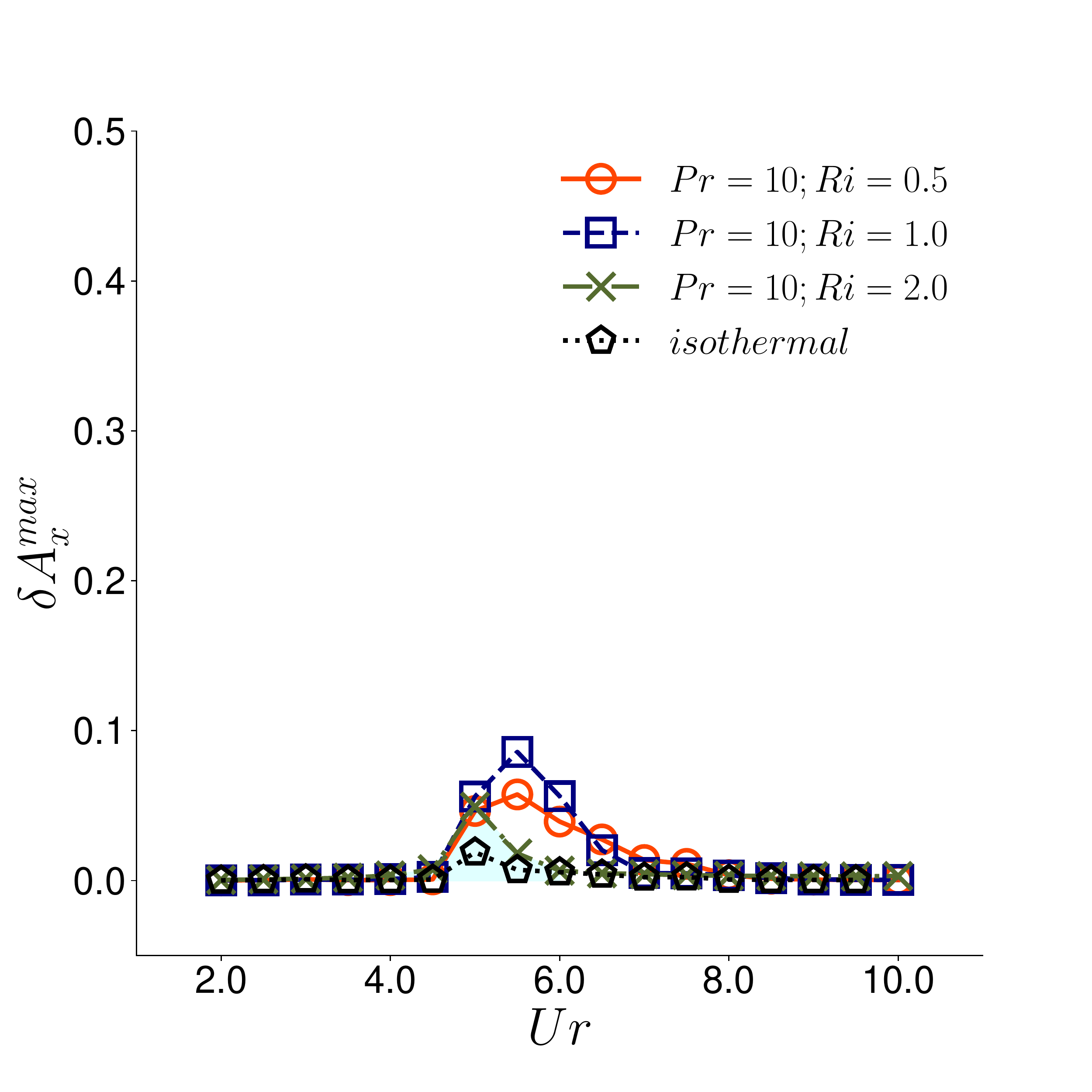}
	\caption{}
	\label{fig:Ax6}
\end{subfigure}
	\caption{Streamwise structural dynamics of a freely-vibrating cylinder at $Re=100$, $m^*=10$, $\zeta = 0.01$, $Ur \in [2, 10]$, $Pr \in [0.7, 10]$ and $Ri \in [0.5, 2.0]$: (a, c, e) the mean streamwise displacement and (b, d, f) the maximum streamwise fluctuation. The VIV lock-in regions for $Ri = 2.0$ are filled with \emph{light cyan} color.}
	\label{fig:Ax}
\end{figure*}
\begin{figure*}
	\centering
	\begin{subfigure}{0.5\textwidth}
	\centering
	\includegraphics[trim=0.0cm 0.1cm 0.1cm 0.1cm,scale=0.25,clip]{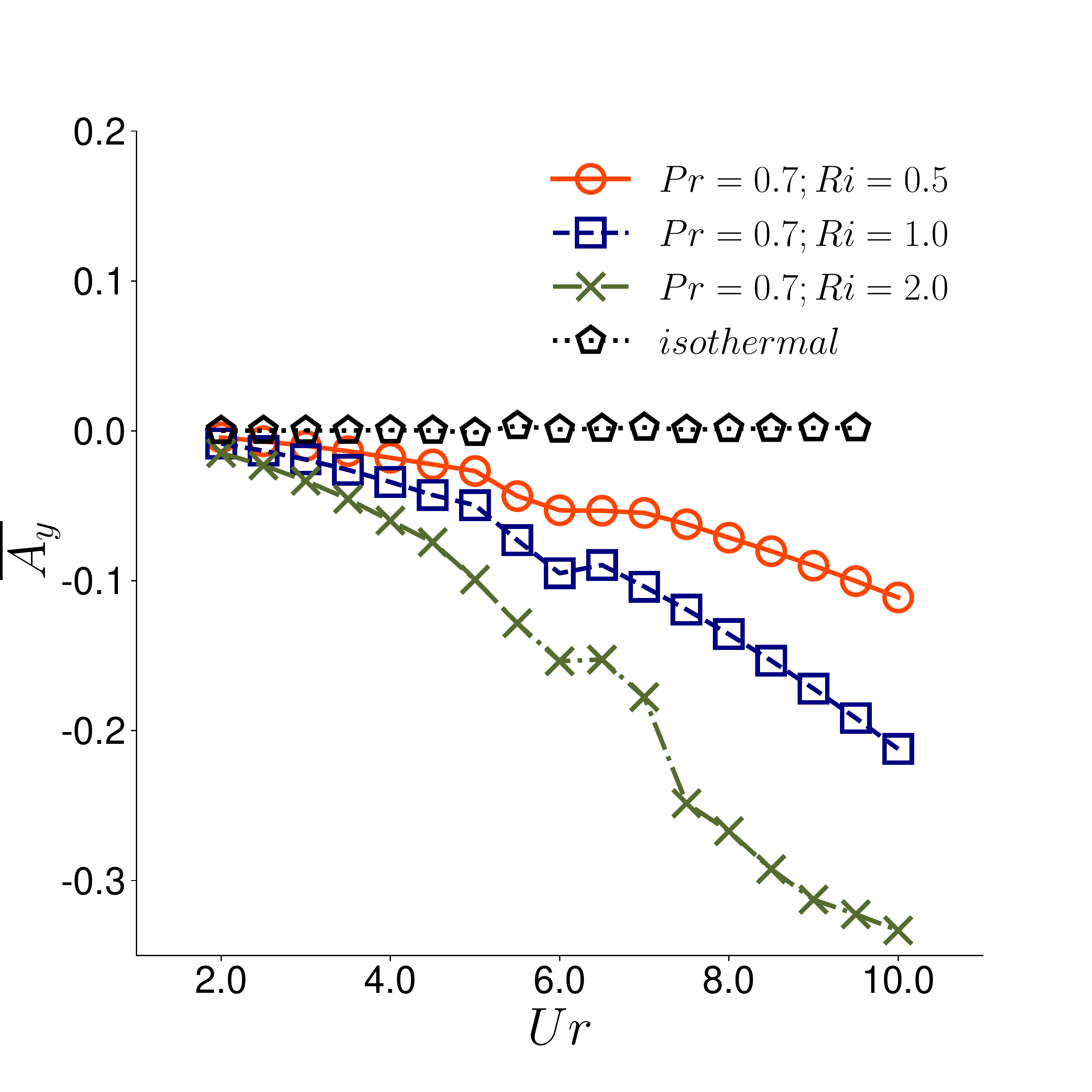}
	\caption{}
	\label{fig:Ay1}
\end{subfigure}%
\begin{subfigure}{0.5\textwidth}
	\centering
	\includegraphics[trim=0.0cm 0.1cm 0.1cm 0.1cm,scale=0.25,clip]{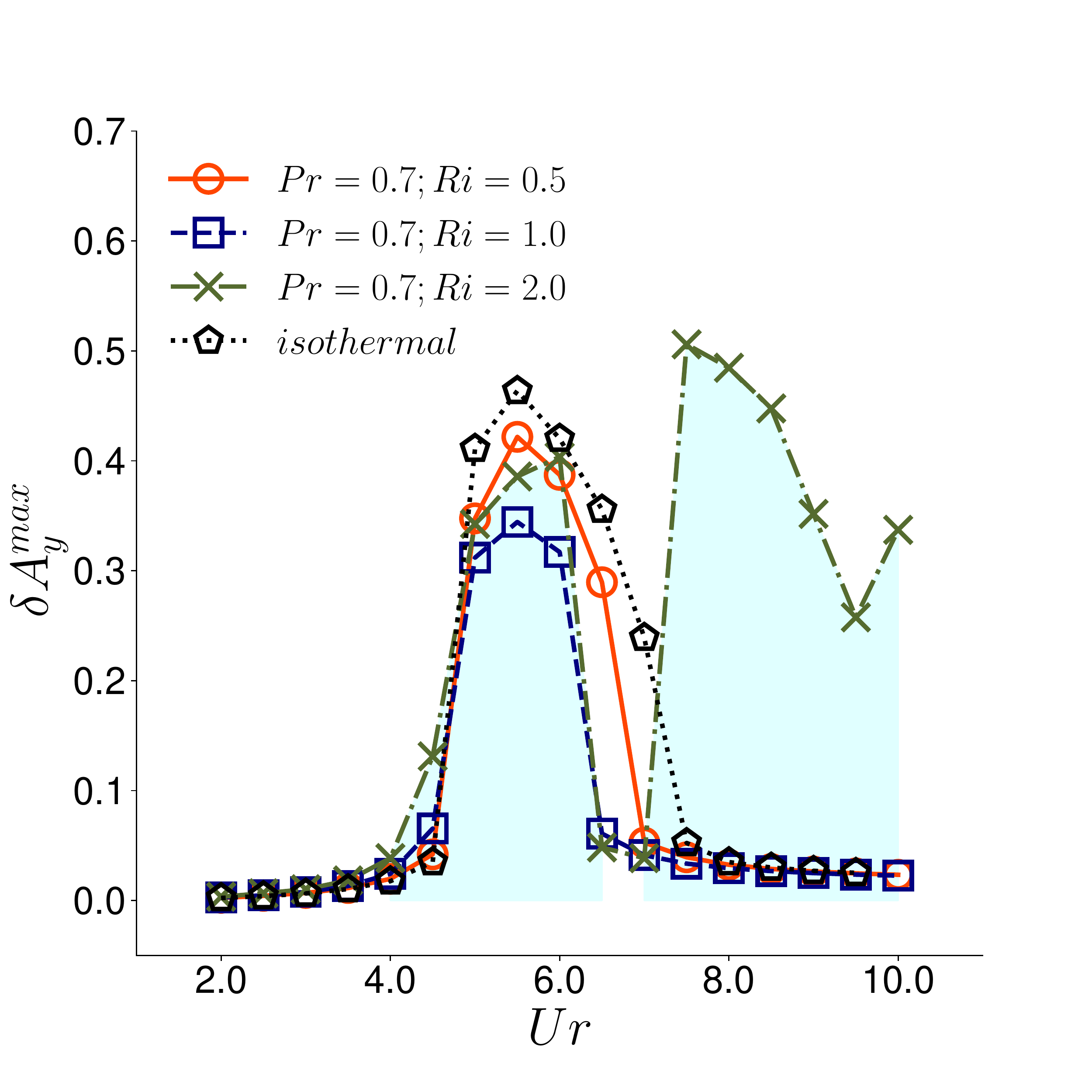}
	\caption{}
	\label{fig:Ay2}
\end{subfigure}
\begin{subfigure}{0.5\textwidth}
	\centering
	\includegraphics[trim=0.0cm 0.1cm 0.1cm 0.1cm,scale=0.25,clip]{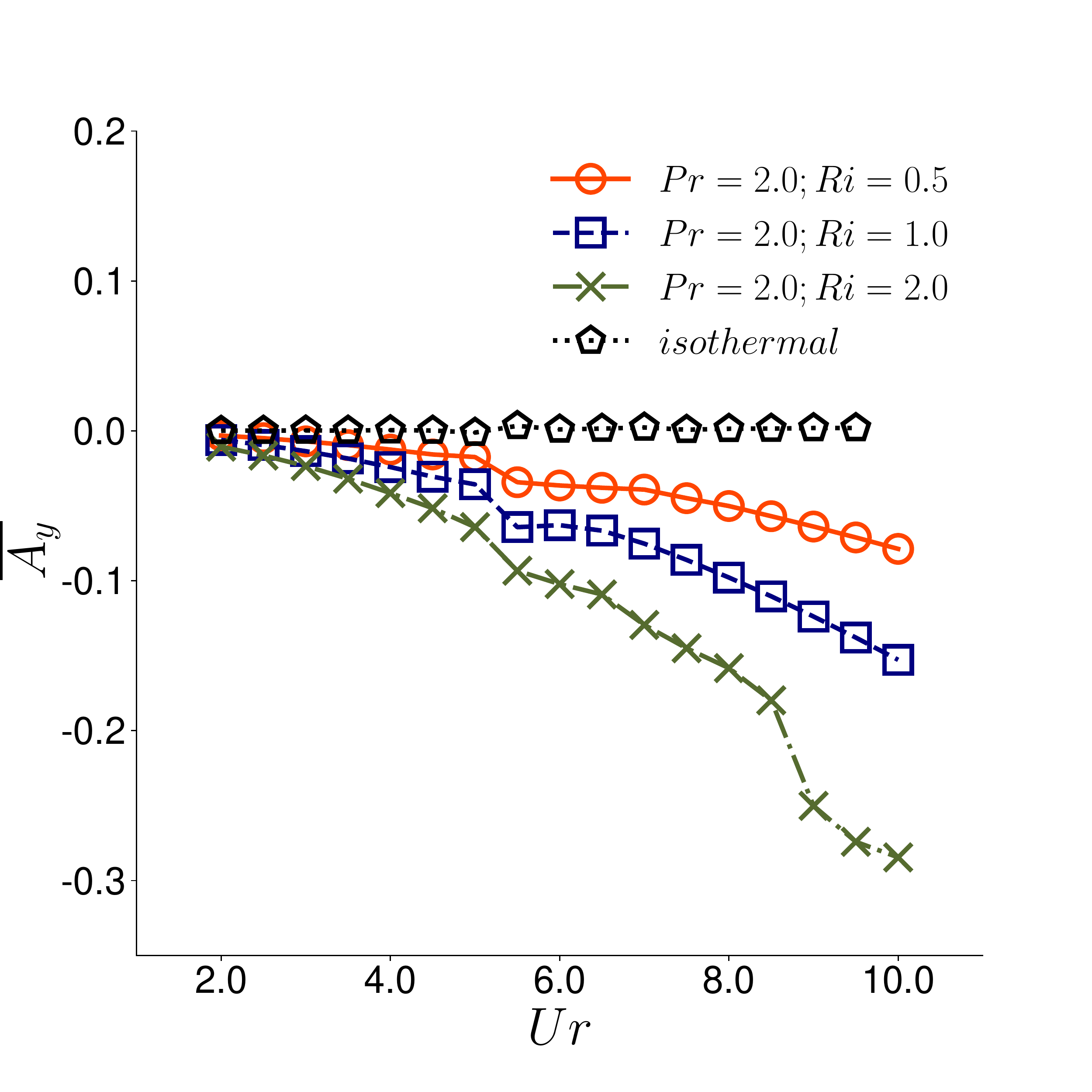}
	\caption{}
	\label{fig:Ay3}
\end{subfigure}%
\begin{subfigure}{0.5\textwidth}
	\centering
	\includegraphics[trim=0.0cm 0.1cm 0.1cm 0.1cm,scale=0.25,clip]{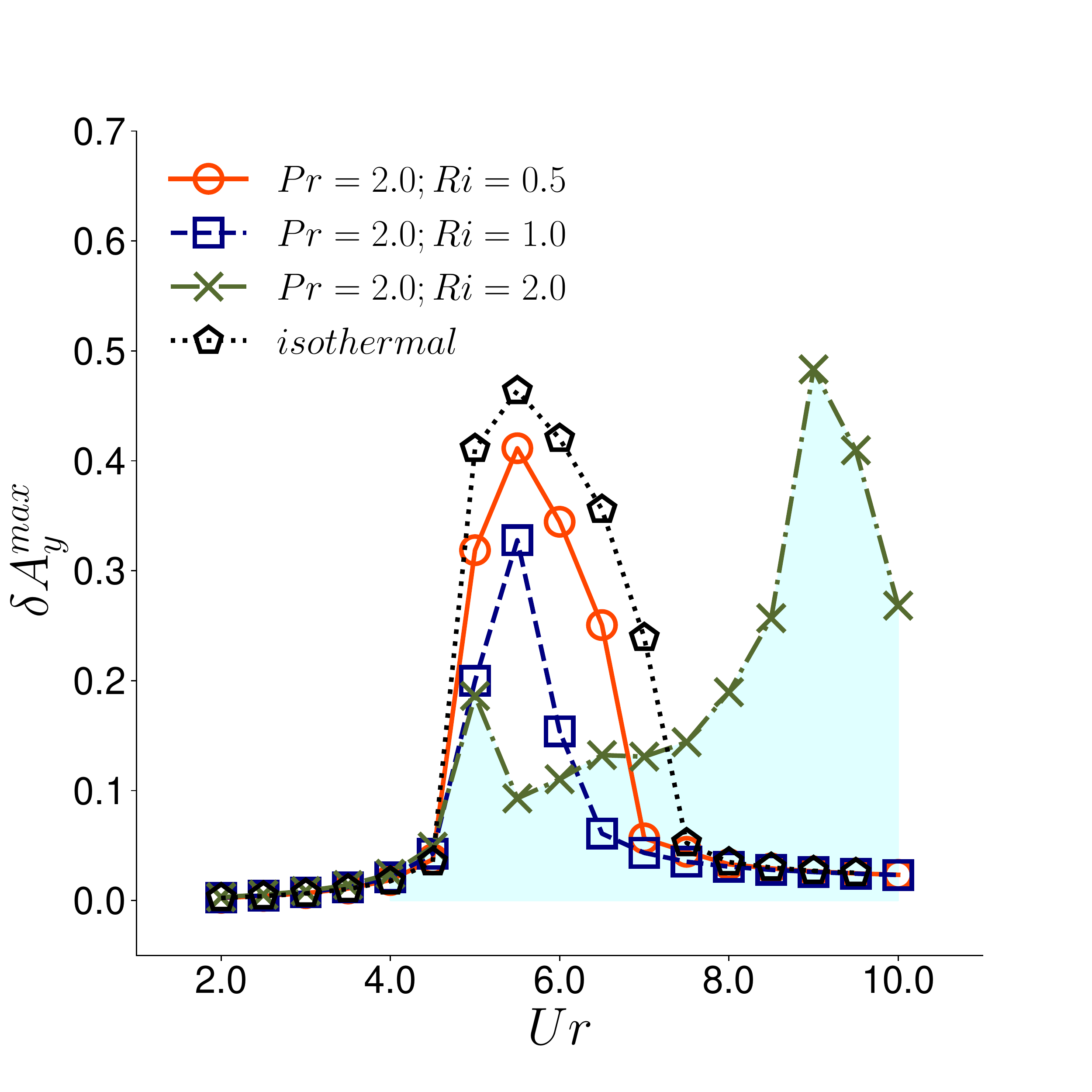}
	\caption{}
	\label{fig:Ay4}
\end{subfigure}
\begin{subfigure}{0.5\textwidth}
	\centering
	\includegraphics[trim=0.0cm 0.1cm 0.1cm 0.1cm,scale=0.25,clip]{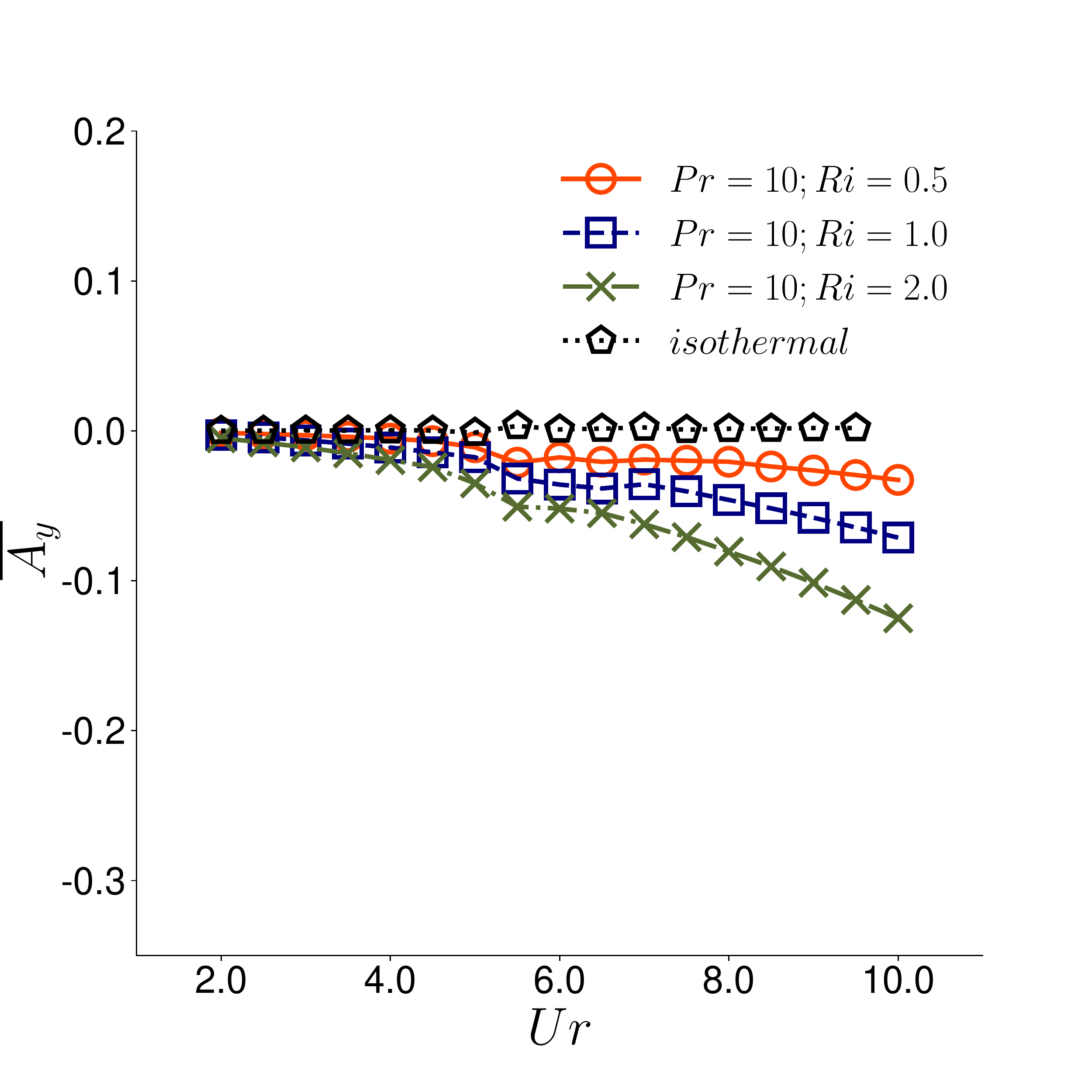}
	\caption{}
	\label{fig:Ay5}
\end{subfigure}%
\begin{subfigure}{0.5\textwidth}
	\centering
	\includegraphics[trim=0.0cm 0.1cm 0.1cm 0.1cm,scale=0.25,clip]{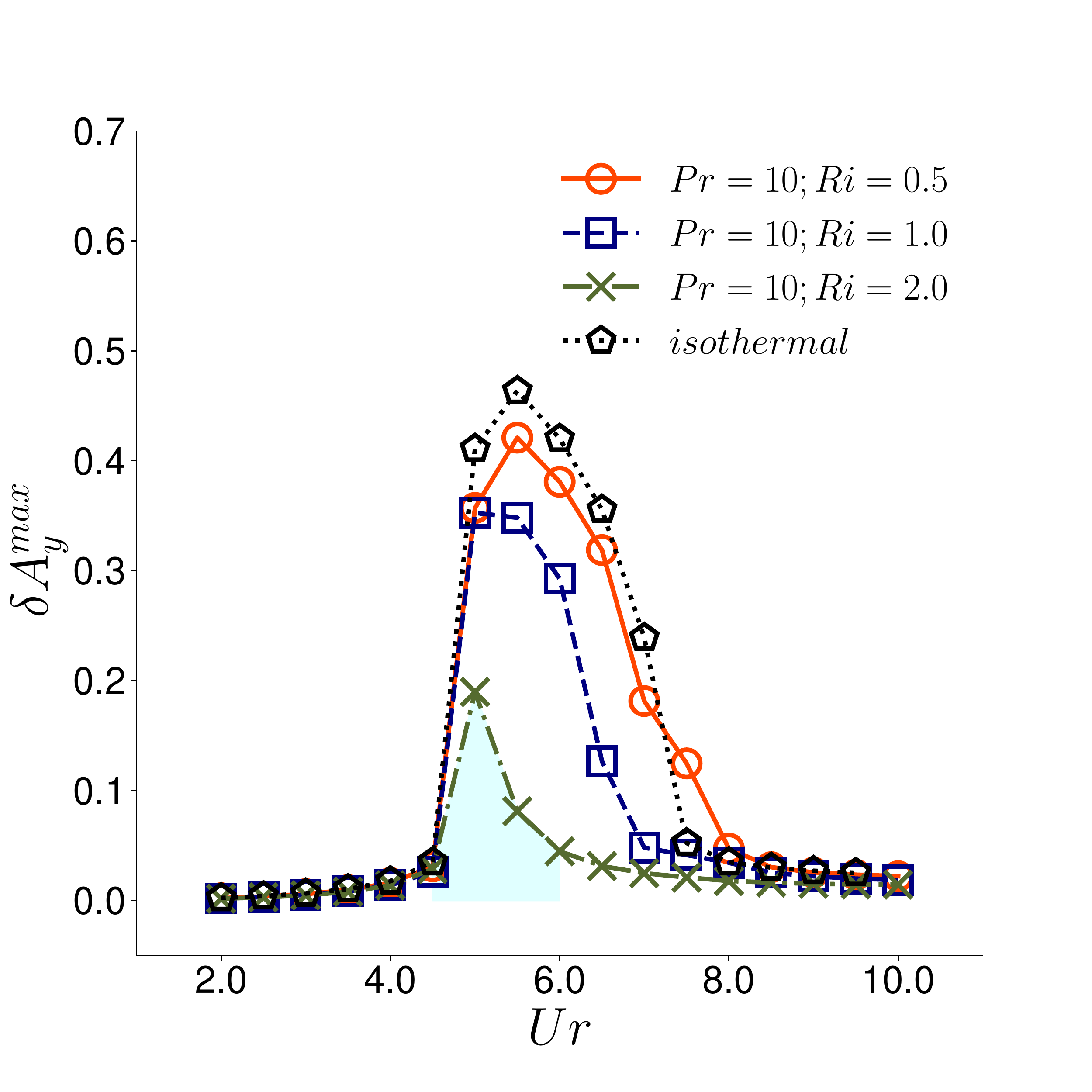}
	\caption{}
	\label{fig:Ay6}
\end{subfigure}
	\caption{Transverse structural dynamics of a freely-vibrating cylinder at $Re=100$, $m^*=10$, $\zeta = 0.01$, $Ur \in [2, 10]$, $Pr \in [0.7, 10]$ and $Ri \in [0.5, 2.0]$: (a, c, e) the mean transverse displacement and (b, d, f) the maximum transverse fluctuation.  The VIV lock-in regions for $Ri = 2.0$ are filled with \emph{light cyan} color.}
	\label{fig:Ay}
\end{figure*}
\begin{figure*}
	\centering
	\begin{subfigure}{0.5\textwidth}
	\centering
	\includegraphics[trim=0.0cm 0.1cm 0.1cm 0.1cm,scale=0.25,clip]{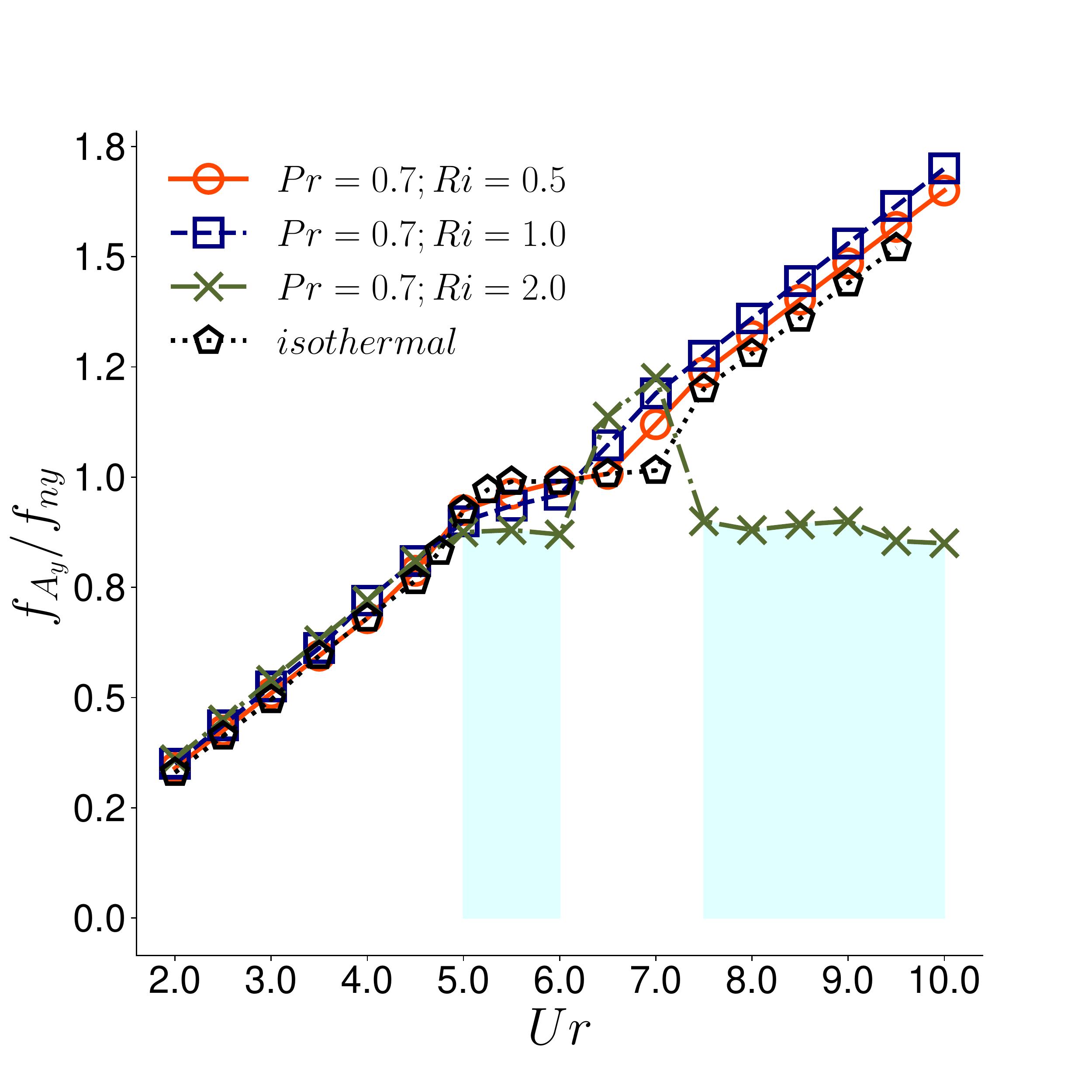}
	\caption{}
	\label{fig:freq1}
\end{subfigure}%
\begin{subfigure}{0.5\textwidth}
	\centering
	\includegraphics[trim=0.0cm 0.1cm 0.1cm 0.1cm,scale=0.25,clip]{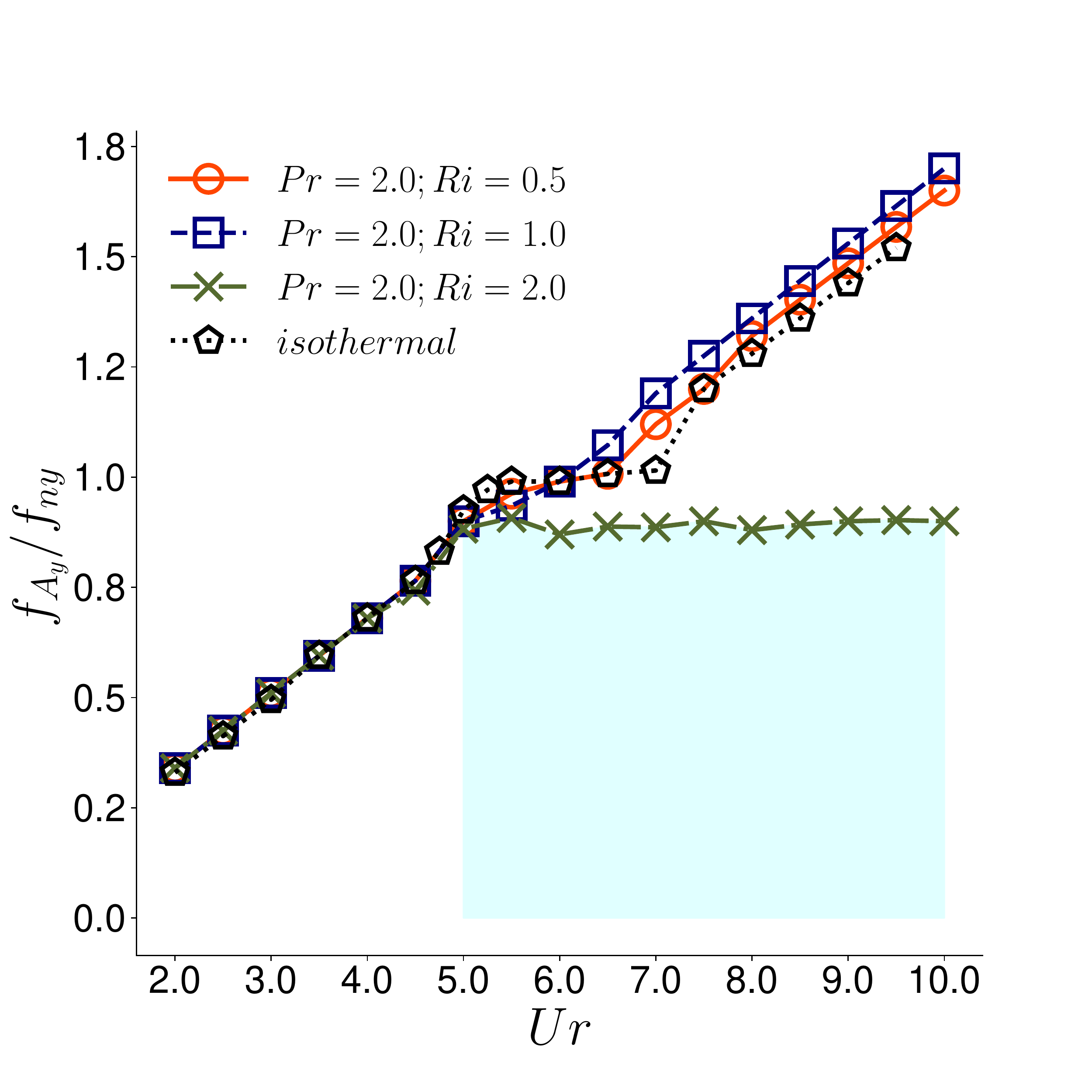}
	\caption{}
	\label{fig:freq2}
\end{subfigure}
\begin{subfigure}{0.5\textwidth}
	\centering
	\includegraphics[trim=0.0cm 0.1cm 0.1cm 0.1cm,scale=0.25,clip]{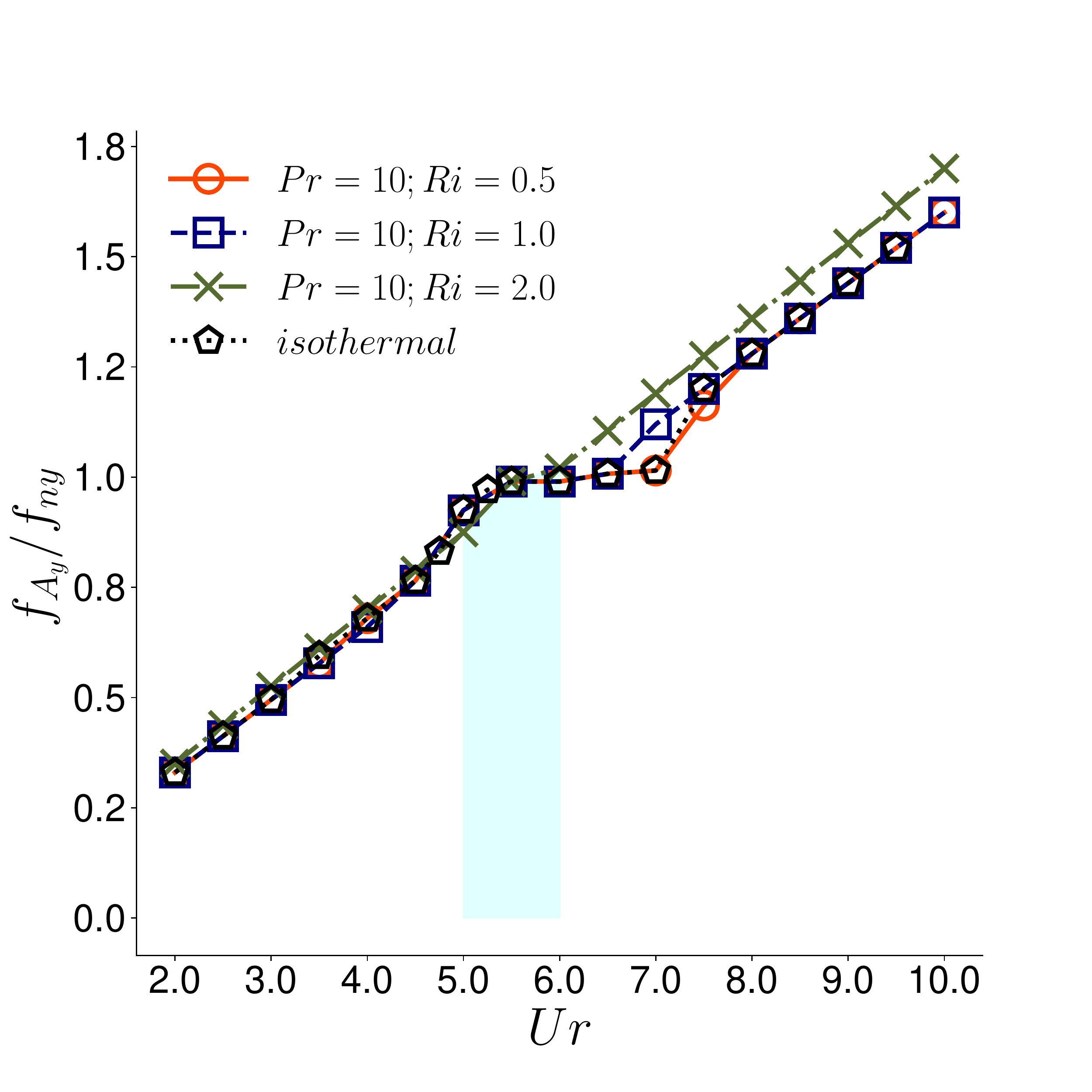}
	\caption{}
	\label{fig:freq3}
\end{subfigure}
	\caption{Frequency ratio of a freely-vibrating cylinder at $Re=100$, $m^*=10$, $\zeta = 0.01$ and $Ur \in [2, 10]$: (a) $Pr = 0.7$ and $Ri \in [0.5, 2.0]$; (b) $Pr = 2.0$ and $Ri \in [0.5, 2.0]$ and (c) $Pr = 10$ and $Ri \in [0.5, 2.0]$. The VIV lock-in regions for $Ri = 2.0$ are filled with \emph{light cyan} color.}
	\label{fig:freq}
\end{figure*}
First of all, we start with the streamwise motions of the freely-vibrating circular cylinder. In Figure~\ref{fig:Ax}, it can be seen the change of mean streamwise displacements ($\overline{A_{x}}$) with respect to reduced velocity ($Ur$) is insensitive to different combinations of $Pr$ and $Ri$ values in mixed convection flow, except the relative amplification, about 40\%, during VIV lock-in. The values of $\overline{A_x}$ surges proportionally with $Ur$ values. However, the maximum streamwise fluctuations ($\delta A^{max}_x$) behave acutely different for various combination of $Pr$ and $Ri$ values. The maximum streamwise fluctuations in mixed convection flow is significantly large compared those in isothermal flow for VIV lock-in. However, as Prandtl number arises, the values of $\delta A^{max}_x$ for VIV lock-in decreases and almost diminishes in the case of $Pr = 10$ and $Ri = 2.0$ in Figure~\ref{fig:Ax6}. It means the slow dissipation of heat energy with respect fluid momentum (high Prandtl number) limits the influence of heat energy field to hydrodynamics in near wake. Consequently the structural dynamics approach to those in isothermal flow. This conclusion is further supported by the subsequent analyses of the transverse structural motions and numerical results in Sections~\ref{sec:hydro} and~\ref{sec:thermal}. On the other hand, the increment of Richardson number means a stronger influence of the buoyancy force on the fluid momentum transportation and simultaneously affects the structural dynamics in transverse direction. In particular, it is observed in Figure~\ref{fig:Ax}(b,d,f) that the VIV lock-in regions are further widened in the cases of $Ri = 2.0$, until much high reduced velocity values, e.g., $Ur \approx 10$. In Figure~\ref{fig:Ax2}, a secondary VIV lock-in region occurs in the case of $Pr = 0.7$ and $Ri= 2.0$ for  $Ur \in (7.0, 10)$. The formation of this secondary VIV lock-in region highlights the influence of strong buoyancy force on the structural dynamics at high $Ur$ values. On the contrary, the structural dynamics is found insensitive to the buoyancy force for the pre-lock-in regions. Furthermore, Figure~\ref{fig:Ax4} also shows a very wide VIV lock-in region formed by coalescence of the primary and secondary VIV lock-in regions in the case of $Pr = 2.0$ and $Ri = 2.0$, the dotted green curve with \emph{cross} markers. As Prandtl number further increases, the VIV lock-in region gets eventually suppressed at $Ri = 2.0$ over a wide range of reduced velocity values, e.g., $Ur \in [0, 10]$.     
   
Similarly, the transverse motions of cylinder are plotted in Figure~\ref{fig:Ay}. Unlike the streamwise displacements, the difference of mean transverse displacements ($\overline{A}_y$) between the mixed convection and isothermal flows is enlarged as the reduced velocity value increases in Figure~\ref{fig:Ay}(a, c, e). Especially, this difference is overall large in the cases of higher Richardson numbers for the same $Ur$ value. Nevertheless, similar to the maximum streamwise fluctuation ($\delta A^{max}_x$), the influence of heat energy field on hydrodynamics for VIV lock-in reduces in the cases of higher Prandtl numbers. As Prandtl number keeps increasing, the flow-induced structural dynamics becomes much more close to those in isothermal flow, except for high $Ur$ and $Ri$ values. The secondary VIV lock-in region is again confirmed in the plots of the transverse motions for high reduced velocity values in Figure~\ref{fig:Ay2} and Figure~\ref{fig:Ay4}. The maximum transverse fluctuation becomes significantly excited for $Ur \approx 5.5$ and $Ur \approx 9.0$ respectively. The primary and secondary VIV lock-in regions coalesce in the case of $Pr = 2.0$ and $Ri = 2.0$, striding over a wide range of $Ur$ values in Figure~\ref{fig:Ay4}. It is also noteworthy that the peak of the secondary VIV lock-in region is shifted further higher reduced velocity values until $Ur \approx 9.25$ and the peak transverse fluctuation during the primary VIV lock-in is simultaneously suppressed, e.g., $\delta A^{max}_y \approx 0.2$ in Figure~\ref{fig:Ay4}. This shift of peak value in secondary VIV lock-in is further confirmed in the subsequent analyses of hydrodynamic forces in Section~\ref{sec:hydro} and energy transfer in Section~\ref{sec:thermal}. If both Prandtl and Richardson numbers are high in mixed convection flow, e.g., the dotted green curve with \emph{cross} markers in Figure~\ref{fig:Ay6}, the secondary VIV lock-in region even disappears in the range of $Ur \in [2.0, 10]$ and the peak transverse fluctuation during primary VIV lock-in region is remarkably diminished. In a nutshell, the hydrodynamics and structural dynamics in the cases of high $Pr$ and low $Ri$ values are close to those in isothermal flow, which represent the flow regimes with slow heat energy dissipation and less influence of buoyancy-driven flow. In contrast, for high $Ri$ values, e.g., $Ri \approx 2.0$ in this study, a secondary VIV lock-in region could be induced, which shifts its location with respect to the $Ur$ values for different $Pr$ numbers. It is found the buoyancy-driven flow has a severe impact on the structural dynamics for high reduced velocity values, especially the behavior of VIV lock-in. 

In the frequency domain, VIV lock-in regions can be identified distinctly. During VIV lock-in, the frequency of transverse vibration induced by lift force ($f_{A_y}$) is locked with the structural frequency. By plotting the frequency ratio ($f_{A_y}/f_{ny}$) of the transverse direction with respect to reduced velocity, the VIV lock-in region for isothermal flow could be identified apparently as the dotted back line with \emph{pentagon} markers in Figure~\ref{fig:freq1} for $Ur \in [5.0, 7.0]$. Correspondingly, the VIV lock-in regions in mixed convection flow become narrower for $Pr = 0.7$ and $Ri \in [0.5, 2.0)$. As Richardson number increases further, the primary VIV lock-in region become even narrow. Again, the aforementioned secondary VIV lock-in region is evidently confirmed in the case of $Pr = 0.7$ and $Ri = 2.0$ for $Ur = [7.5,10]$, the region filled with \emph{light cyan} color in Figure~\ref{fig:freq1}. For higher Prandtl number ($Pr = 2.0$), the coalescence of primary and secondary VIV lock-in regions is also confirmed in the frequency domain, as shown by the \emph{light cyan} region for $Ur > 5.0$ in Figure~\ref{fig:freq2}. This observation is extremely meaningful for the structural dynamics, since VIV lock-in is an indication of intensive vibrations and high energy transfer between fluid and structure. It implies that the interference from buoyancy force becomes evident and detrimental to the structural stability for low $Pr$ and high $Ri$ values. In contrast, similar to the observations in Figure~\ref{fig:Ax} and Figure~\ref{fig:Ay}, the hydrodynamic and structural responses of a freely-vibrating cylinder becomes more close to those in isothermal flow in the cases of high $Pr$ and low $Ri$ values, instead. Figure~\ref{fig:freq3} shows the frequency ratios for $Ri = 0.5$ and $1.0$ behave almost identical to the responses in isothermal flow, except the case of $Ri = 2.0$, in which the VIV lock-in is almost suppressed by the strong interference of buoyancy force.   

\begin{figure}[!htp]
	\centering
	\begin{subfigure}{0.5\textwidth}
	\centering
	\includegraphics[trim=2cm 0.1cm 0.1cm 0.1cm,scale=0.4,clip]{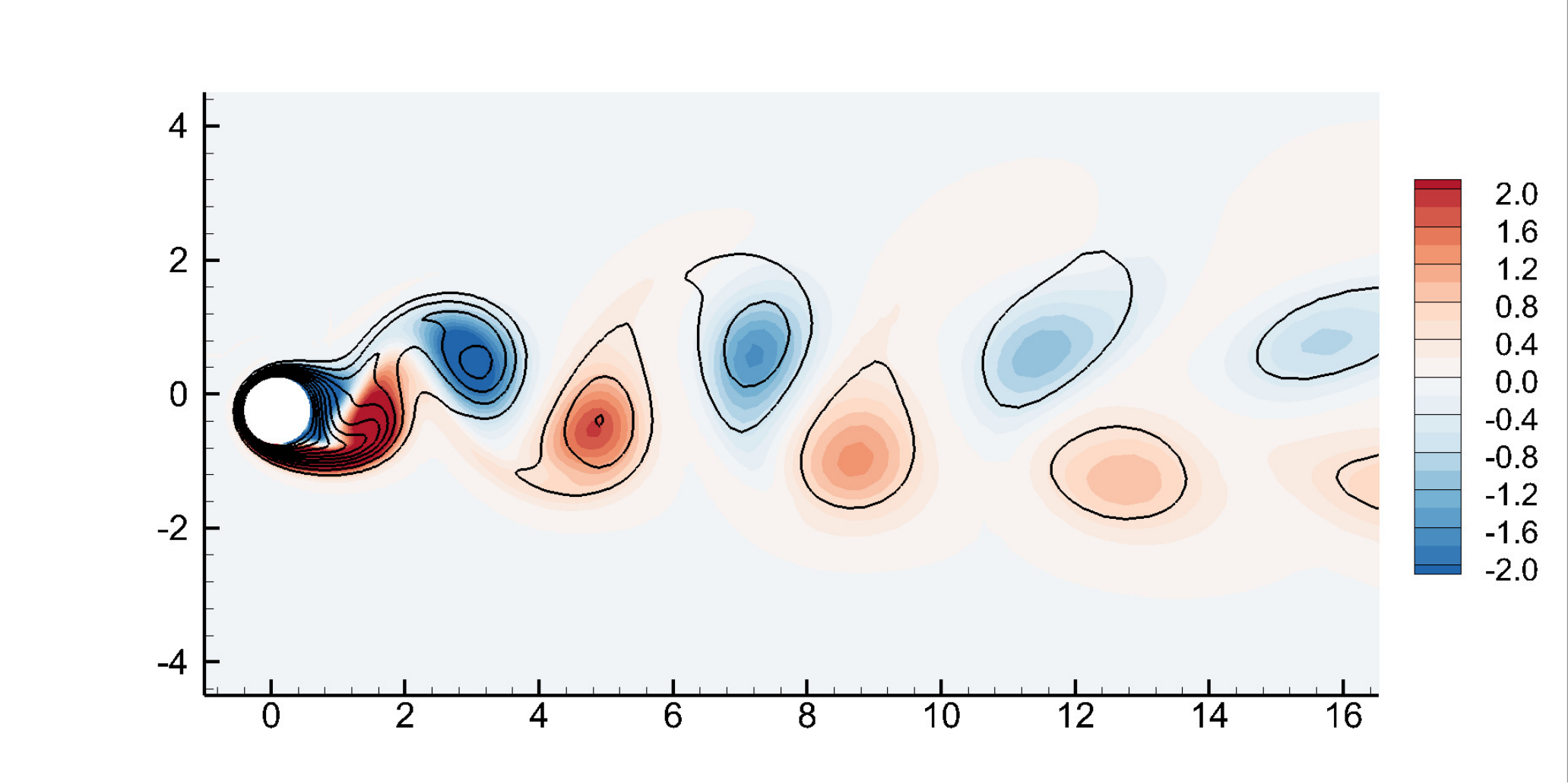}
	\caption{}
	\label{fig:re100r5P07R05}
\end{subfigure}
\begin{subfigure}{0.5\textwidth}
	\centering
	\includegraphics[trim=2cm 0.1cm 0.1cm 0.1cm,scale=0.4,clip]{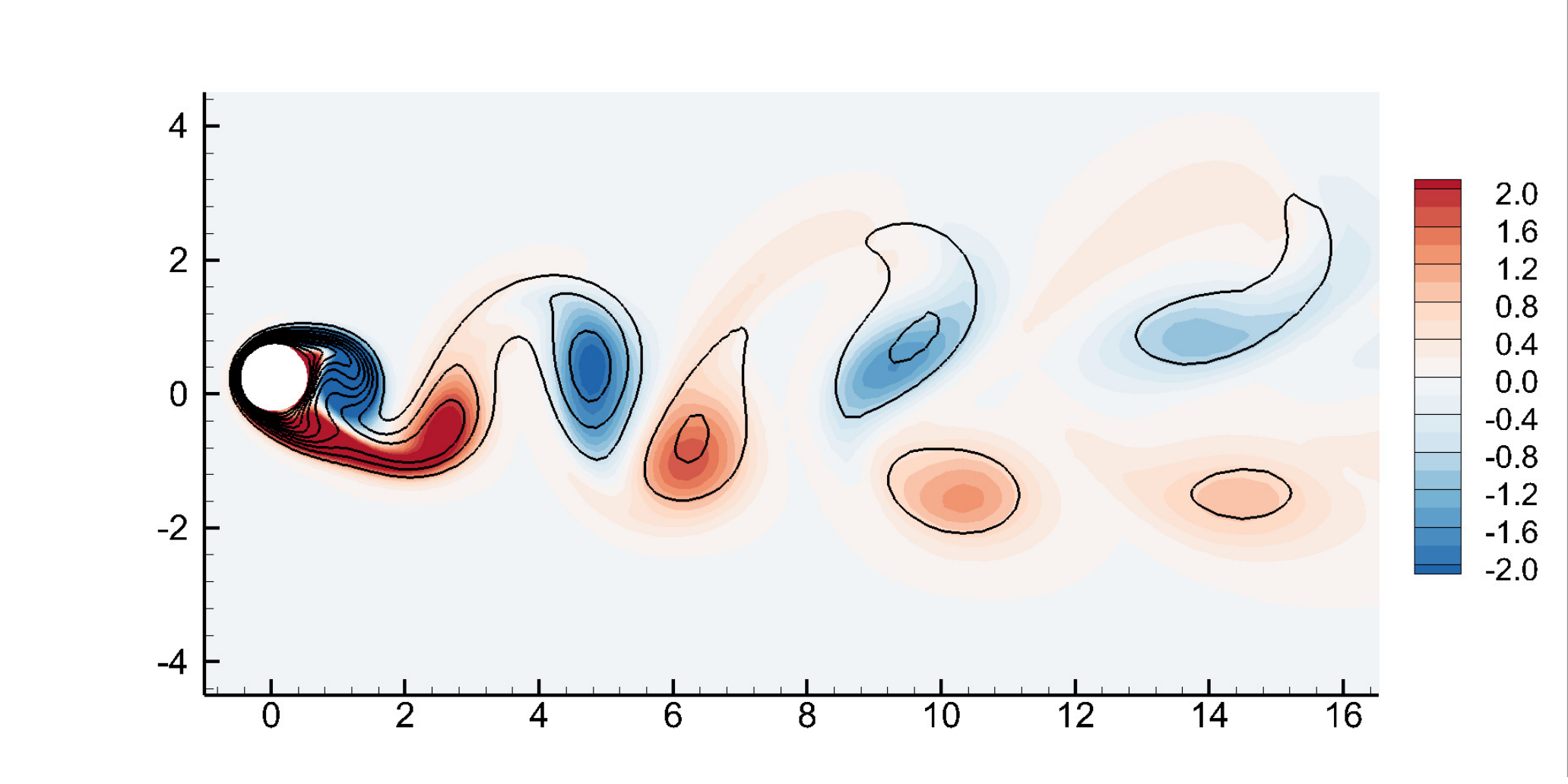}
	\caption{}
	\label{fig:re100r5P07R1}
\end{subfigure}
\begin{subfigure}{0.5\textwidth}
	\centering
	\includegraphics[trim=2cm 0.1cm 0.1cm 0.1cm,scale=0.4,clip]{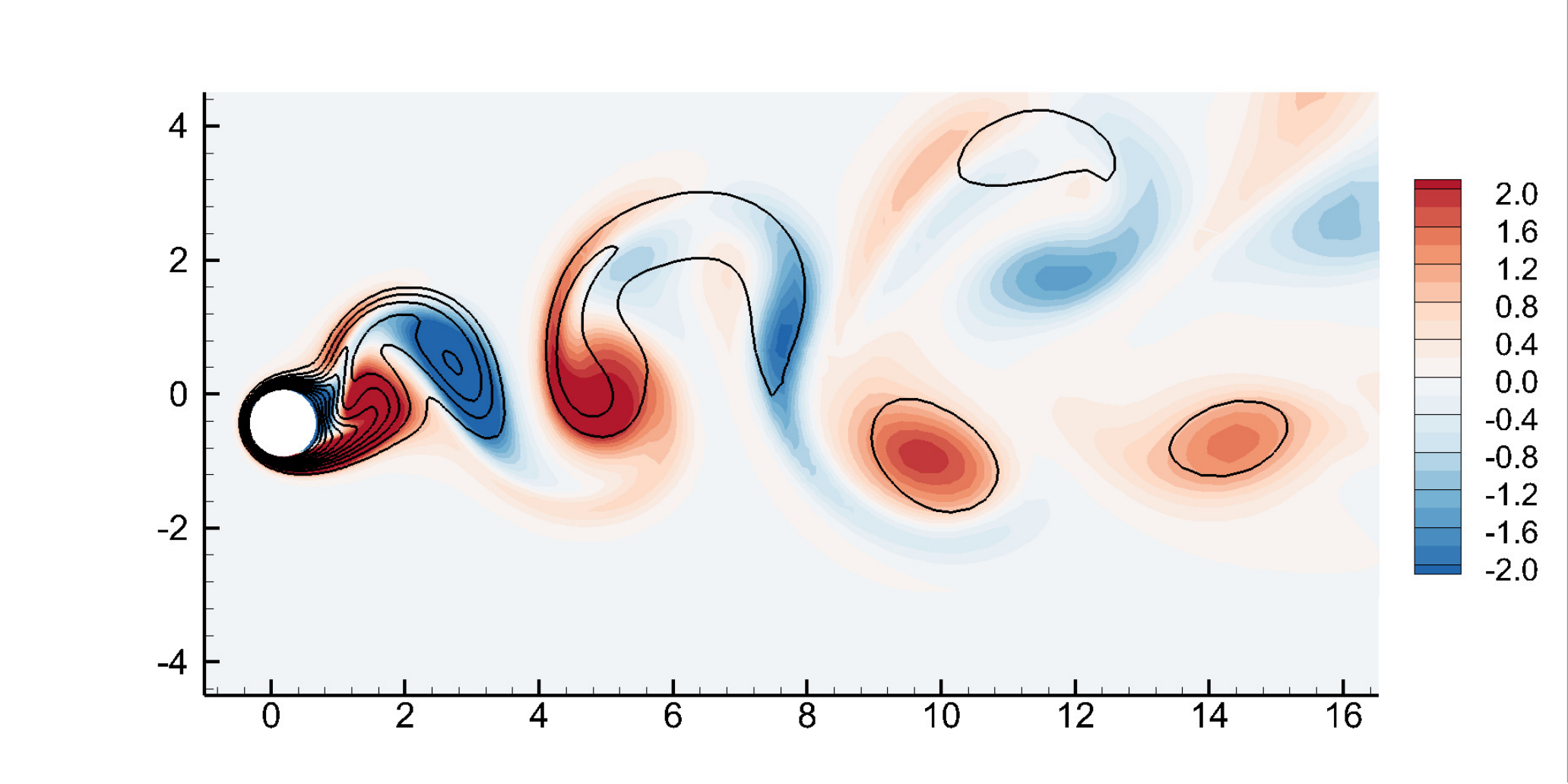}
	\caption{}
	\label{fig:re100r5P07R2}
\end{subfigure}
	\caption{Spanwise vorticity $\omega$ contour of a freely-vibrating cylinder at $Re=100$, $Pr = 0.7$, $m^*=10$, $\zeta = 0.01$, $Ur = 5.0$ and $\tau = 350$: (a) $Ri = 0.5$; (b) $Ri = 1.0$ and (c) $Ri = 2.0$. The solid lines are heat energy field for $\theta \in [0.08, 0.9]$.}
	\label{fig:re100r5P07}
\end{figure}
\begin{figure}[!htp]
	\centering
	\begin{subfigure}{0.5\textwidth}
	\centering
	\includegraphics[trim=2cm 0.1cm 0.1cm 0.1cm,scale=0.4,clip]{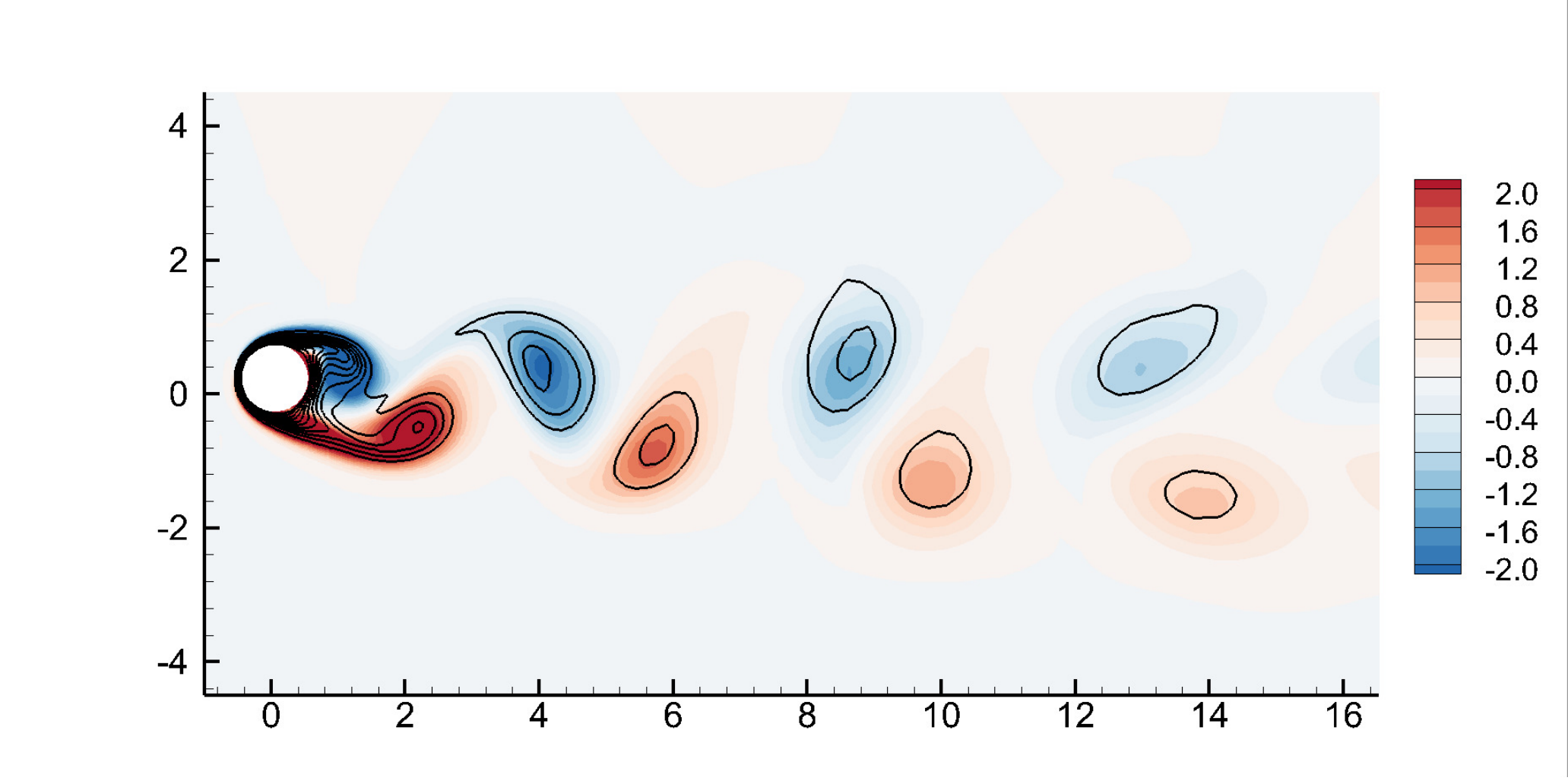}
	\caption{}
	\label{fig:re100r5P2R05}
\end{subfigure}
\begin{subfigure}{0.5\textwidth}
	\centering
	\includegraphics[trim=2cm 0.1cm 0.1cm 0.1cm,scale=0.4,clip]{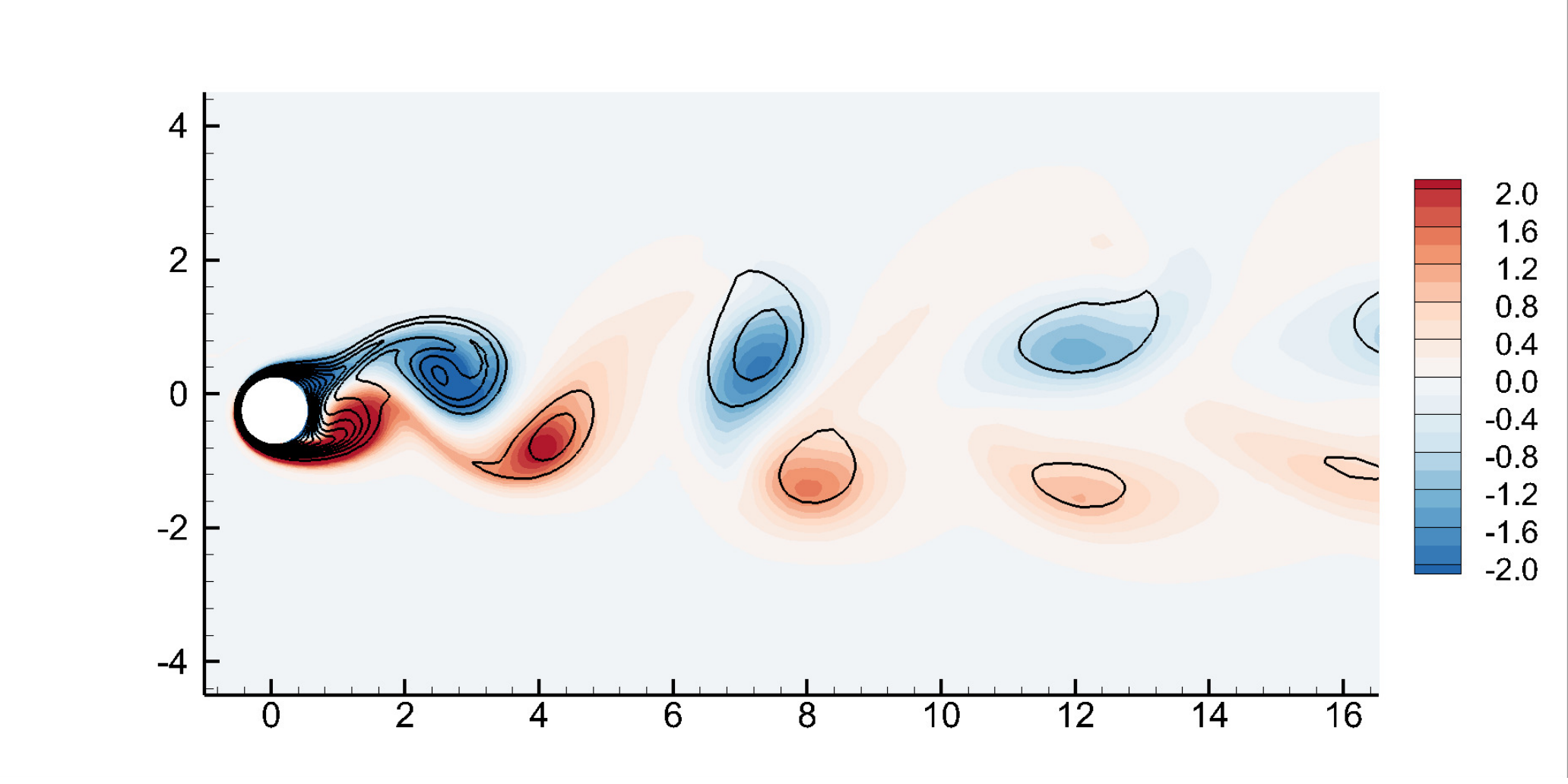}
	\caption{}
	\label{fig:re100r5P2R1}
\end{subfigure}
\begin{subfigure}{0.5\textwidth}
	\centering
	\includegraphics[trim=2cm 0.1cm 0.1cm 0.1cm,scale=0.4,clip]{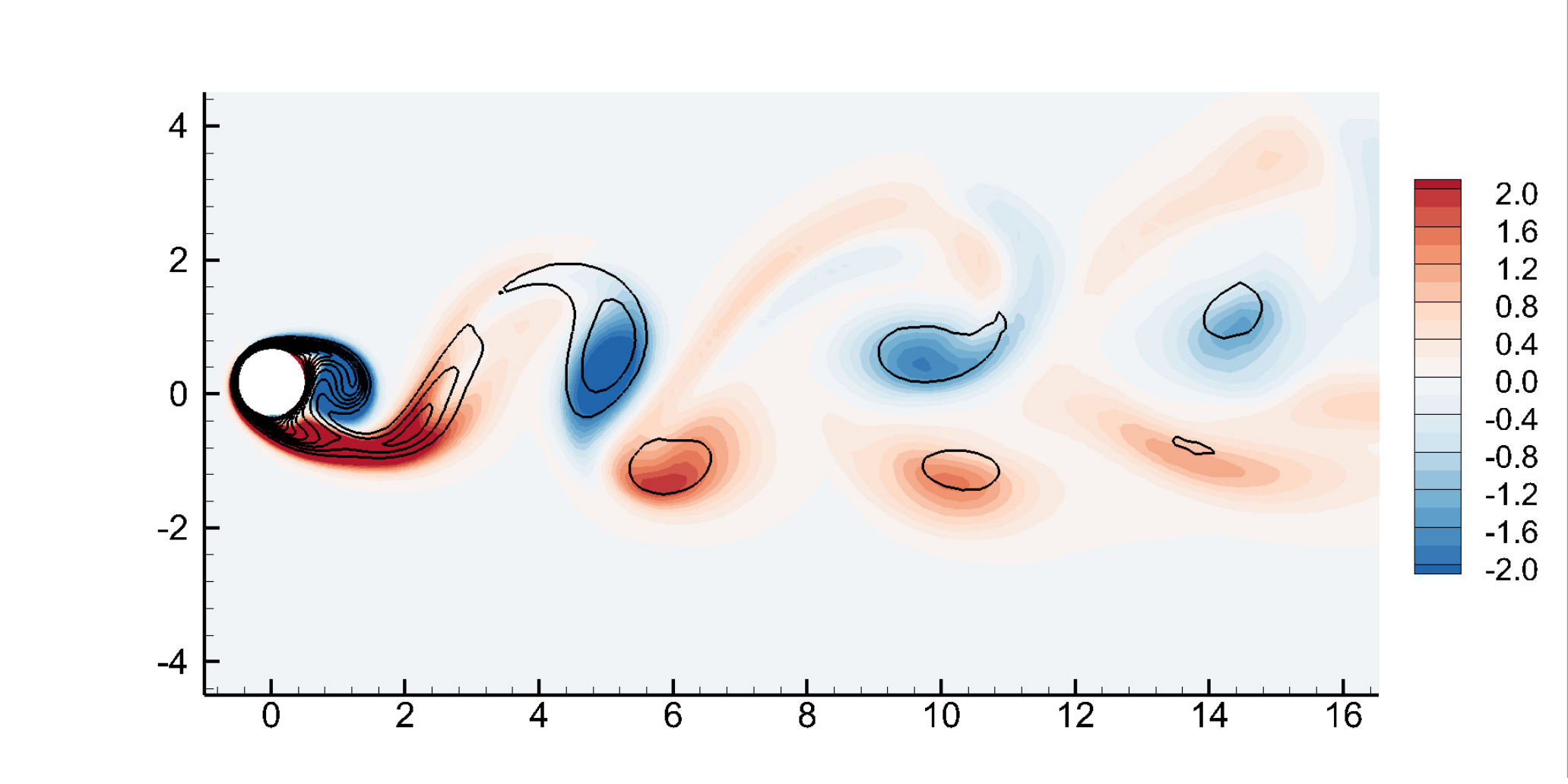}
	\caption{}
	\label{fig:re100r5P2R2}
\end{subfigure}
	\caption{Spanwise vorticity $\omega$ contour of a freely-vibrating cylinder at $Re=100$, $Pr = 2.0$, $m^*=10$, $\zeta = 0.01$, $Ur = 5.0$ and $\tau = 350$: (a) $Ri = 0.5$; (b) $Ri = 1.0$ and (c) $Ri = 2.0$. The solid lines are heat energy field for $\theta \in [0.08, 0.9]$.}
	\label{fig:re100r5P2}
\end{figure}

According to Equation~\eqref{eq:ns_dim}, it is known that Richardson number indicates the significance of buoyancy-driven flow in fluid momentum transpiration. A high Richardson number is linked the dominance of natural convection flow in wake. As shown in Figure~\ref{fig:re100r5P07}, the wake behind cylinder is apparently deflected upward for high Richardson numbers, e.g., $Ri = 2.0$ in this study, due to the influence of natural convection. The temperature contour closely follows the vorticity clusters and its strength dissipates gradually downstream. For the cases of low Richardson number, e.g., Figure~\ref{fig:re100r5P07R05} and~\ref{fig:re100r5P07R1}, the strength of heat energy field can sustain much farther downstream, since the forced convection is more dominant over natural convection. As Prandtl number increases until $Pr =2.0$, the size of temperature contour shrinks significantlywith respect to the vorticity clusters in Figure~\ref{fig:re100r5P2}, because the heat energy field is less diffusive. Similar to the cases of $Pr = 0.7$, the wake deflects upward at higher Richardson numbers, in which intensive and complex shear-layer mixing is observed in the upper portion of wake. Furthermore, it is also found that the temperature gradient become strong in the near wake right behind the cylinder, e.g., the dense temperature-contour layers, in Figure~\ref{fig:re100r5P2}, compared with the cases of lower Prandtl number in Figure~\ref{fig:re100r5P07}. 

\begin{figure}[!htp]
	\centering
	\begin{subfigure}{0.5\textwidth}
	\centering
	\includegraphics[trim=2cm 0.1cm 0.1cm 0.1cm,scale=0.4,clip]{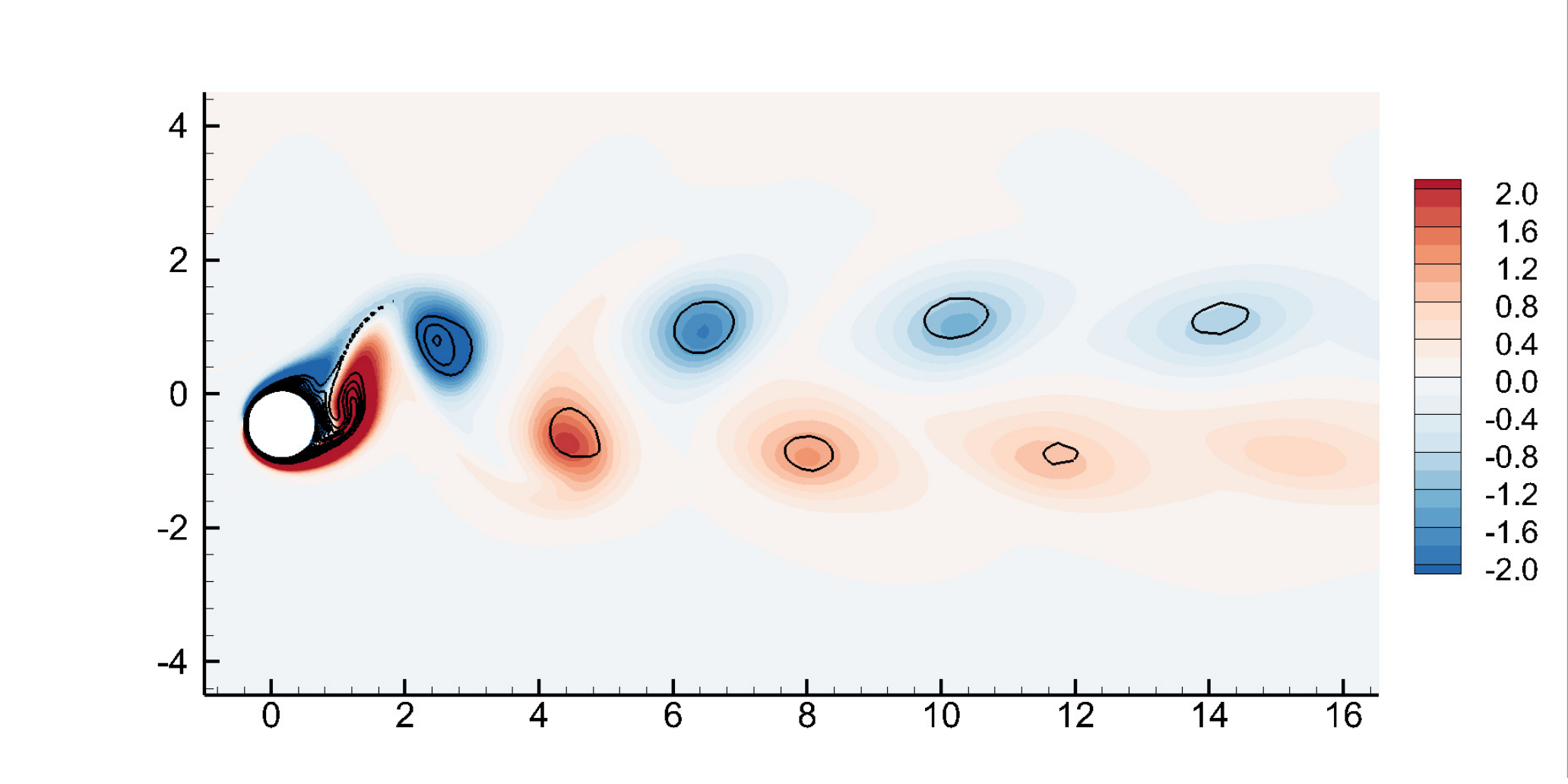}
	\caption{}
	\label{fig:re100r5P10R05}
\end{subfigure}
\begin{subfigure}{0.5\textwidth}
	\centering
	\includegraphics[trim=2cm 0.1cm 0.1cm 0.1cm,scale=0.4,clip]{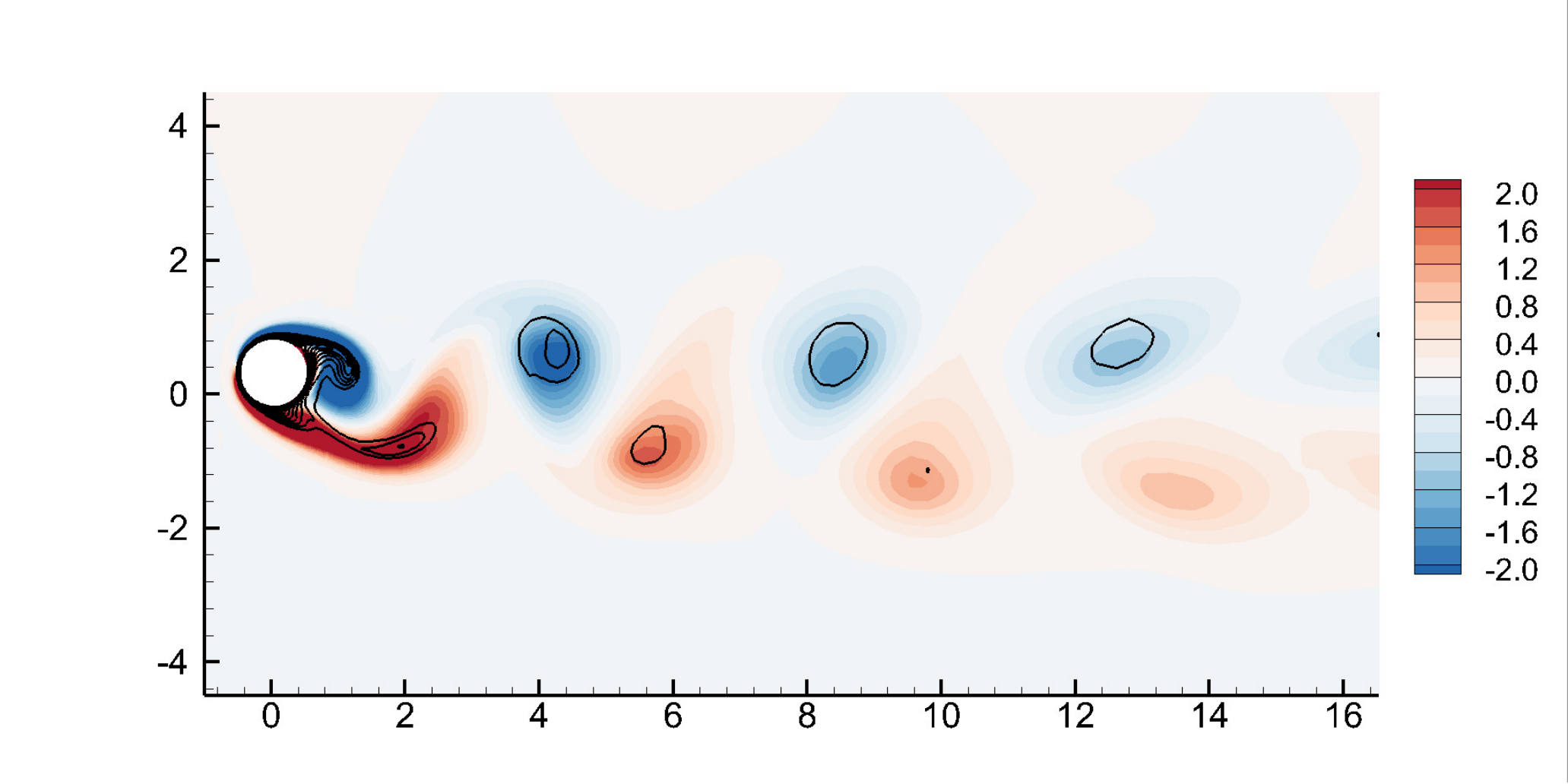}
	\caption{}
	\label{fig:re100r5P10R1}
\end{subfigure}
\begin{subfigure}{0.5\textwidth}
	\centering
	\includegraphics[trim=2cm 0.1cm 0.1cm 0.1cm,scale=0.4,clip]{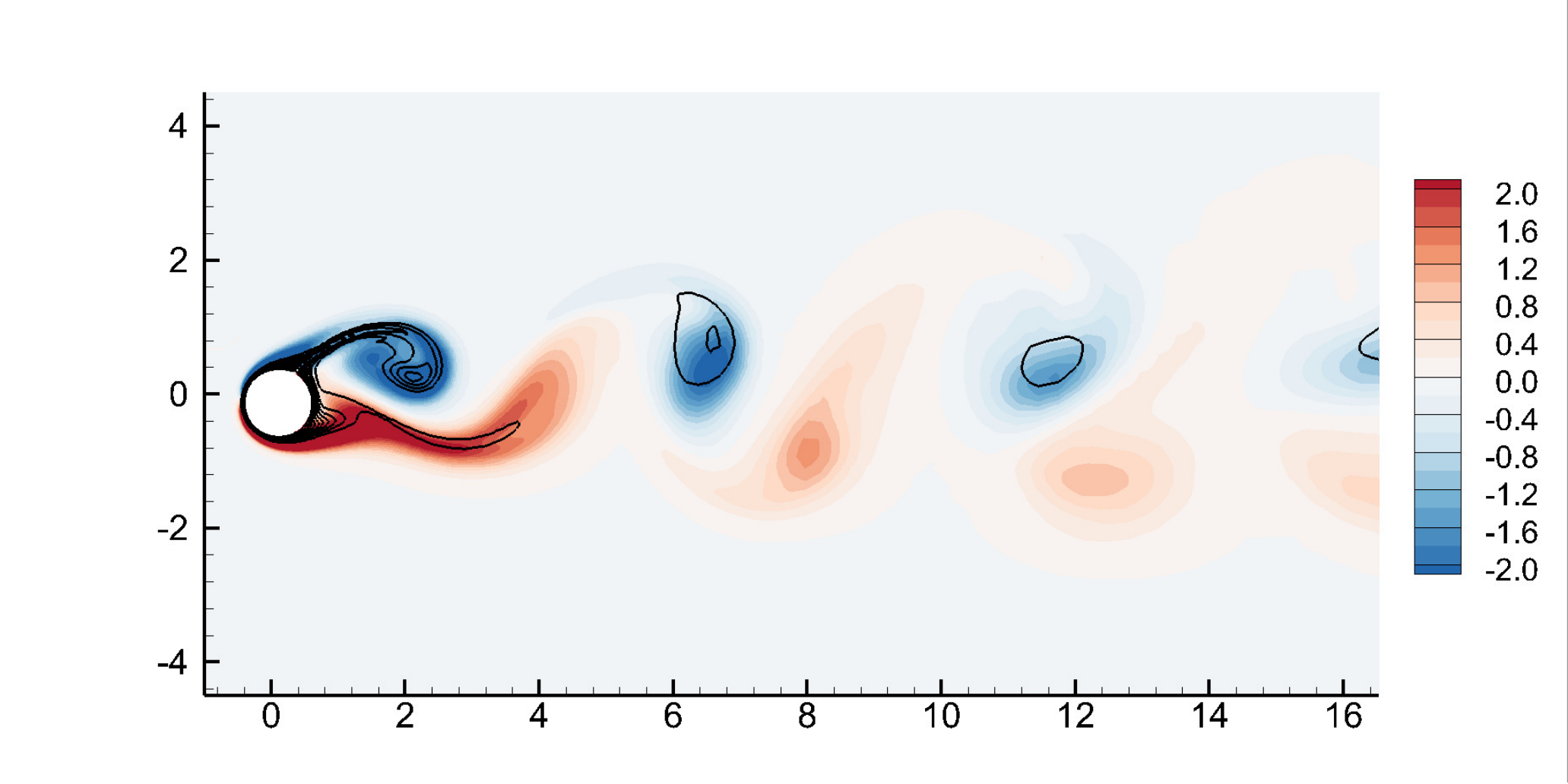}
	\caption{}
	\label{fig:re100r5P10R2}
\end{subfigure}
	\caption{Spanwise vorticity $\omega$ contour of a freely-vibrating cylinder at $Re=100$, $Pr = 10$, $m^*=10$, $\zeta = 0.01$, $Ur = 5.0$ and $\tau = 350$: (a) $Ri = 0.5$; (b) $Ri = 1.0$ and (c) $Ri = 2.0$. The solid lines are temperature field for $\theta \in [0.08, 0.9]$.}
	\label{fig:re100r5P10}
\end{figure}
\begin{figure}[!htp]
	\centering
	\begin{subfigure}{0.5\textwidth}
	\centering
	\includegraphics[trim=1cm 1cm 1cm 1cm,scale=0.35,clip]{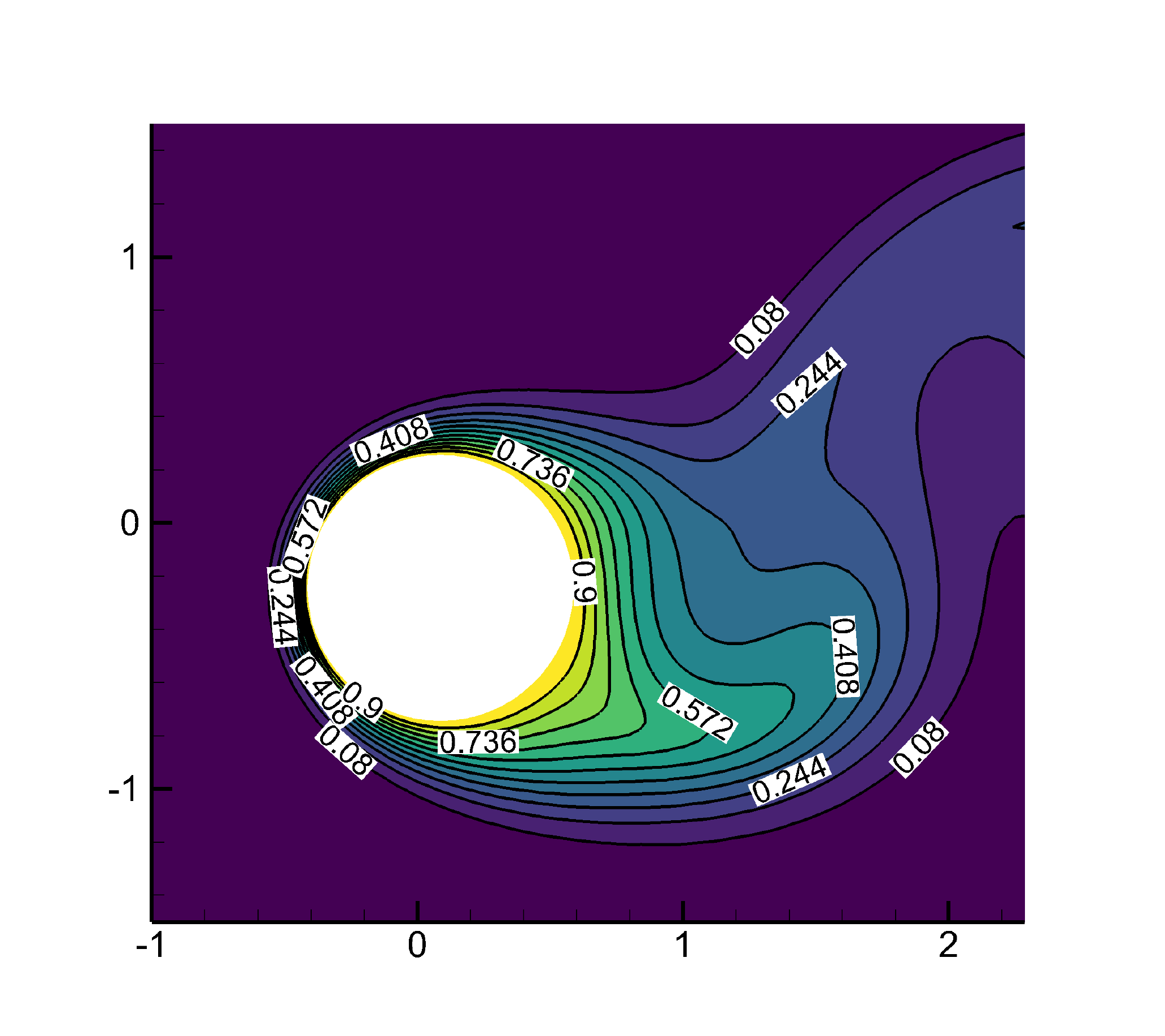}
	\caption{}
	\label{fig:re100r5P07R05_z}
\end{subfigure}
\begin{subfigure}{0.5\textwidth}
	\centering
	\includegraphics[trim=1cm 1cm 1cm 1cm,scale=0.35,clip]{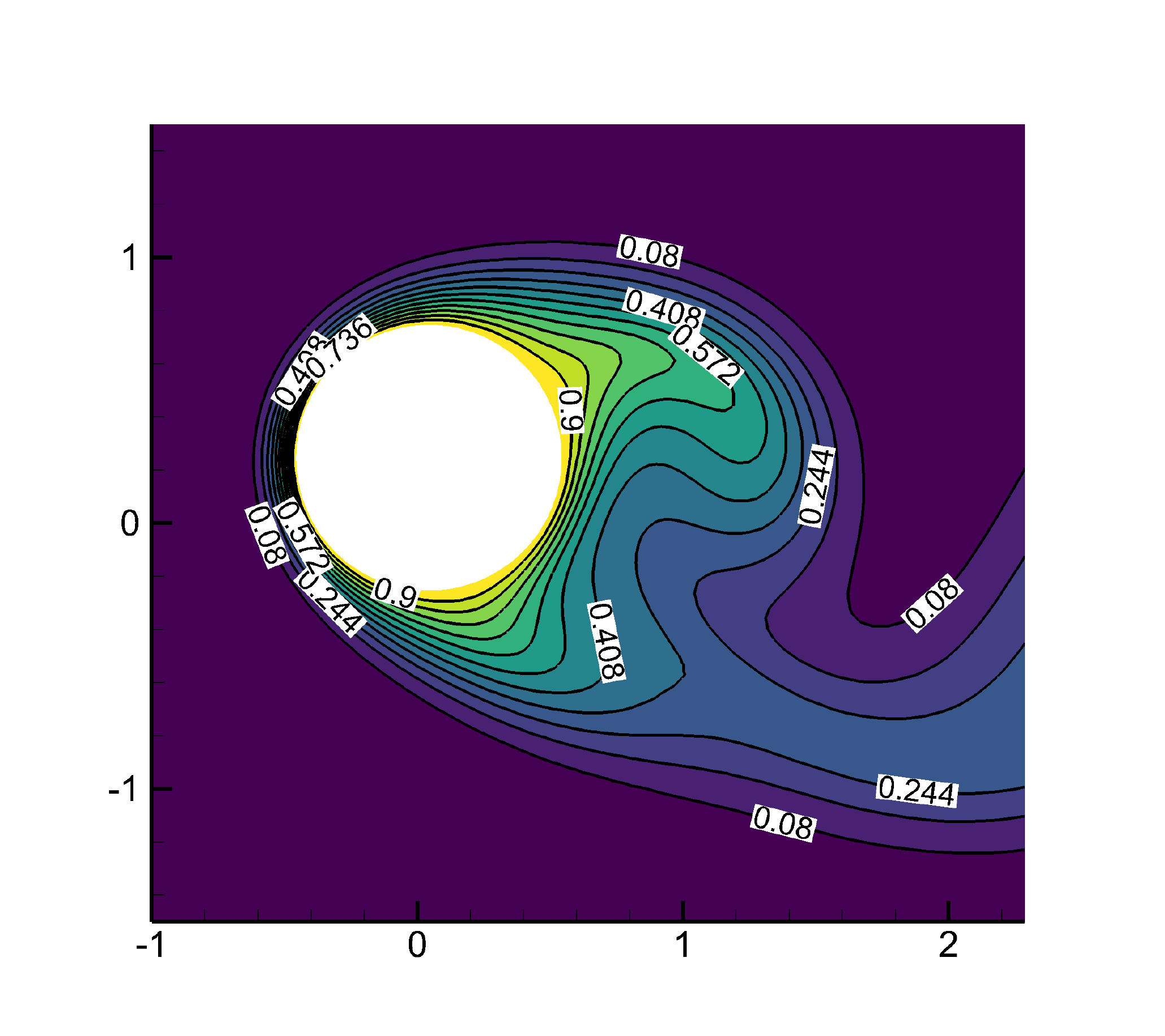}
	\caption{}
	\label{fig:re100r5P07R1_z}
\end{subfigure}
\begin{subfigure}{0.5\textwidth}
	\centering
	\includegraphics[trim=1cm 1cm 1cm 1cm,scale=0.35,clip]{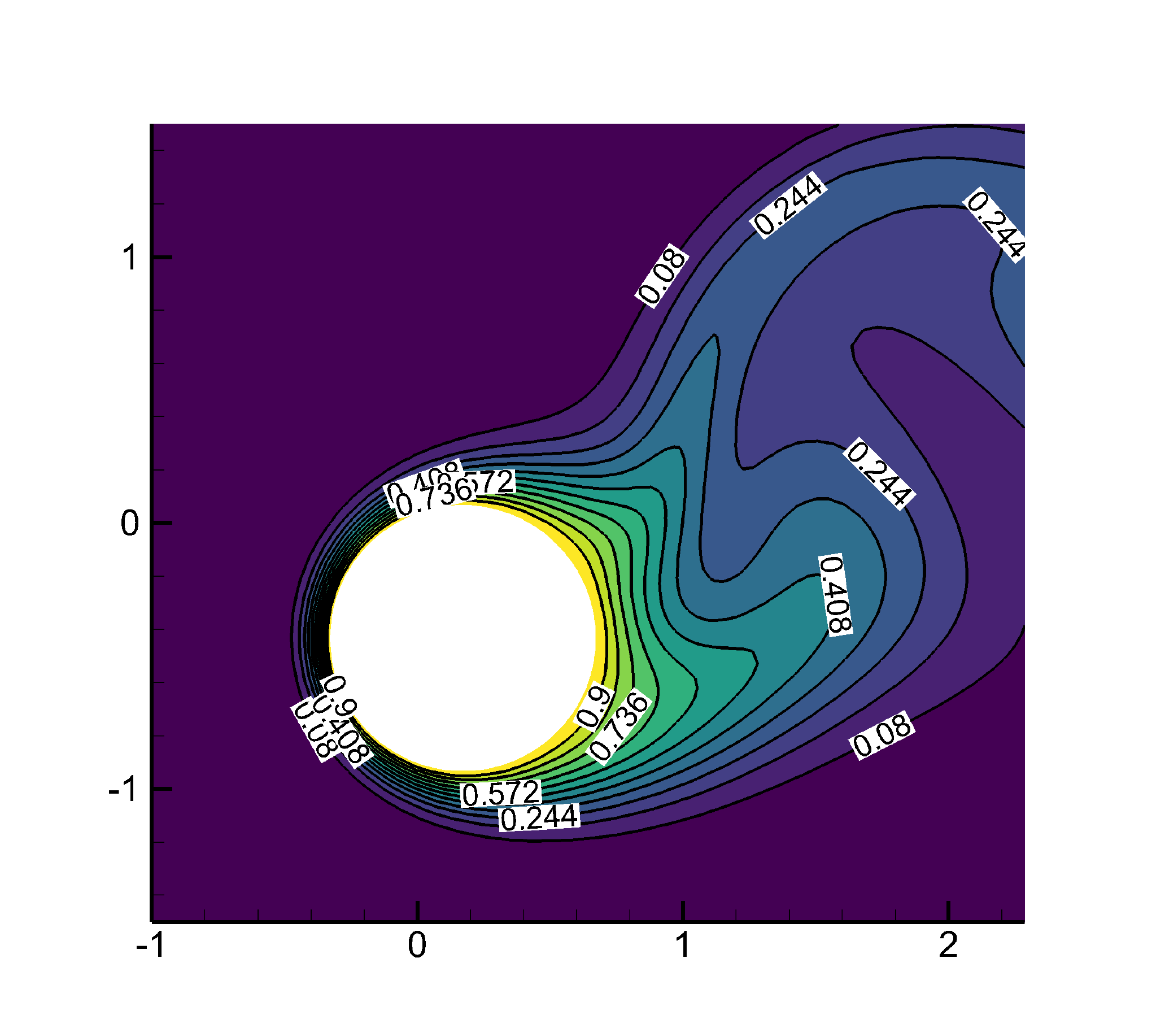}
	\caption{}
	\label{fig:re100r5P07R2_z}
\end{subfigure}
	\caption{Thermal contour of a freely-vibrating cylinder at $Re=100$, $Pr = 0.7$, $m^*=10$, $\zeta = 0.01$, $Ur = 5.0$ and $\tau = 350$: (a) $Ri = 0.5$; (b) $Ri = 1.0$ and (c) $Ri = 2.0$}
	\label{fig:re100r5P07_z}
\end{figure}

\begin{figure}[!htp]
	\centering
	\begin{subfigure}{0.5\textwidth}
	\centering
	\includegraphics[trim=1cm 1cm 1cm 1cm,scale=0.35,clip]{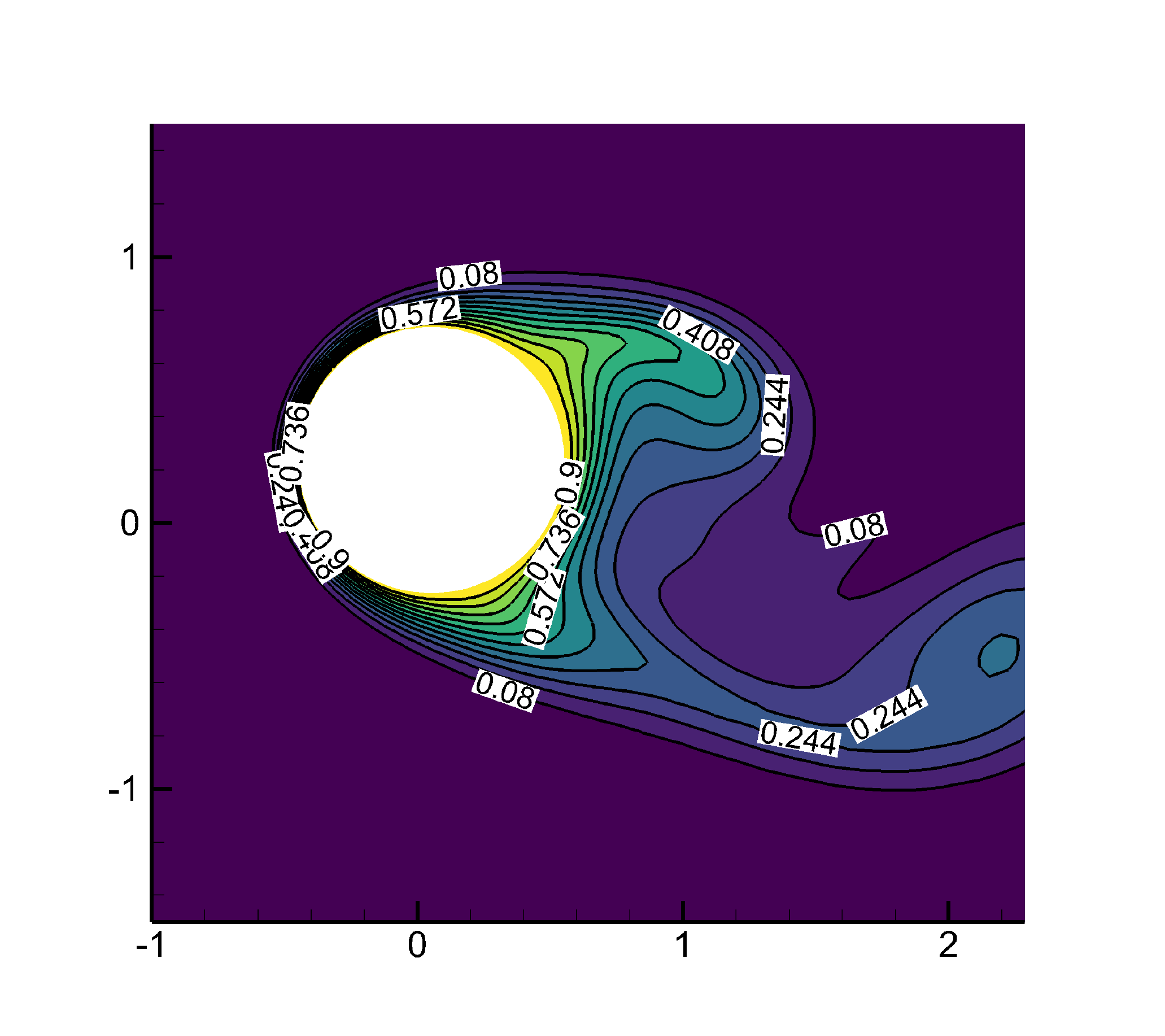}
	\caption{}
	\label{fig:re100r5P2R05_z}
\end{subfigure}
\begin{subfigure}{0.5\textwidth}
	\centering
	\includegraphics[trim=1cm 1cm 1cm 1cm,scale=0.35,clip]{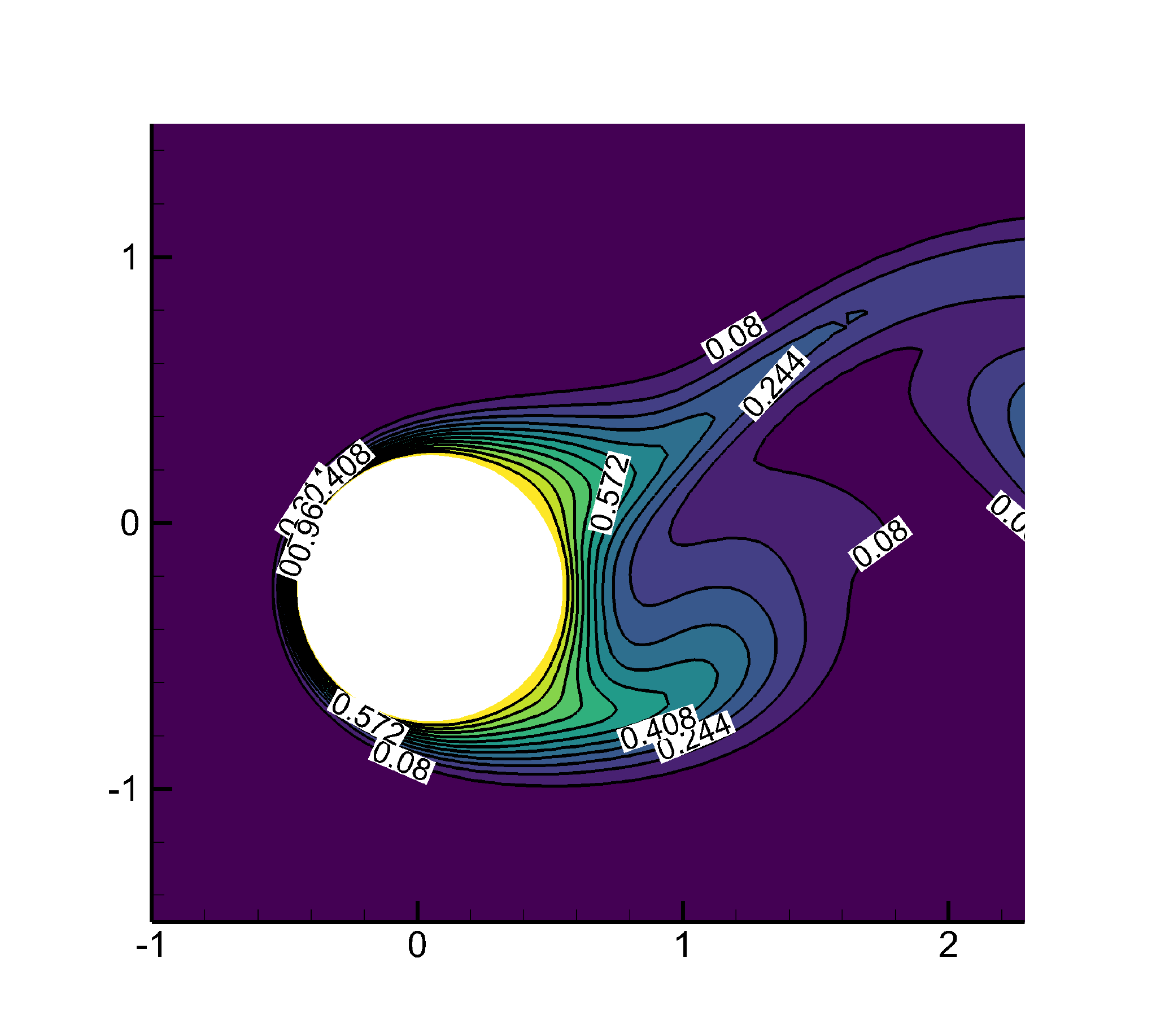}
	\caption{}
	\label{fig:re100r5P2R1_z}
\end{subfigure}
\begin{subfigure}{0.5\textwidth}
	\centering
	\includegraphics[trim=1cm 1cm 1cm 1cm,scale=0.35,clip]{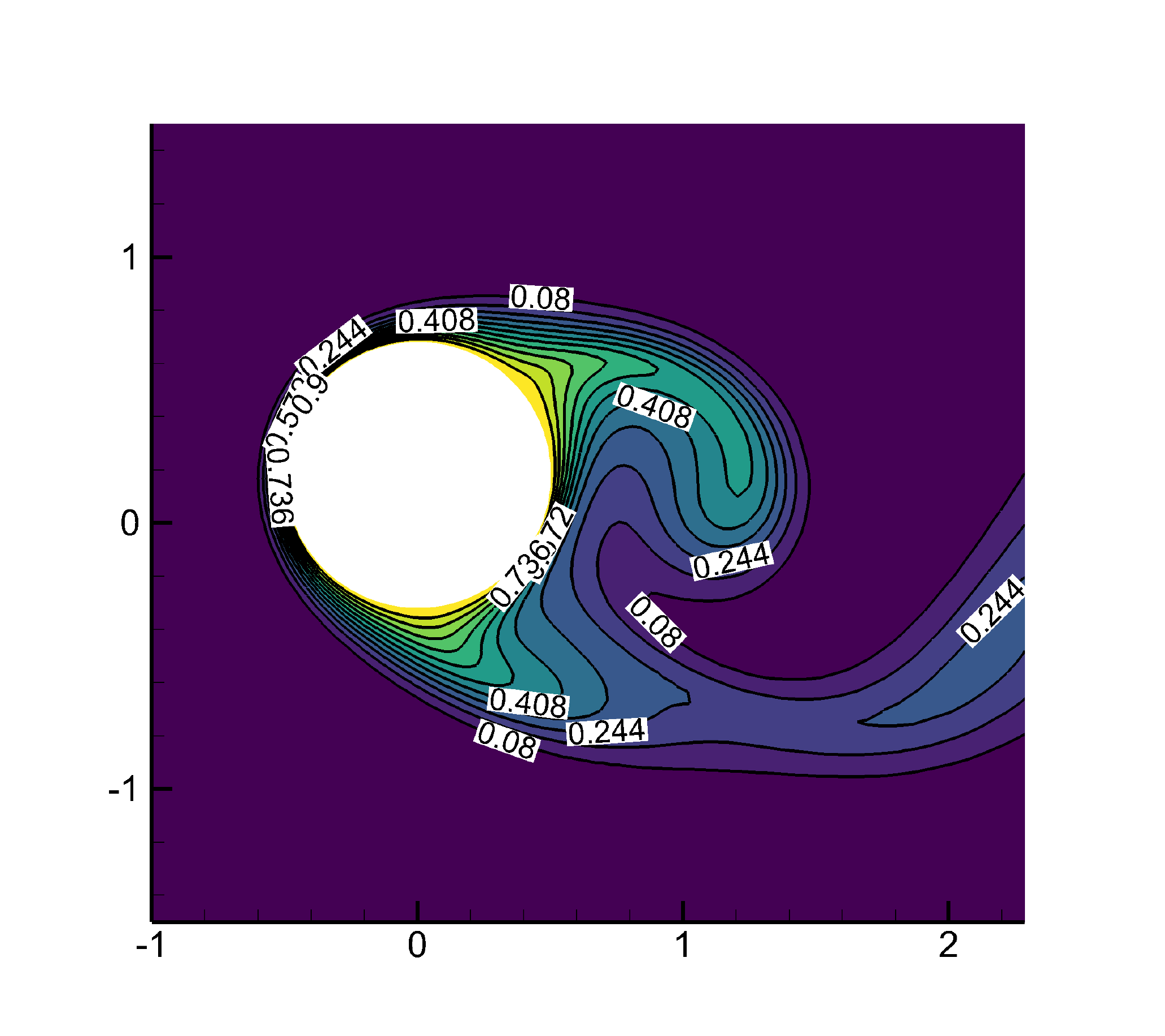}
	\caption{}
	\label{fig:re100r5P2R2_z}
\end{subfigure}
	\caption{Thermal contour of a freely-vibrating cylinder at $Re=100$, $Pr = 2.0$, $m^*=10$, $\zeta = 0.01$, $Ur = 5.0$ and $\tau = 350$: (a) $Ri = 0.5$; (b) $Ri = 1.0$ and (c) $Ri = 2.0$}
	\label{fig:re100r5P2_z}
\end{figure}

\begin{figure}[!htp]
	\centering
	\begin{subfigure}{0.5\textwidth}
	\centering
	\includegraphics[trim=1cm 1cm 1cm 1cm,scale=0.35,clip]{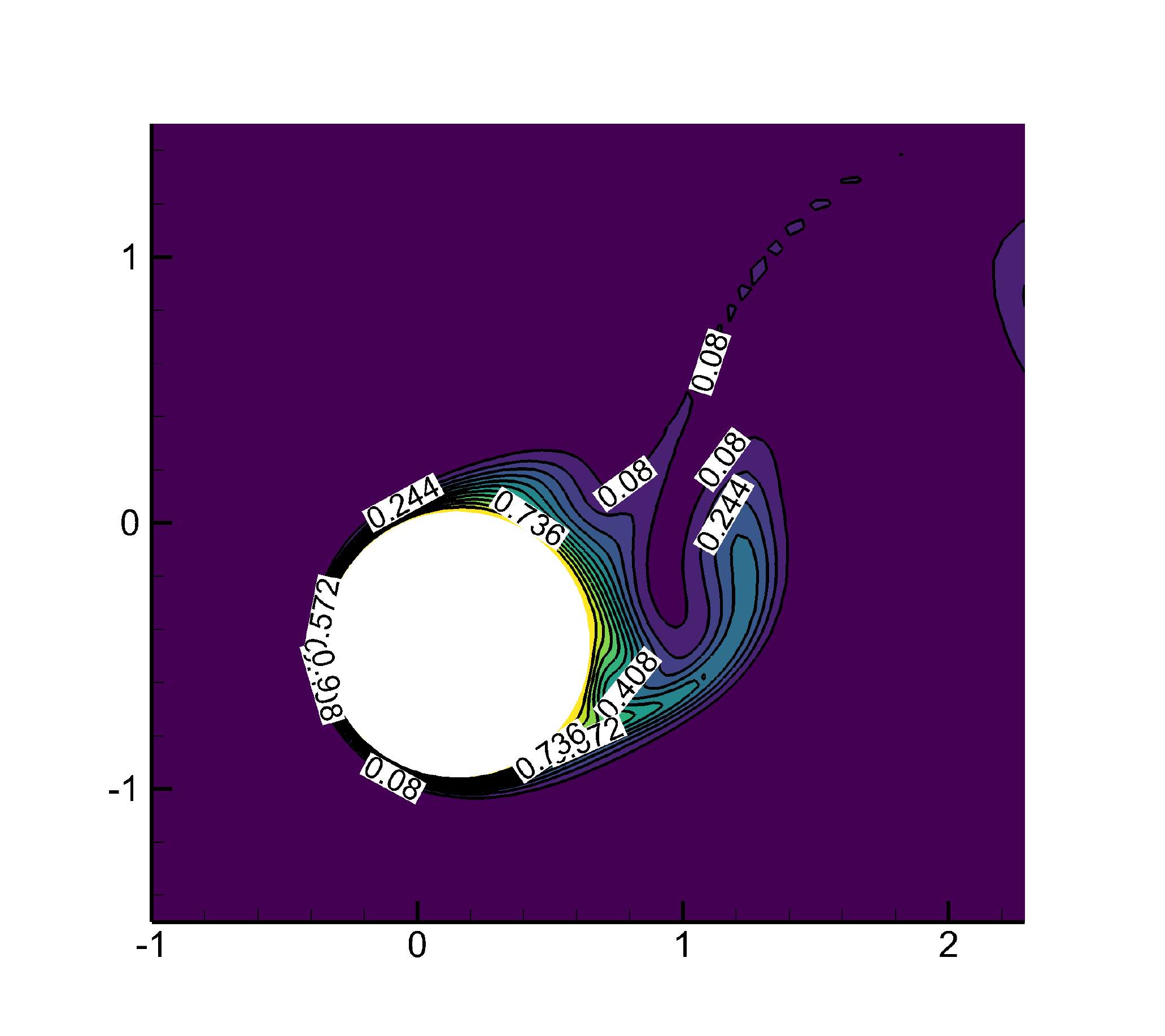}
	\caption{}
	\label{fig:re100r5P10R05_z}
\end{subfigure}
\begin{subfigure}{0.5\textwidth}
	\centering
	\includegraphics[trim=1cm 1cm 1cm 1cm,scale=0.35,clip]{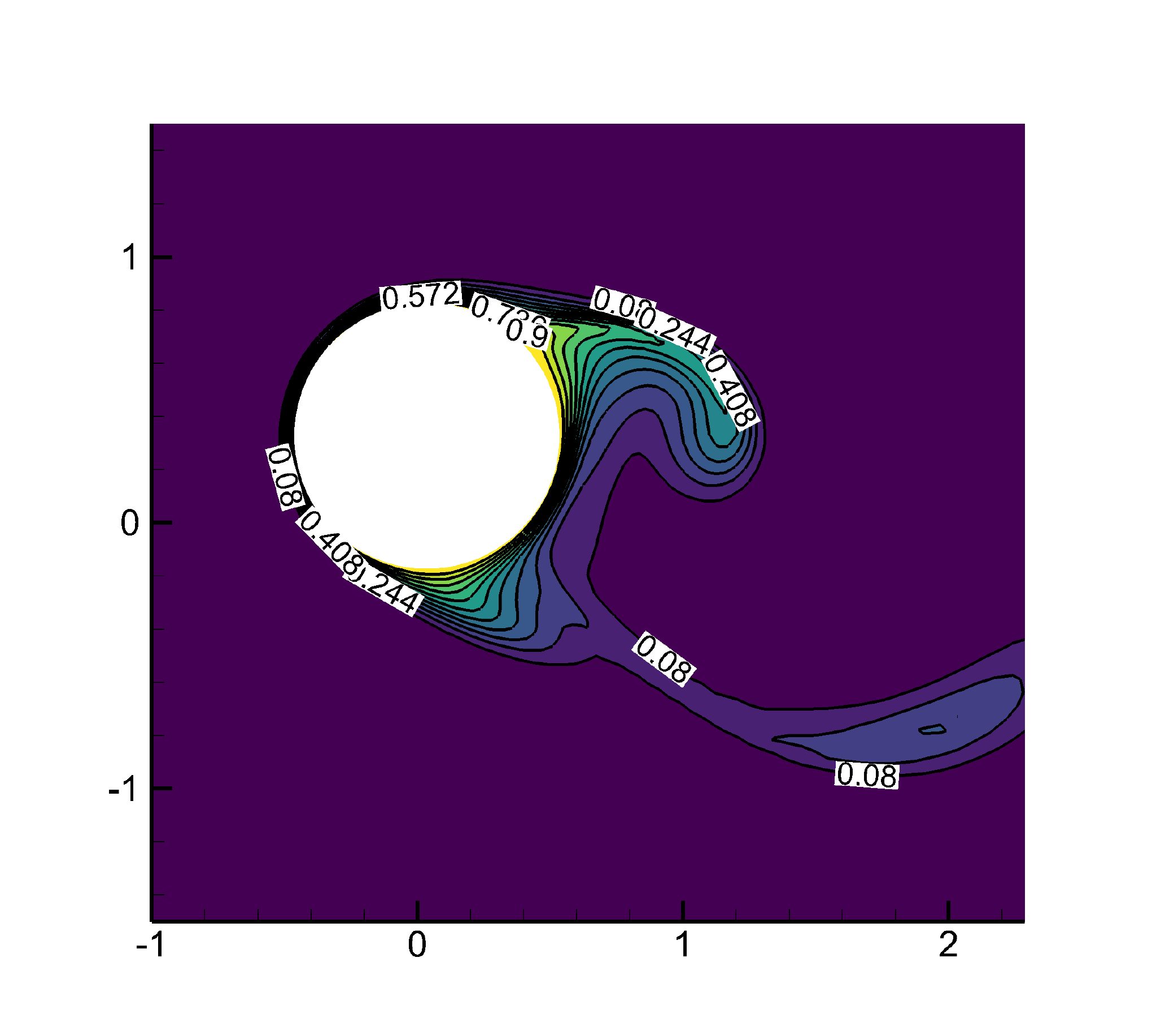}
	\caption{}
	\label{fig:re100r5P10R1_z}
\end{subfigure}
\begin{subfigure}{0.5\textwidth}
	\centering
	\includegraphics[trim=1cm 1cm 1cm 1cm,scale=0.35,clip]{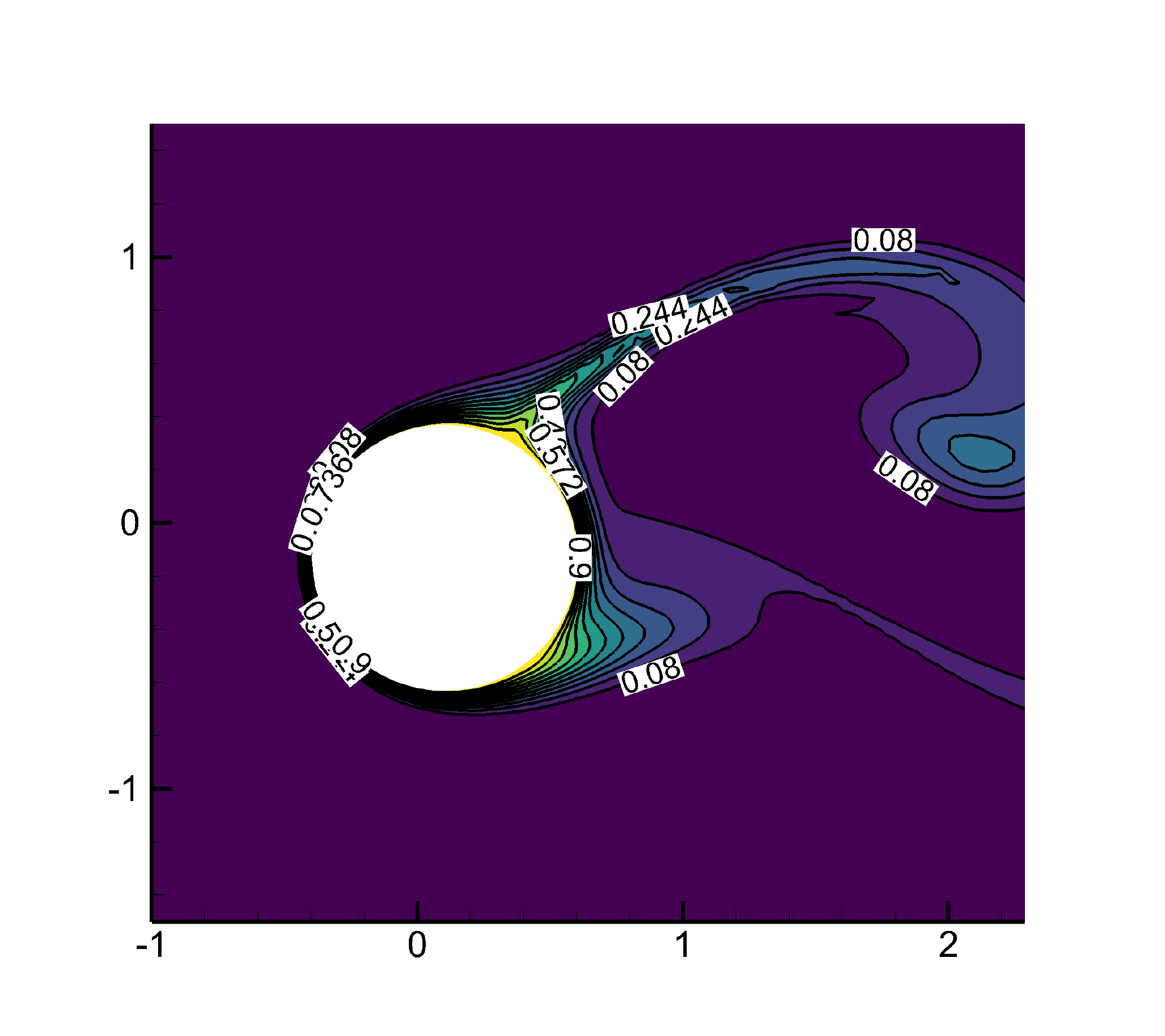}
	\caption{}
	\label{fig:re100r5P10R2_z}
\end{subfigure}
	\caption{Thermal contour of a freely-vibrating cylinder at $Re=100$, $Pr = 10$, $m^*=10$, $\zeta = 0.01$, $Ur = 5.0$ and $\tau = 350$: (a) $Ri = 0.5$; (b) $Ri = 1.0$ and (c) $Ri = 2.0$}
	\label{fig:re100r5P10_z}
\end{figure}
For the cases of $Pr = 10$ in Figure~\ref{fig:re100r5P10}, the temperature contour shrinks even further and disappears rapidly in wake. Due to this limited influence of heat energy field on hydrodynamics in near wake, the vorticity clusters do not behave very oddly from those in isothermal flow, in contrast to the cases of lower Prandtl numbers in Figure~\ref{fig:re100r5P07} and Figure~\ref{fig:re100r5P2}. Apparently, the heat energy is transported with the vorticity clusters downstream, but concentrated in the vortex cores.

Based on the aforementioned investigations, it is realized that the distribution of heat energy field has a direct interference to the structural dynamics and hydrodynamics in near wake right behind cylinder. Hence, in Figure~\ref{fig:re100r5P07_z}, we zoom in and focus on the temperature contours close to the cylinder surface and its near wake during VIV lock-in for $Ur = 5.0$. It is found that the temperature contour is stretched further and forms clusters shedding downstream in the cases of higher Richardson numbers, e.g., the temperature contour $\theta = 0.244$ in Figure~\ref{fig:re100r5P07_z}. Moreover, following the intensive vibration of cylinder, the temperature contour becomes extremely concentrated over the frontal surface of cylinder, toward where it is swiftly moving, e.g., the upper and lower surfaces in Figure~\ref{fig:re100r5P07R05_z} and Figure~\ref{fig:re100r5P07R2_z} respectively.

In Figure~\ref{fig:re100r5P2_z}, the shrunk temperature contours are confirmed again in the cases of $Pr = 2.0$. It can be clearly observed that the density of temperature contour is significantly high in the cases of higher Prandtl numbers, because of the less diffusive heat energy field. It implies a rapid heat convection over the cylinder's surface, which will be confirmed and discussed in Section~\ref{sec:thermal} later on. Again, the high concentration of temperature contour is noticed over the cylinder's frontal surface during large vibration. Due to the instantaneous changes of heat energy field with respect to structural dynamics, intensive fluctuation of heat flux could be expected during VIV lock-in, together with relatively high Nusselt numbers. The performance of heat transfer will be further discussed in Section~\ref{sec:thermal}. The temperature contour retains a high density around the entire surface of cylinder for the cases of $Pr = 10$ for different $Ri$ values in Figure~\ref{fig:re100r5P10_z}. In those cases, the heat energy field is highly concentrated around the cylinder's surface and the temperature in wake rapidly approaches toward the referential value downstream. In addition, it if found the temperature contour is relatively stretched for high Richardson numbers, as shown in Figure~\ref{fig:re100r5P10R2_z}.     

\subsection{hydrodynamic responses in mixed convection flow} \label{sec:hydro}
\begin{figure*}[!thp]
	\centering
	\begin{subfigure}{0.5\textwidth}
	\centering
	\includegraphics[trim=0.0cm 0.1cm 0.1cm 0.1cm,scale=0.25,clip]{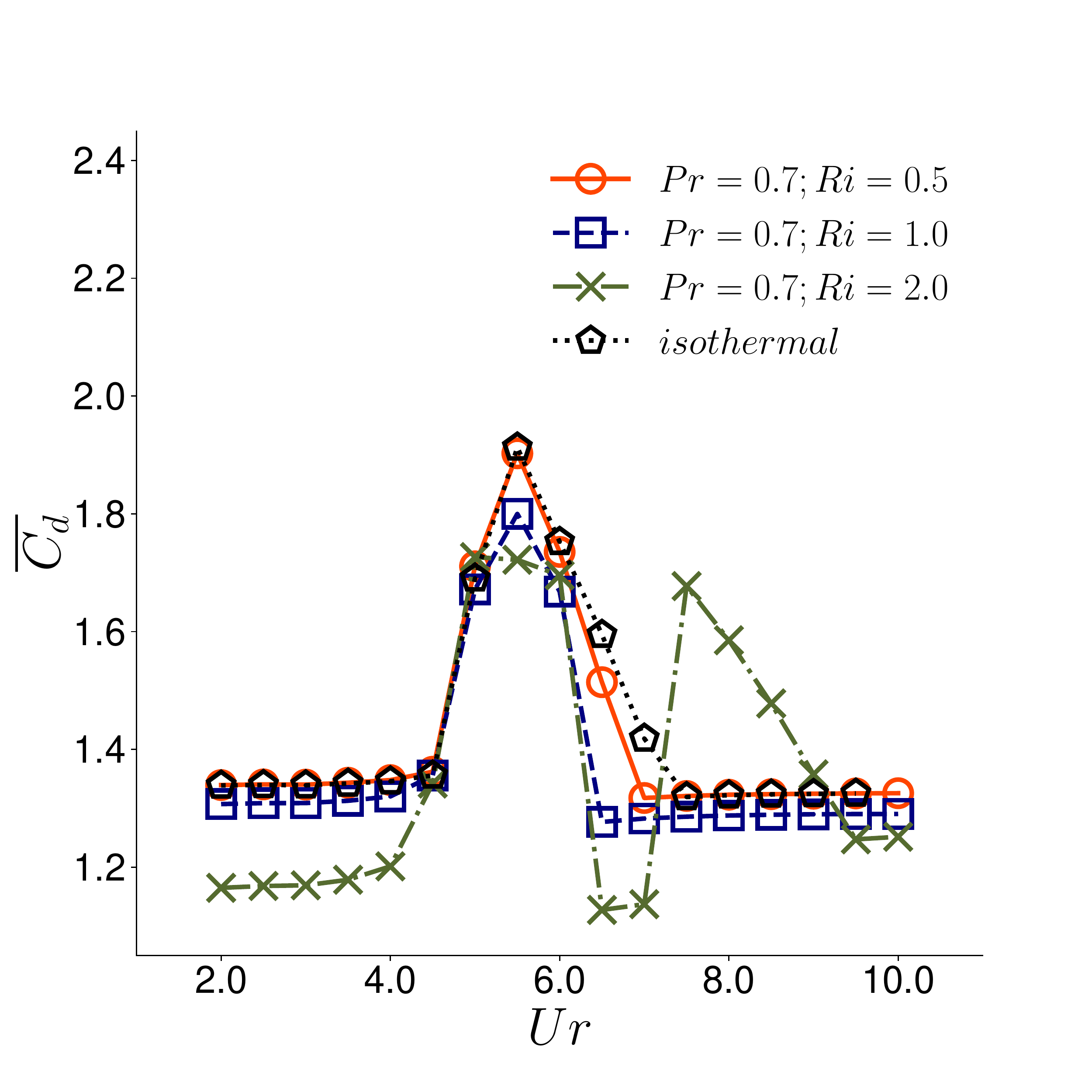}
	\caption{}
	\label{fig:Cd1}
\end{subfigure}%
\begin{subfigure}{0.5\textwidth}
	\centering
	\includegraphics[trim=0.0cm 0.1cm 0.1cm 0.1cm,scale=0.25,clip]{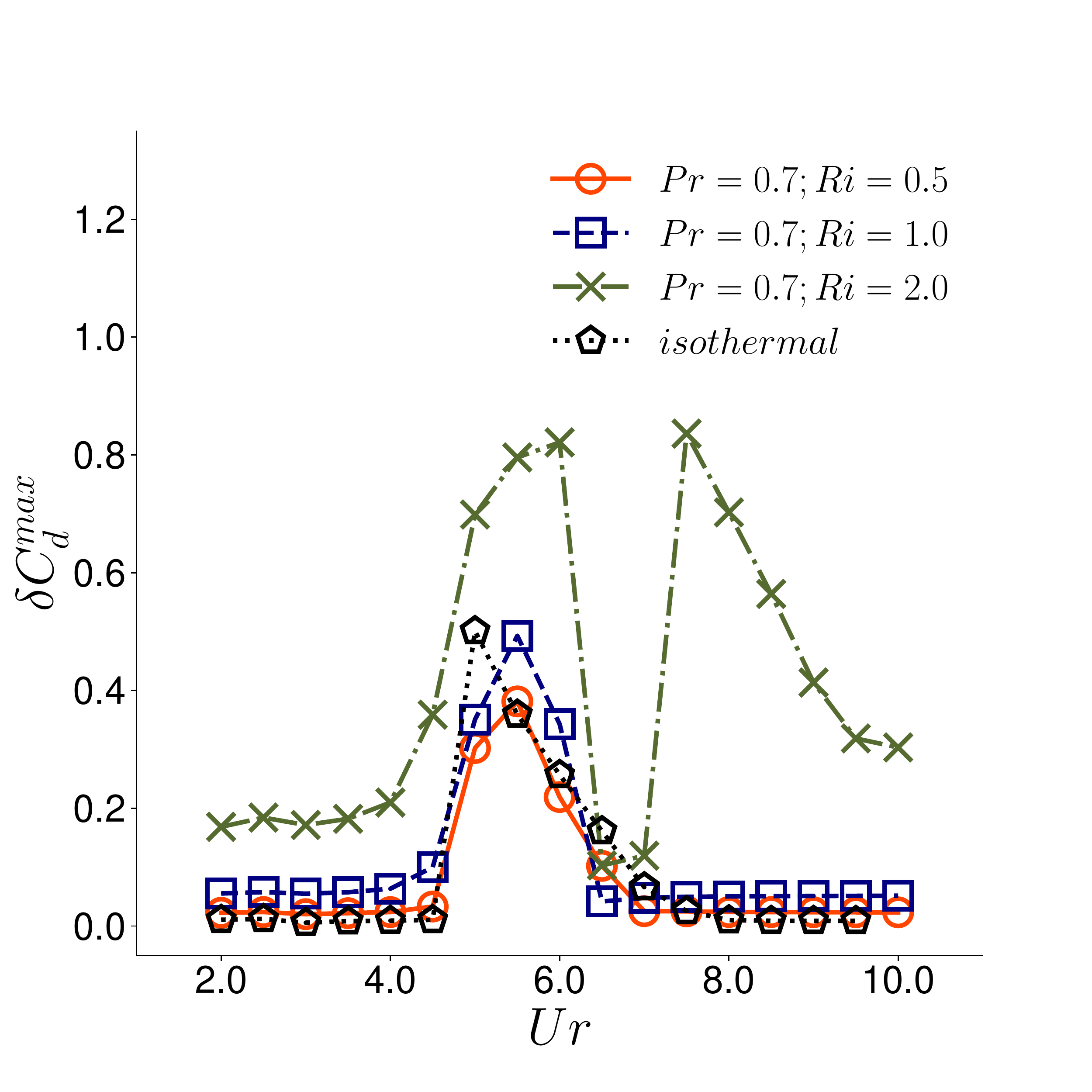}
	\caption{}
	\label{fig:Cd2}
\end{subfigure}
\begin{subfigure}{0.5\textwidth}
	\centering
	\includegraphics[trim=0.0cm 0.1cm 0.1cm 0.1cm,scale=0.25,clip]{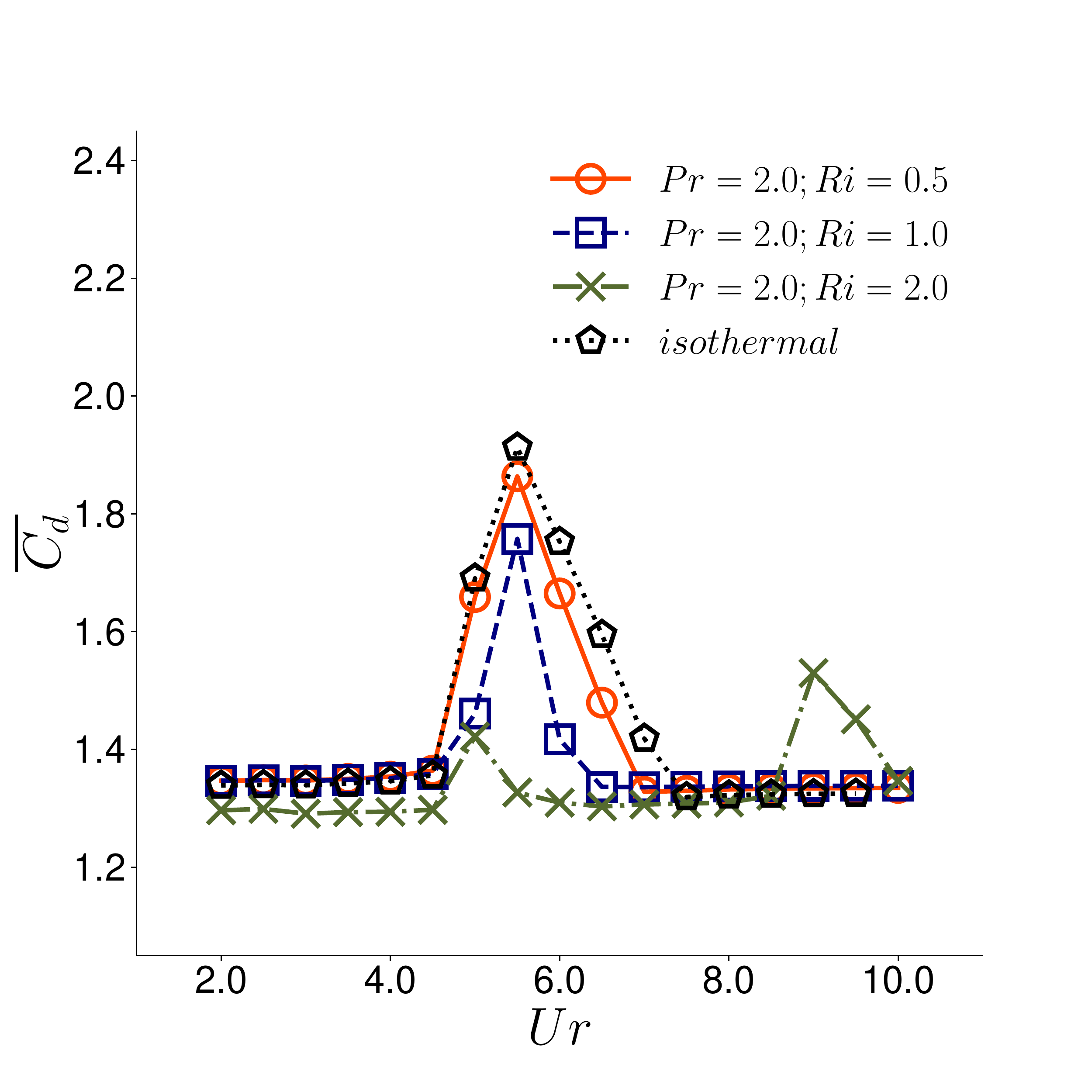}
	\caption{}
	\label{fig:Cd3}
\end{subfigure}%
\begin{subfigure}{0.5\textwidth}
	\centering
	\includegraphics[trim=0.0cm 0.1cm 0.1cm 0.1cm,scale=0.25,clip]{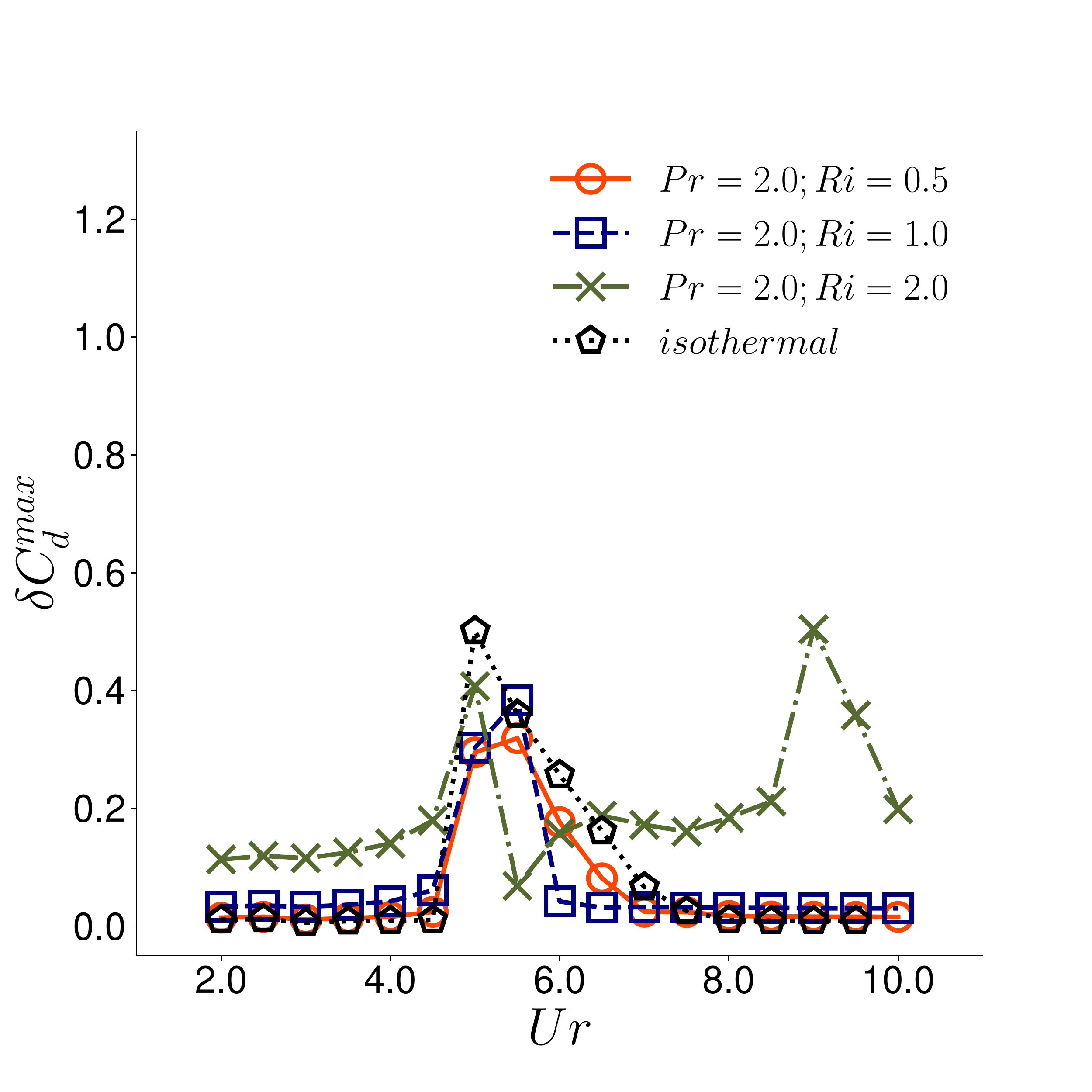}
	\caption{}
	\label{fig:Cd4}
\end{subfigure}
\begin{subfigure}{0.5\textwidth}
	\centering
	\includegraphics[trim=0.0cm 0.1cm 0.1cm 0.1cm,scale=0.25,clip]{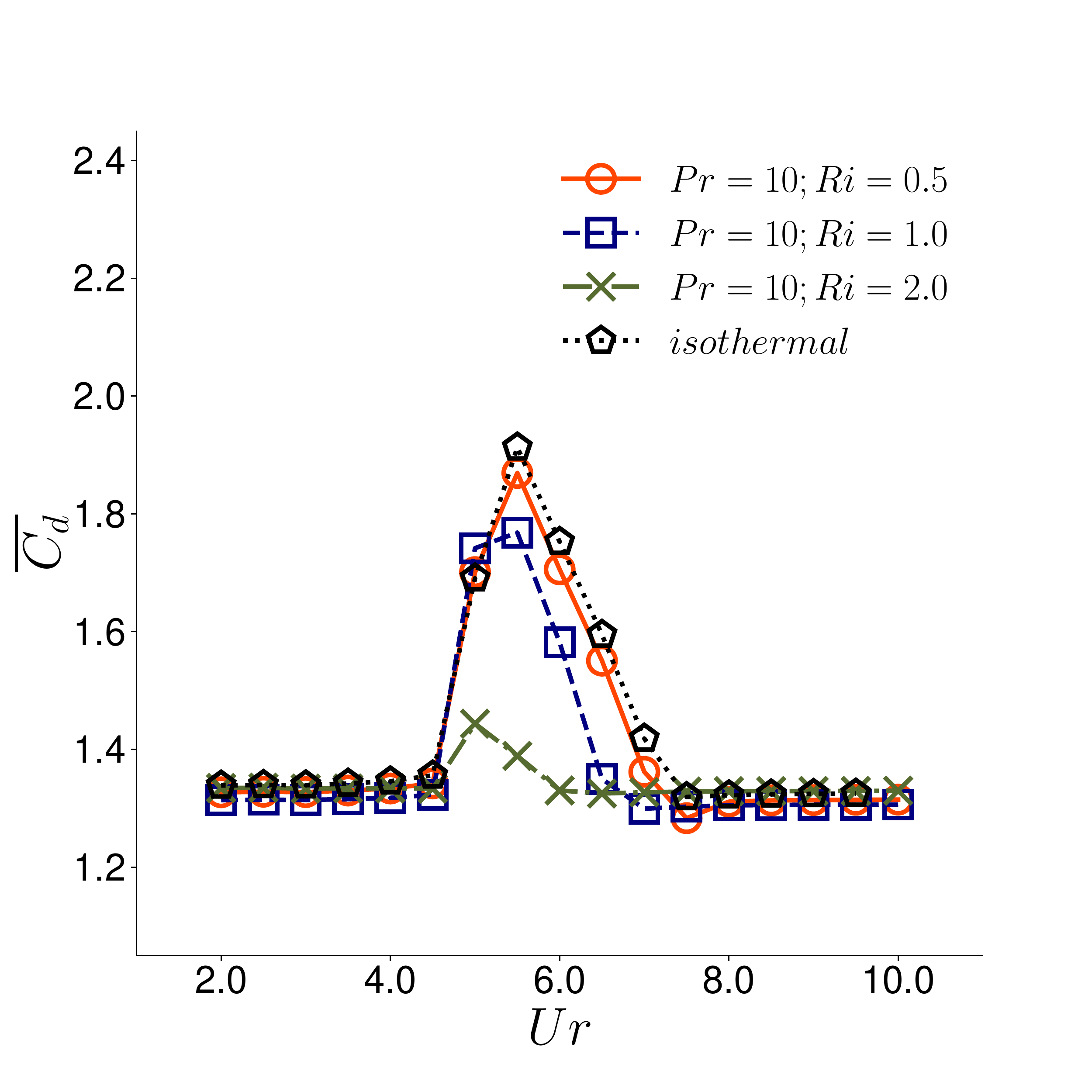}
	\caption{}
	\label{fig:Cd5}
\end{subfigure}%
\begin{subfigure}{0.5\textwidth}
	\centering
	\includegraphics[trim=0.0cm 0.1cm 0.1cm 0.1cm,scale=0.25,clip]{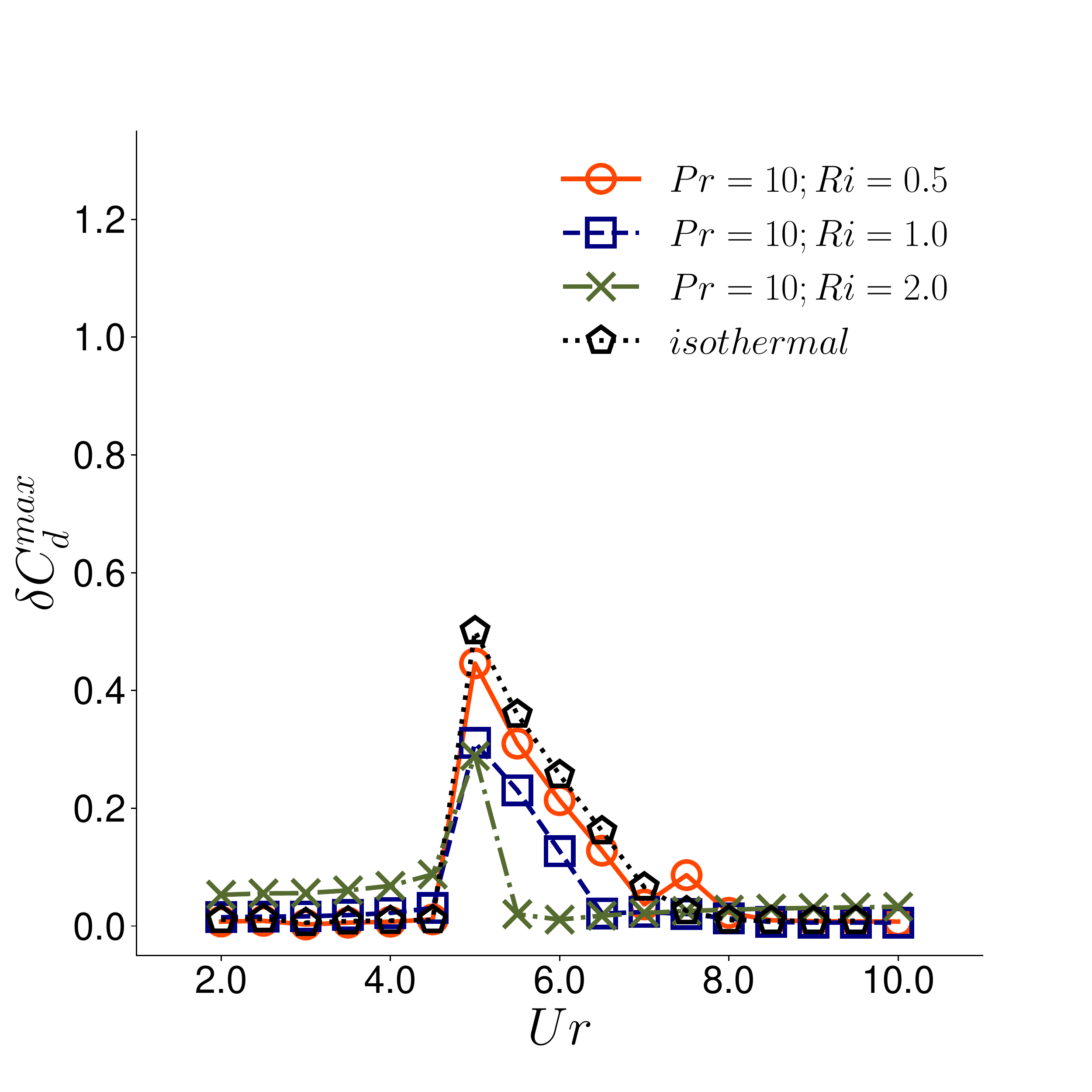}
	\caption{}
	\label{fig:Cd6}
\end{subfigure}
	\caption{Drag coefficients of a freely-vibrating cylinder at $Re=100$, $m^*=10$, $\zeta = 0.01$, $Ur \in [2, 10]$, $Pr \in [0.7, 10]$ and $Ri \in [0.5, 2.0]$: (a, c, e) the mean drag coefficient and (b, d, f) the maximum fluctuation of drag coefficient}
	\label{fig:Cd}
\end{figure*}
It was extensively reported in literature that a balanced state is achieved between the hydrodynamic forces and structural dynamics in isothermal and incompressible flow, especially during the self-limiting VIV lock-in. In mixed convection flow, the non-trivial buoyancy force in transverse direction is expected to perturb this balance and elevates the complexity of shear-layer mixing. As illustrated by the dotted black line with \emph{pentagon} markers in Figure~\ref{fig:Cd1}, the mean drag coefficient ($\overline{C_d}$) in isothermal flow is about $1.32$ for a wide range of $Ur$ values, except for the VIV lock-in. On the other hand, in mixed convection flow, as Richardson number surges, the peak value of $\overline{C_d}$ during VIV lock-in reduces significantly. In cases of low Richardson numbers $Ri < 2.0$ in Figure~\ref{fig:Cd}(a, c, e), the value of $\overline{C_d}$ becomes very insensitive to the values of $Ur$, $Pr$ and $Ri$ during off lock-in regions. Here, the off lock-in region refers to the ranges of $Ur$ values outside VIV lock-in regions. In contrast, an apparent interference of buoyancy-driven flow happens for $Ri = 2.0$. In the case of $Pr = 0.7$ and $Ri = 2.0$, the value of $\overline{C_d}$ gets excited within the secondary VIV lock-in region ($Ur \in [7.0, 9.5]$) and becomes approximately $1.18$ for the off lock-in regions, as depicted in Figure~\ref{fig:Cd1}. The increment of Prandtl number remarkably weakens the influence of buoyancy force on $\overline{C_d}$ for off lock-in regions in Figure~\ref{fig:Cd} (a, c, e). However, overall, the peaks of $\overline{C_d}$ during VIV lock-in are apparently suppressed in cases of high Richardson numbers. 

Similar to the observation in $\overline{C_d}$, the maximum fluctuations of drag coefficient ($\delta C^{max}_d$) are close to those in isothermal flow and insensitive to the values of $Ur$, $Pr$ and $Ri$ in the cases of $Ri < 2.0$ during off lock-in regions. Nonetheless, the suppression of $\delta C^{max}_d$ can still be observed during VIV lock-in, because the influence of natural convection in Figure~\ref{fig:Cd}(b, d, f). Again, the secondary VIV lock-in region is confirmed for $Ri = 2.0$ in Figure~\ref{fig:Cd2} and Figure~\ref{fig:Cd4}, within which the values of $\delta C^{max}_d$ are excited tremendously. Figure~\ref{fig:Cd6} also confirms the conclusion drawn above, which says the VIV lock-in almost disappears in the cases of $Pr = 10.0$ and $Ri=2.0$. 

\begin{figure*}[!htp]
	\centering
	\begin{subfigure}{0.5\textwidth}
	\centering
	\includegraphics[trim=0.0cm 0.1cm 0.1cm 0.1cm,scale=0.25,clip]{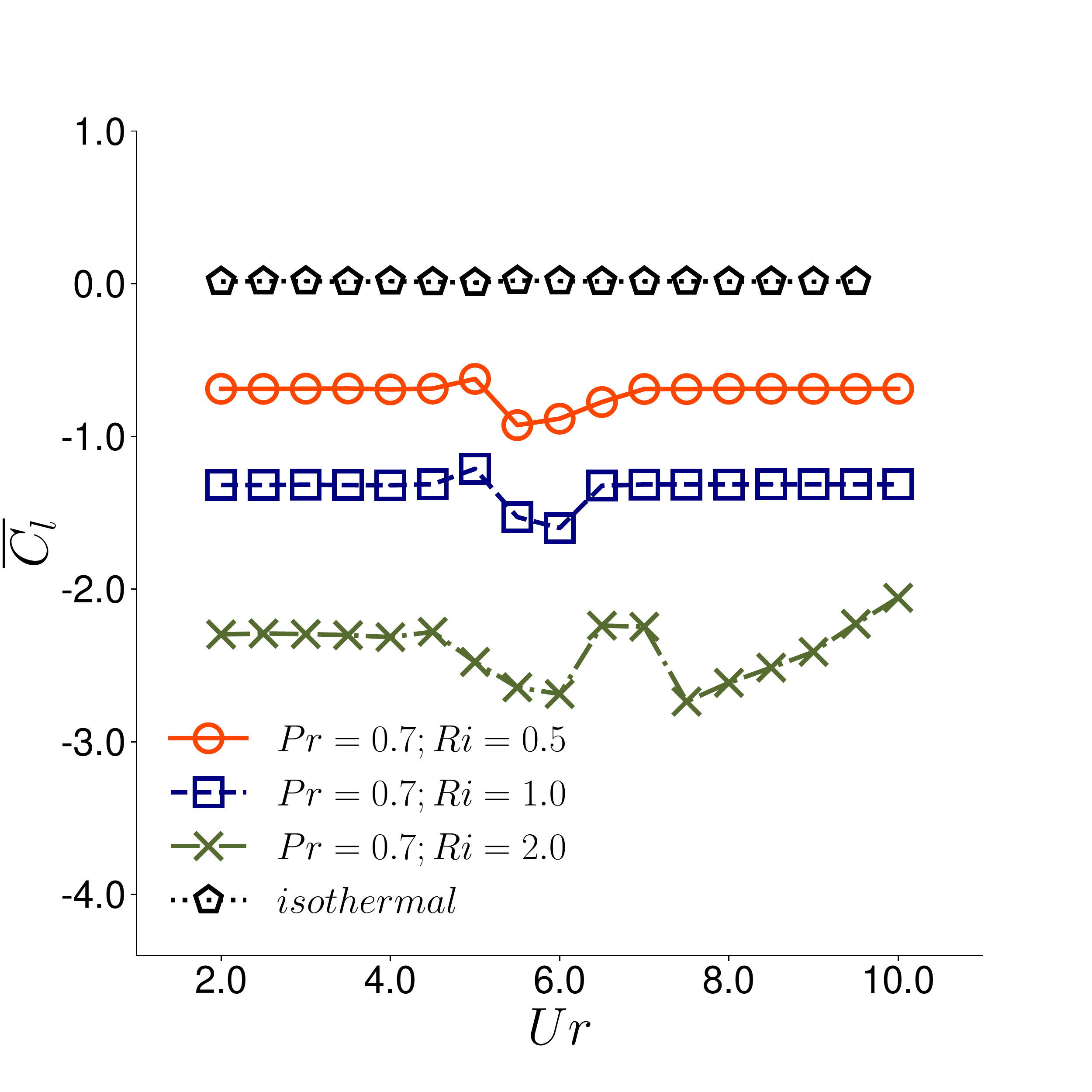}
	\caption{}
	\label{fig:Cl1}
\end{subfigure}%
\begin{subfigure}{0.5\textwidth}
	\centering
	\includegraphics[trim=0.0cm 0.1cm 0.1cm 0.1cm,scale=0.25,clip]{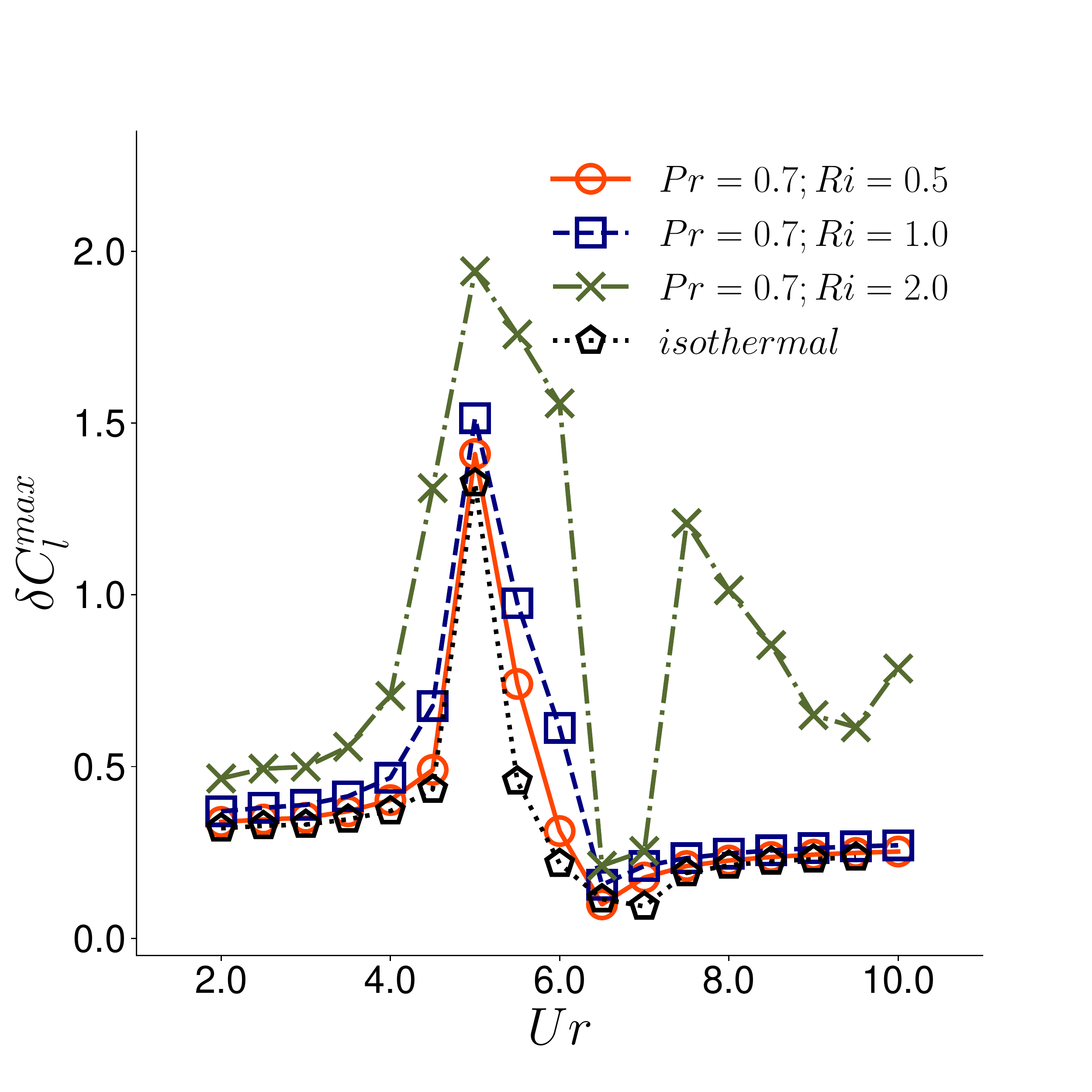}
	\caption{}
	\label{fig:Cl2}
\end{subfigure}
\begin{subfigure}{0.5\textwidth}
	\centering
	\includegraphics[trim=0.0cm 0.1cm 0.1cm 0.1cm,scale=0.25,clip]{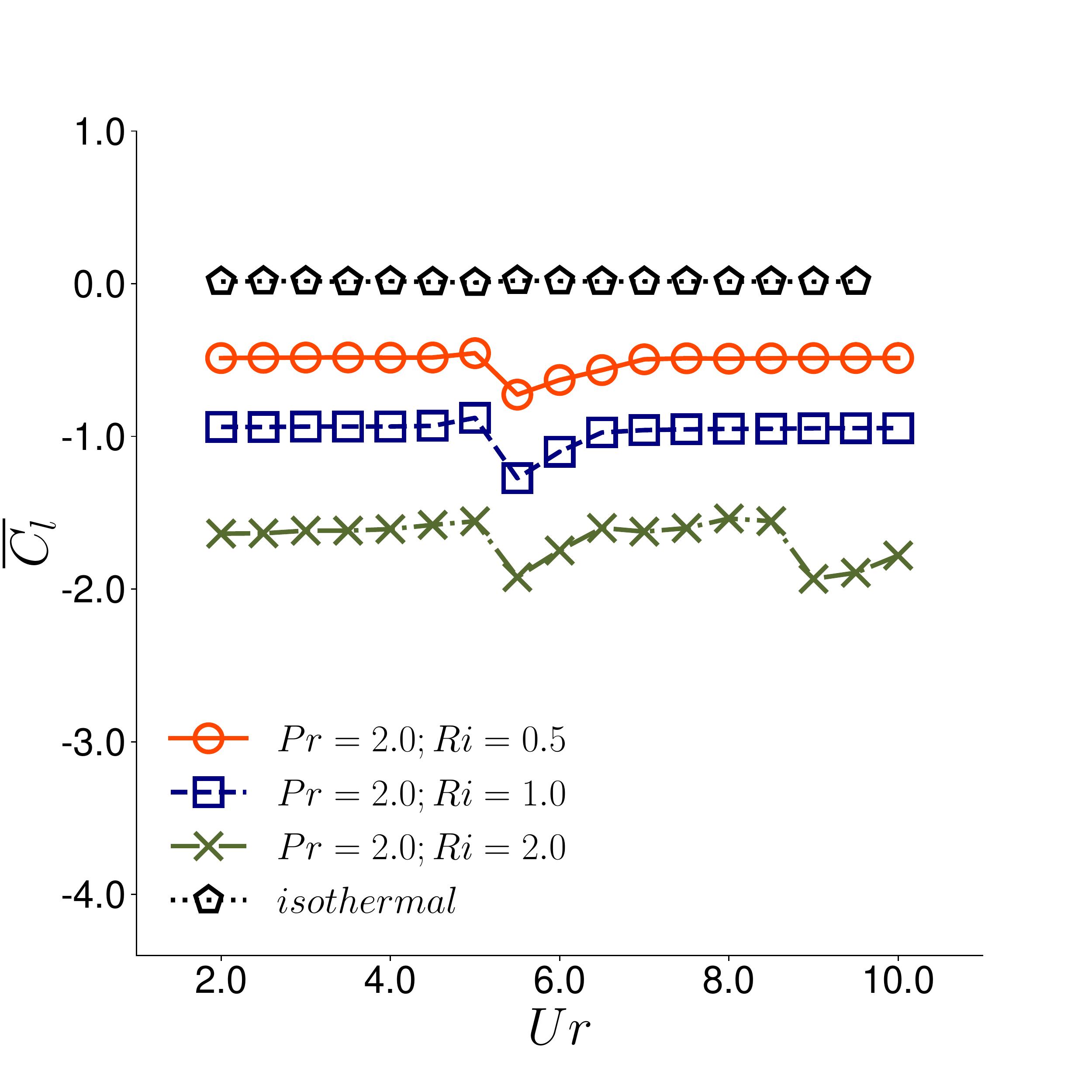}
	\caption{}
	\label{fig:Cl3}
\end{subfigure}%
\begin{subfigure}{0.5\textwidth}
	\centering
	\includegraphics[trim=0.0cm 0.1cm 0.1cm 0.1cm,scale=0.25,clip]{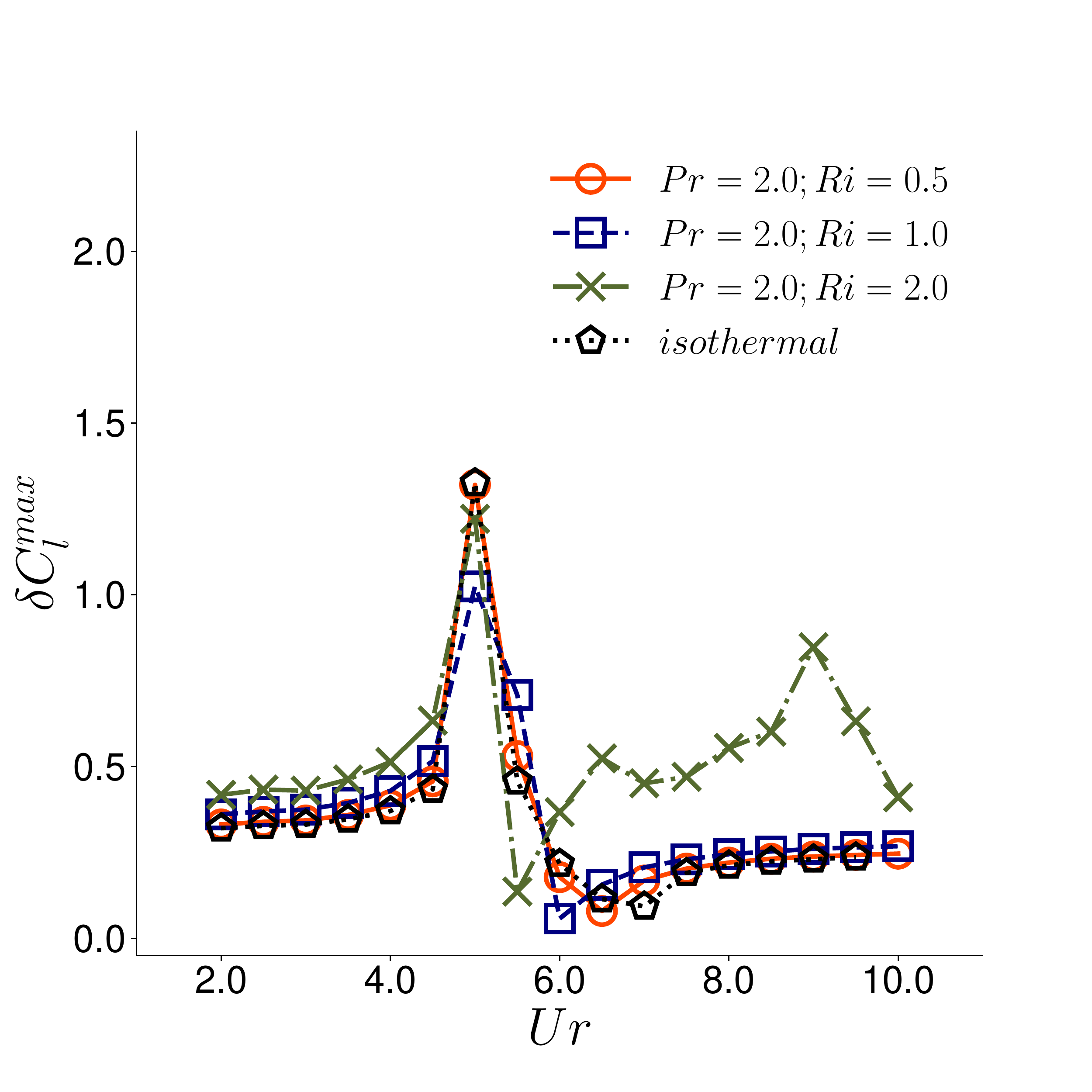}
	\caption{}
	\label{fig:Cl4}
\end{subfigure}
\begin{subfigure}{0.5\textwidth}
	\centering
	\includegraphics[trim=0.0cm 0.1cm 0.1cm 0.1cm,scale=0.25,clip]{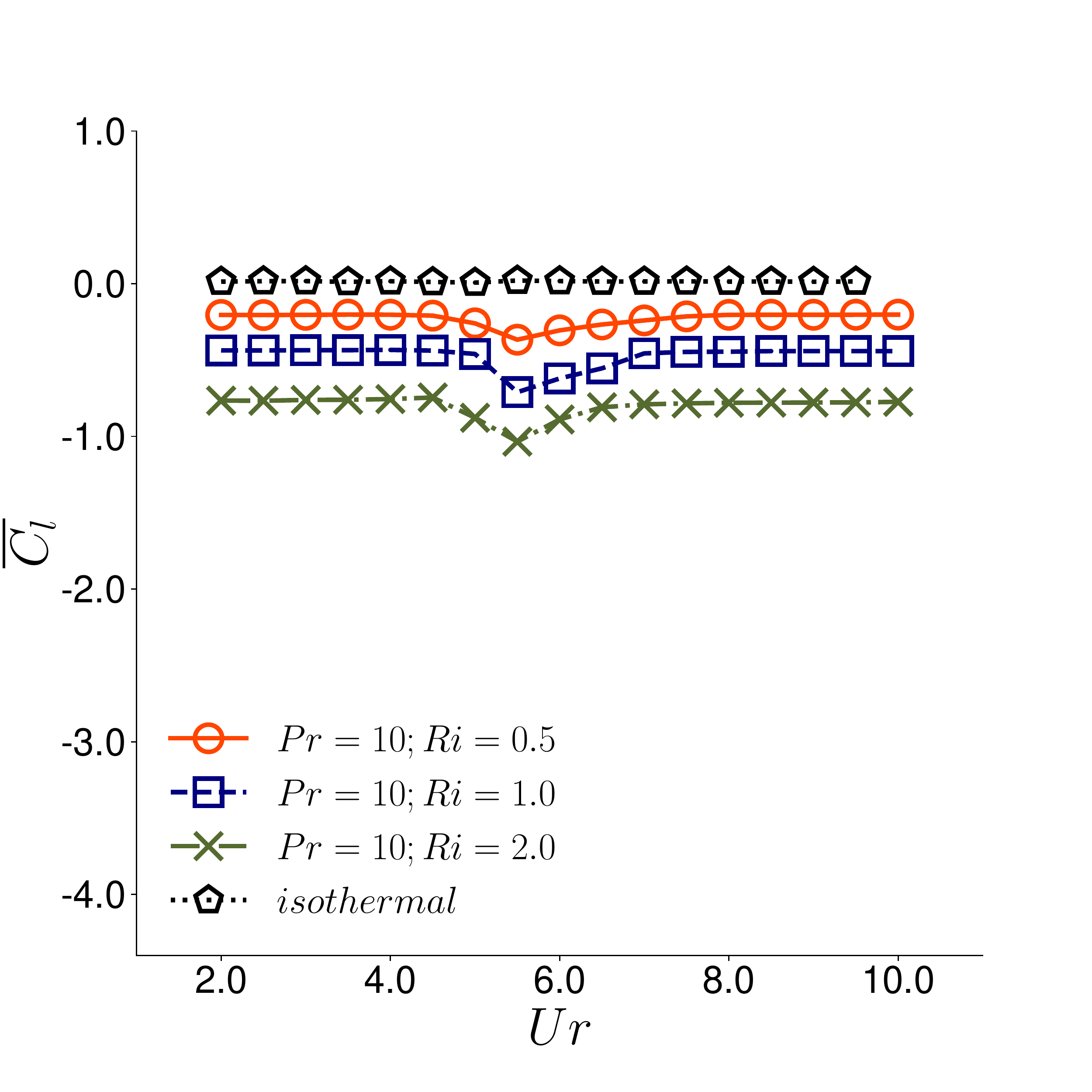}
	\caption{}
	\label{fig:Cl5}
\end{subfigure}%
\begin{subfigure}{0.5\textwidth}
	\centering
	\includegraphics[trim=0.0cm 0.1cm 0.1cm 0.1cm,scale=0.25,clip]{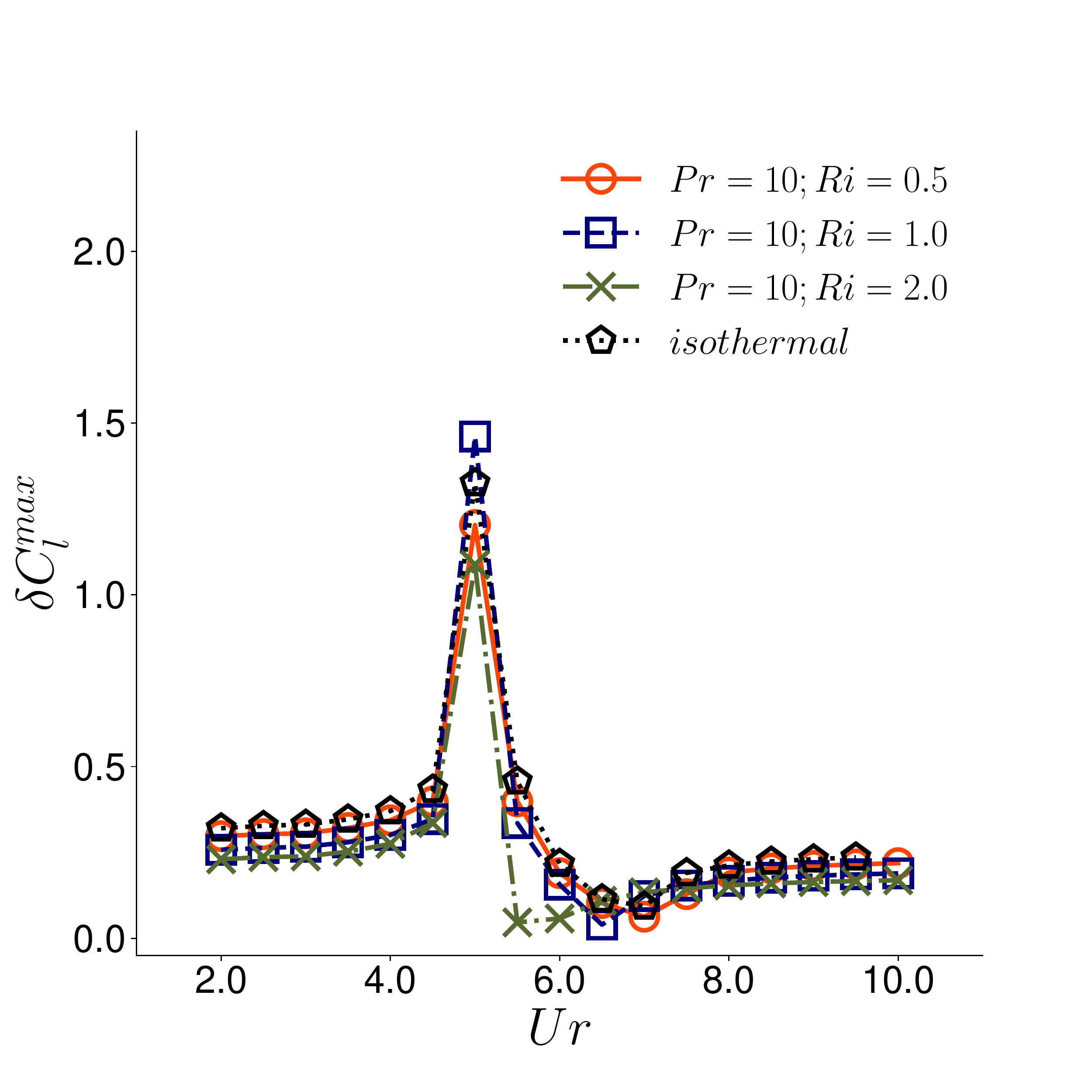}
	\caption{}
	\label{fig:Cl6}
\end{subfigure}
	\caption{Lift coefficients of a freely-vibrating cylinder at $Re=100$, $m^*=10$, $\zeta = 0.01$, $Ur \in [2, 10]$, $Pr \in [0.7, 10]$ and $Ri \in [0.5, 2.0]$: (a, c, e) the mean lift coefficient and (b, d, f) the maximum fluctuation of lift coefficient}
	\label{fig:Cl}
\end{figure*}
\begin{figure}[!htp]
	\centering
	\begin{subfigure}{0.5\textwidth}
	\centering
	\includegraphics[trim=0.cm 0.1cm 0.1cm 0.1cm,scale=0.25,clip]{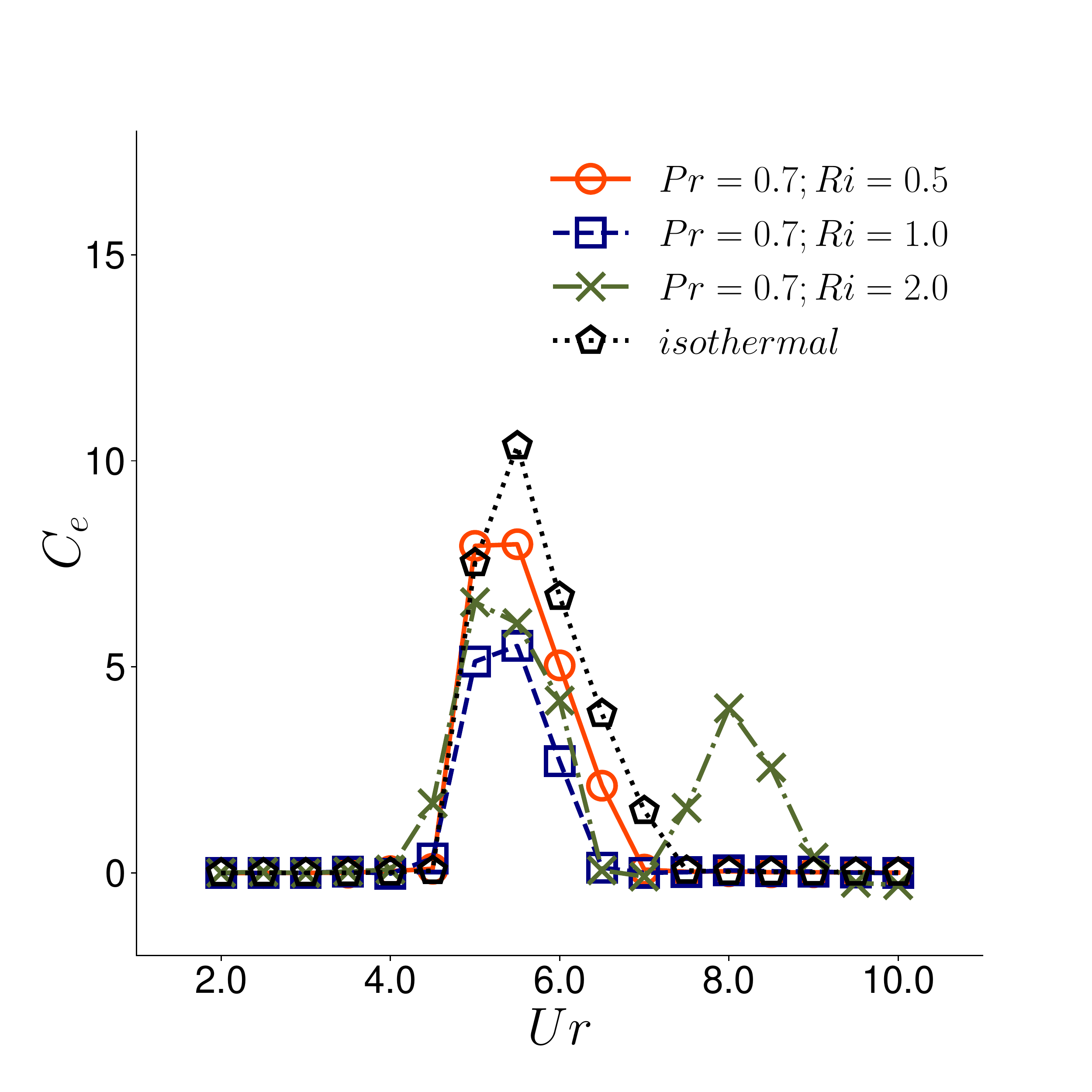}
	\caption{}
	\label{fig:Ce1}
\end{subfigure}
\begin{subfigure}{0.5\textwidth}
	\centering
	\includegraphics[trim=0.cm 0.1cm 0.1cm 0.1cm,scale=0.25,clip]{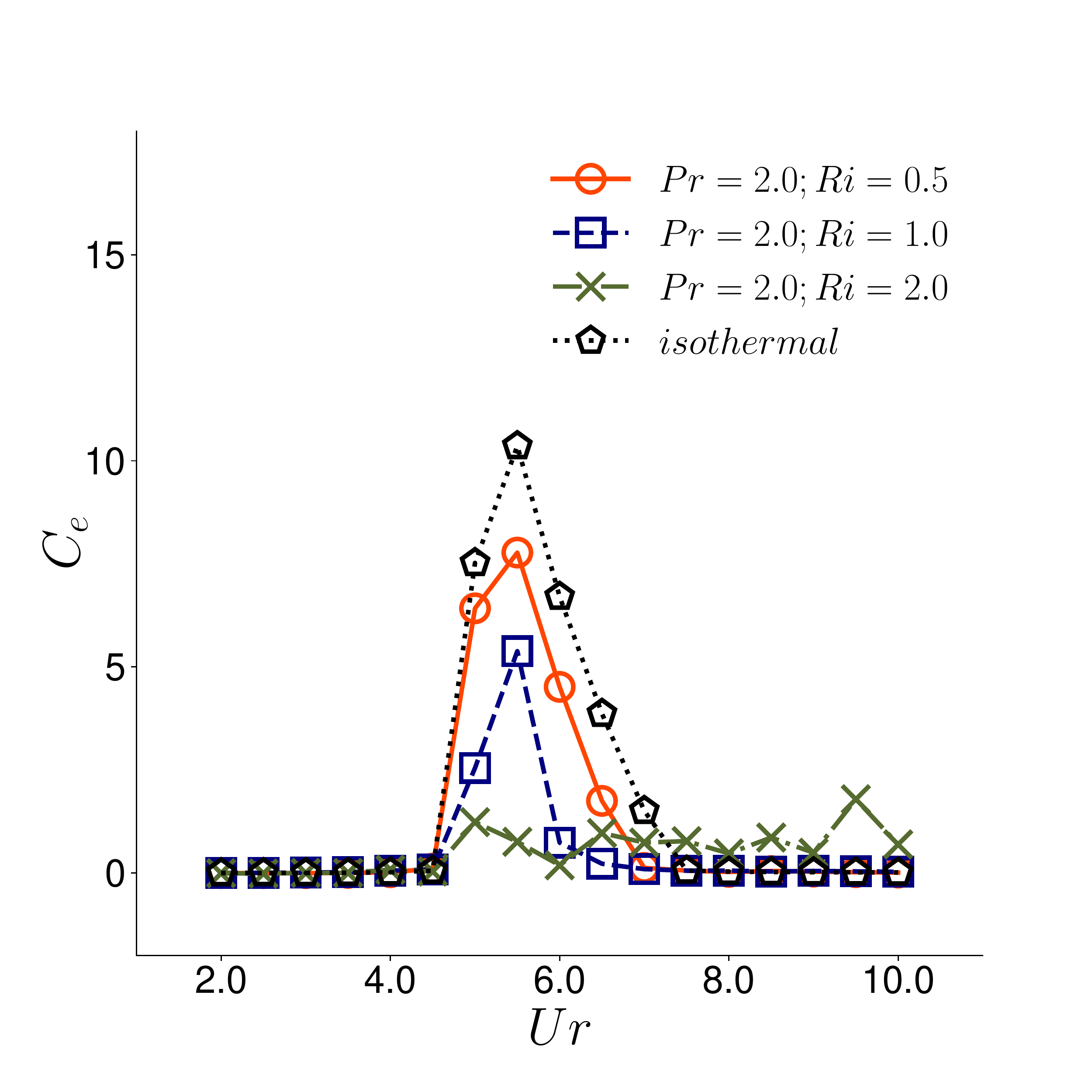}
	\caption{}
	\label{fig:Ce2}
\end{subfigure}
\begin{subfigure}{0.5\textwidth}
	\centering
	\includegraphics[trim=0.cm 0.1cm 0.1cm 0.1cm,scale=0.25,clip]{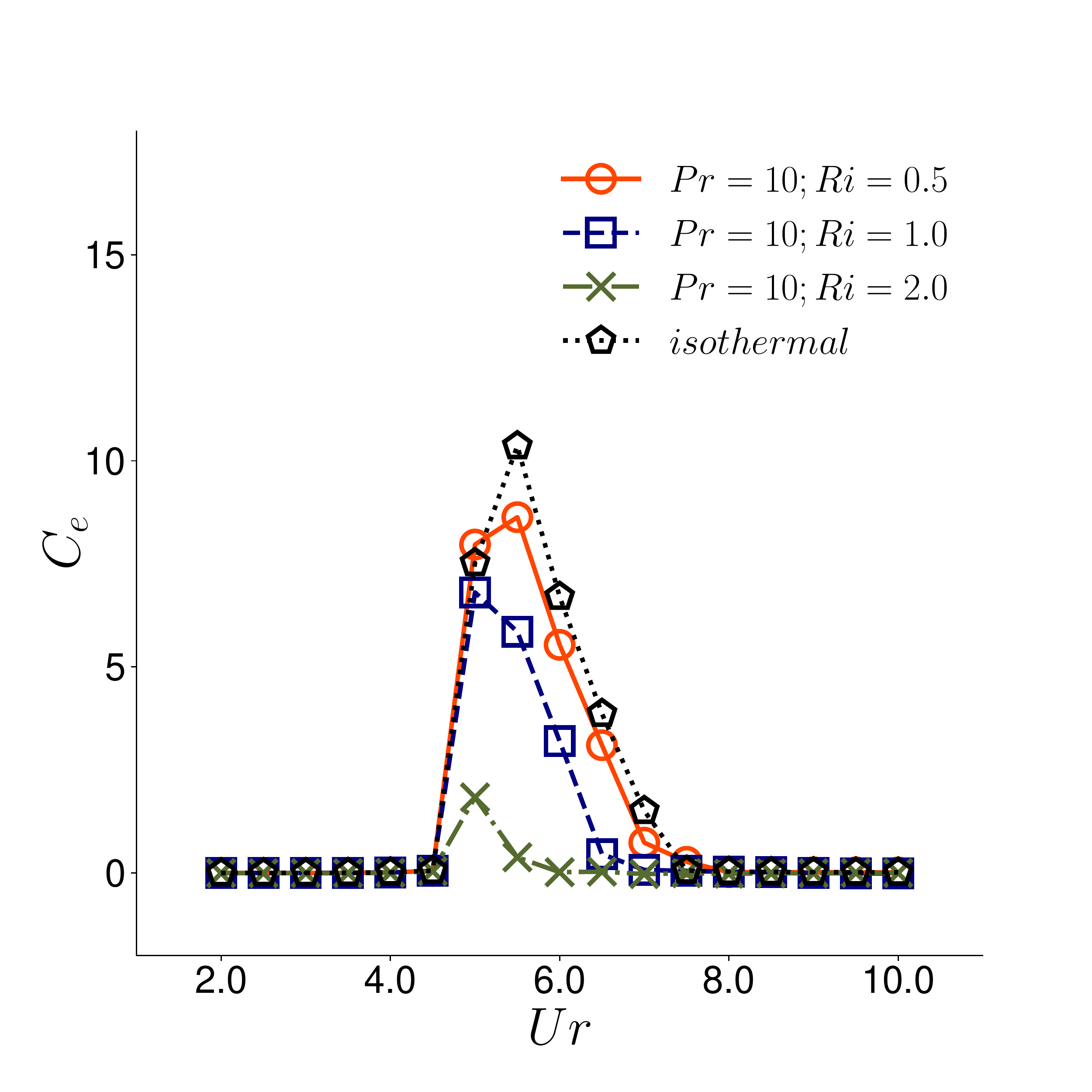}
	\caption{}
	\label{fig:Ce3}
\end{subfigure}
	\caption{Energy transfer between fluid and structure of a freely-vibrating cylinder in transverse direction for $\tau \in [200, 400]$ at $Re=100$, $m^*=10$, $\zeta = 0.01$ and $Ur \in [2, 10]$: (a) $Pr = 0.7$ and $Ri \in [0.5, 2.0]$; (b) $Pr = 2.0$ and $Ri \in [0.5, 2.0]$ and (c) $Pr = 10$ and $Ri \in [0.5, 2.0]$}
	\label{fig:Ce}
\end{figure}
\begin{figure}[!htp]
	\centering
	\begin{subfigure}{0.5\textwidth}
	\centering
	\includegraphics[trim=0.cm 0.1cm 0.1cm 0.1cm,scale=0.25,clip]{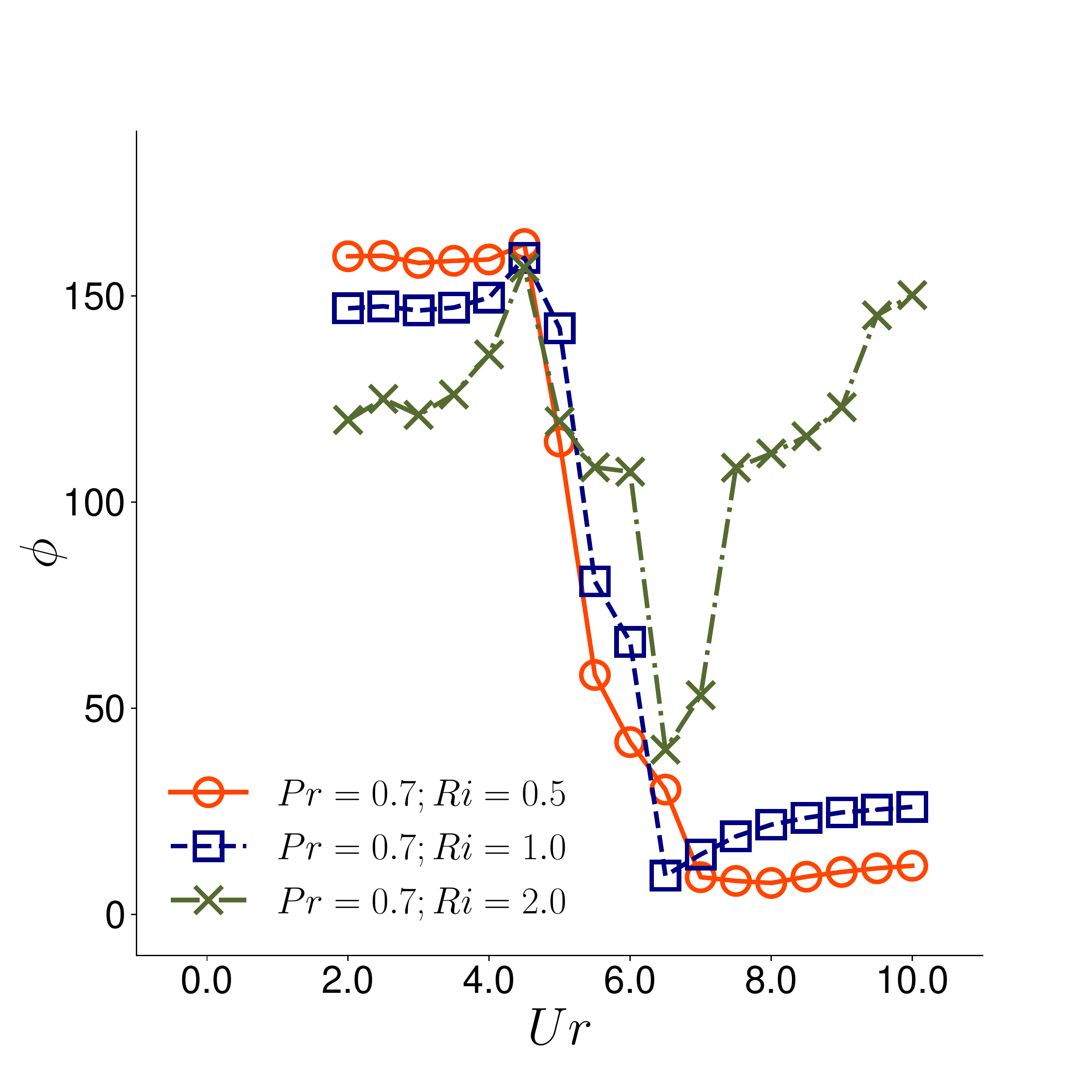}
	\caption{}
	\label{fig:phase1}
\end{subfigure}
\begin{subfigure}{0.5\textwidth}
	\centering
	\includegraphics[trim=0.cm 0.1cm 0.1cm 0.1cm,scale=0.25,clip]{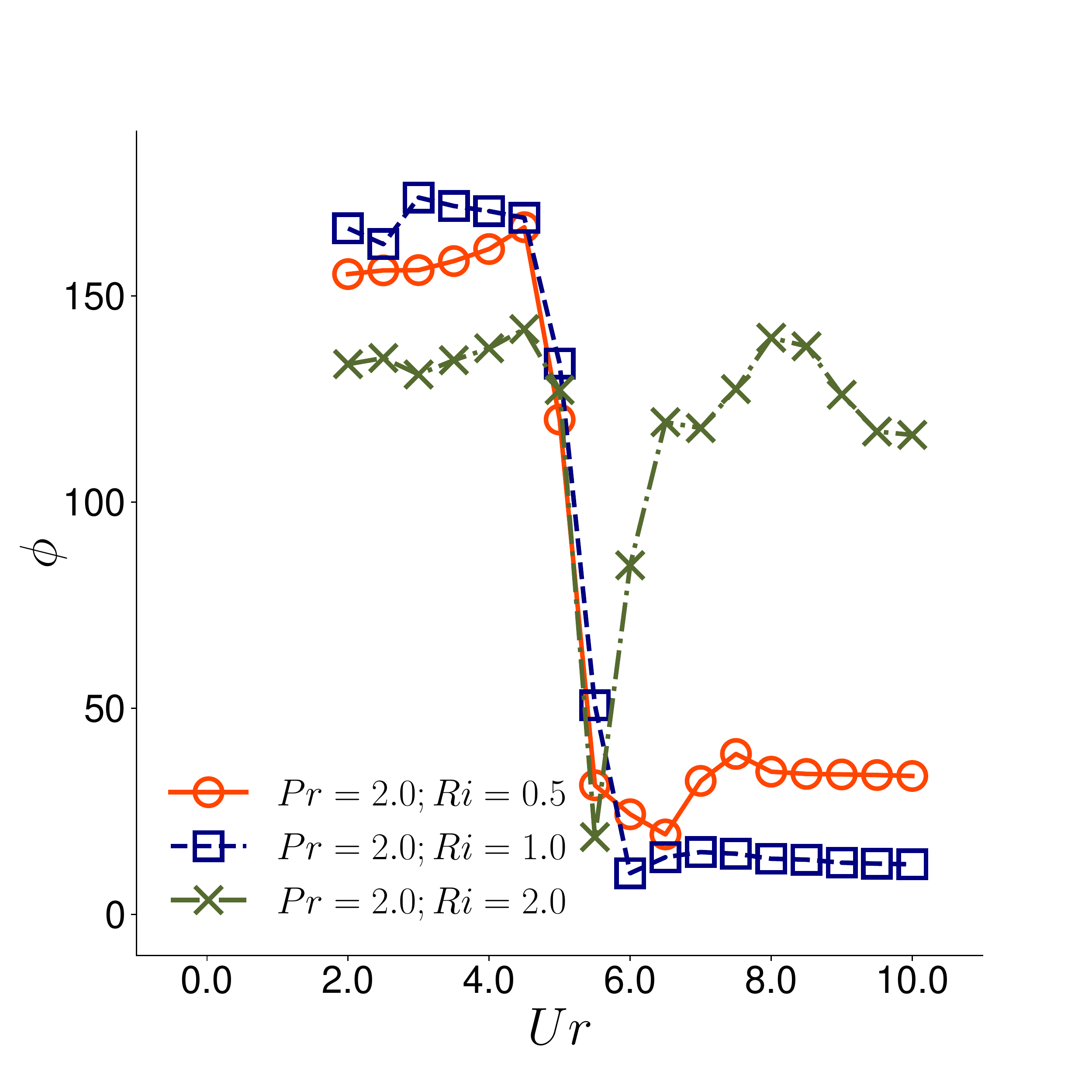}
	\caption{}
	\label{fig:phase2}
\end{subfigure}
\begin{subfigure}{0.5\textwidth}
	\centering
	\includegraphics[trim=0.cm 0.1cm 0.1cm 0.1cm,scale=0.25,clip]{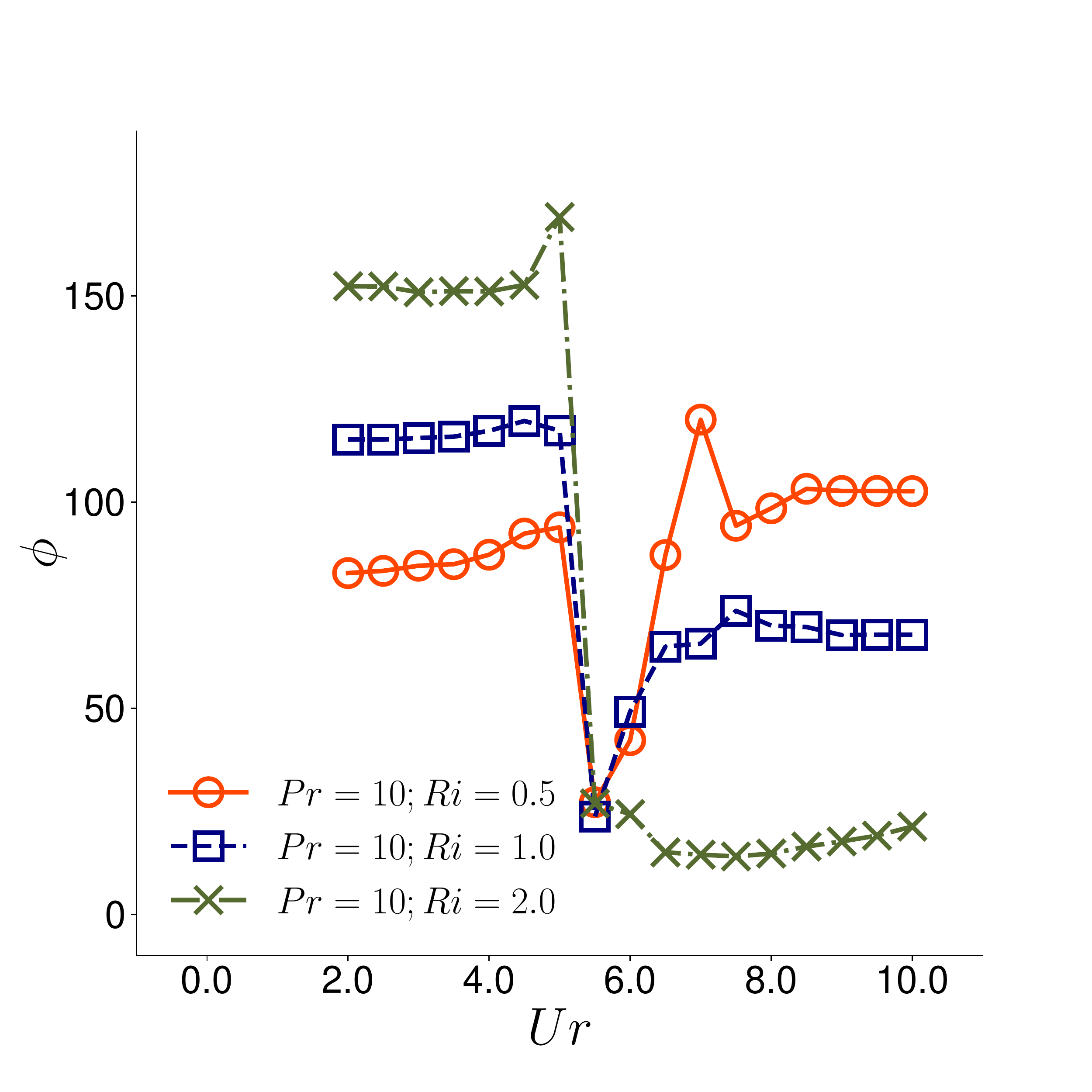}
	\caption{}
	\label{fig:phase3}
\end{subfigure}
	\caption{Phase angle difference between transverse displacement ($A_y$) and lift coefficient ($C_l$) for a freely-vibrating cylinder at $Re=100$, $m^*=10$, $\zeta = 0.01$ and $Ur \in [2, 10]$: (a) $Pr = 0.7$ and $Ri \in [0.5, 2.0]$; (b) $Pr = 2.0$ and $Ri \in [0.5, 2.0]$ and (c) $Pr = 10$ and $Ri \in [0.5, 2.0]$}
	\label{fig:phase}
\end{figure}
The buoyancy force induces fluid flow in the direction opposite to the gravity. Consequently, the responses of lift force is primarily perturbed by the buoyancy-driven flow  in wake. On the other hand, it is also known that the lift force is directly linked with vortex-shedding process, which potentially causes the onset of VIV lock-in. Hence, it is expected that the responses of lift force could be altered significantly for high Richardson numbers. Indeed, Figure~\ref{fig:Cl}(a,c,e) show that the increased influence of buoyancy-driven flow causes evident deviations of mean lift forces ($\overline{C_l}$) over a wide range of $Ur$ values, where the magnitude of $\overline{C_l}$ increases proportionally with Richardson number. Reversely, the values of $\overline{C_l}$ have a prominent tendency to recover the responses in isothermal flow for high Prandtl numbers. Recollecting the conclusion drawn in Section~\ref{sec:struct}, it is understood that high Prandtl number indicates shrunk temperature contours (less diffusive energy field over fluid momentum) and limited influence of heat energy field on hydrodynamics in wake. Hence this observed tendency of $\overline{C_l}$ with respect to Prandtl number should be appreciated, in which the responses of $\overline{C_l}$ get close to those in isothermal flow for higher Prandtl numbers. Overall, the values of $\overline{C_l}$ become relatively excited during VIV lock-in. However, a second amplification of $\overline{C_l}$ is found during the secondary VIV lock-in over  $Ur \in [8.0, 10]$ in the cases of $Pr < 10$ and $Ri = 2.0$ in Figure~\ref{fig:Cl2} and Figure~\ref{fig:Cl4}.     

Compared with the value or $\overline{C_l}$, the magnitude of maximum fluctuation ($\delta C^{max}_l$) is of a primary concern in the analyses of structural and hydrodynamic stability. In Figure~\ref{fig:Cl}(b, d, f), similar to those in isothermal flow, the value of $\delta C^{max}_l$ is excited during VIV lock-in and retains at a very low values during off lock-in regions, especially the post lock-in region. Furthermore, we found that the change of Prandtl number has no significant influence to the value of $\delta C^{max}_l$ over a wide range of $Ur$ values, regardless VIV lock-in or off lock-in regions. Overall, no suppression of $\delta C^{max}_l$ is observed in cases of different Prandtl and Richardson numbers. Instead, the values of $\delta C^{max}_l$ is amplified in cases of $Ri = 2.0$ in Figure~\ref{fig:Cl2} and Figure~\ref{fig:Cl4} for high $Ur$ values exceeding the primary VIV lock-in, except the case of $Pr = 10$  in Figure~\ref{fig:Cl6}. 

\subsection{fluid and heat energy transfer in mixed convection flow} \label{sec:thermal}
\begin{figure*}[!htp]
	\centering
	\begin{subfigure}{0.5\textwidth}
	\centering
	\includegraphics[trim=0.0cm 0.1cm 0.1cm 0.1cm,scale=0.25,clip]{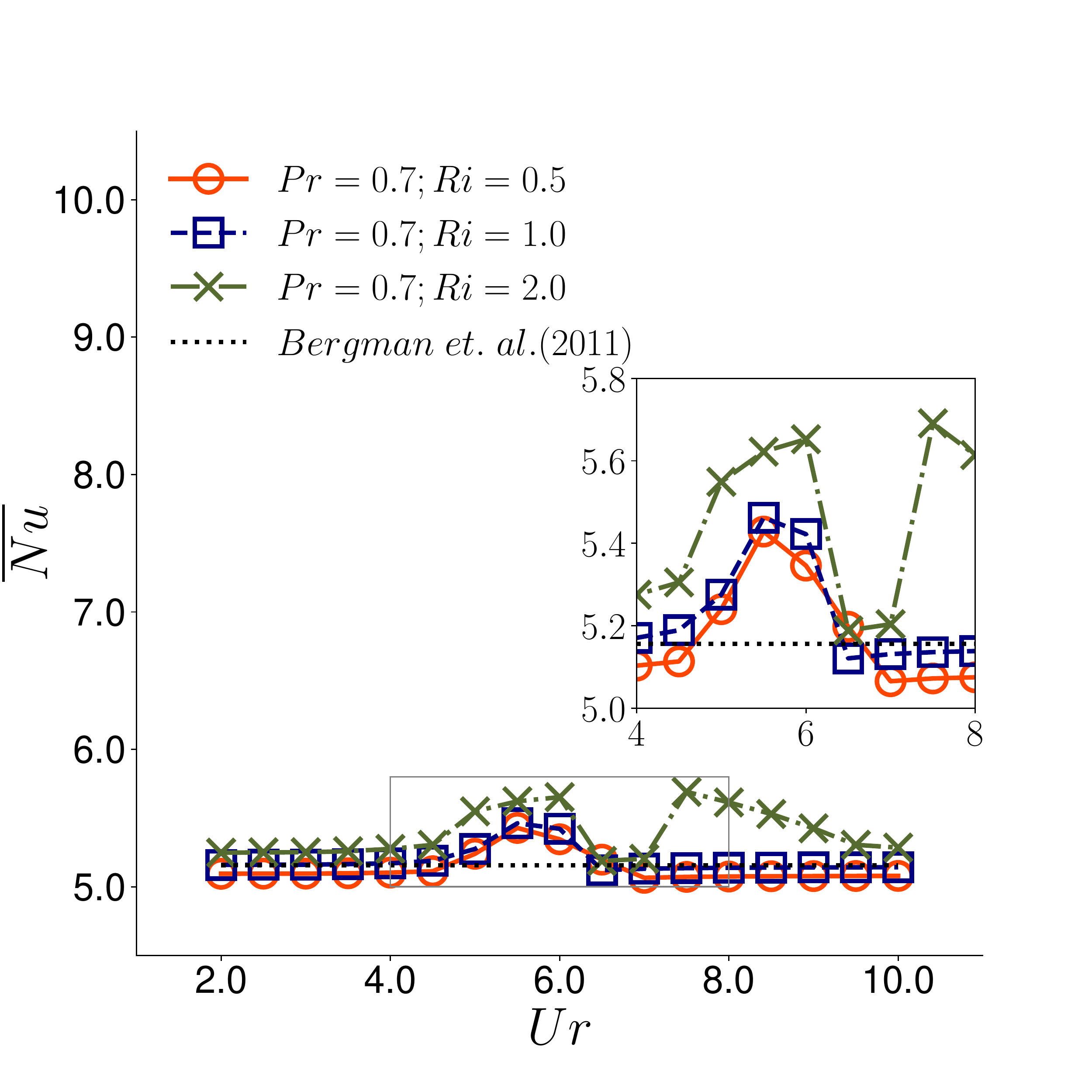}
	\caption{}
	\label{fig:Nu1}
\end{subfigure}%
\begin{subfigure}{0.5\textwidth}
	\centering
	\includegraphics[trim=0.0cm 0.1cm 0.1cm 0.1cm,scale=0.25,clip]{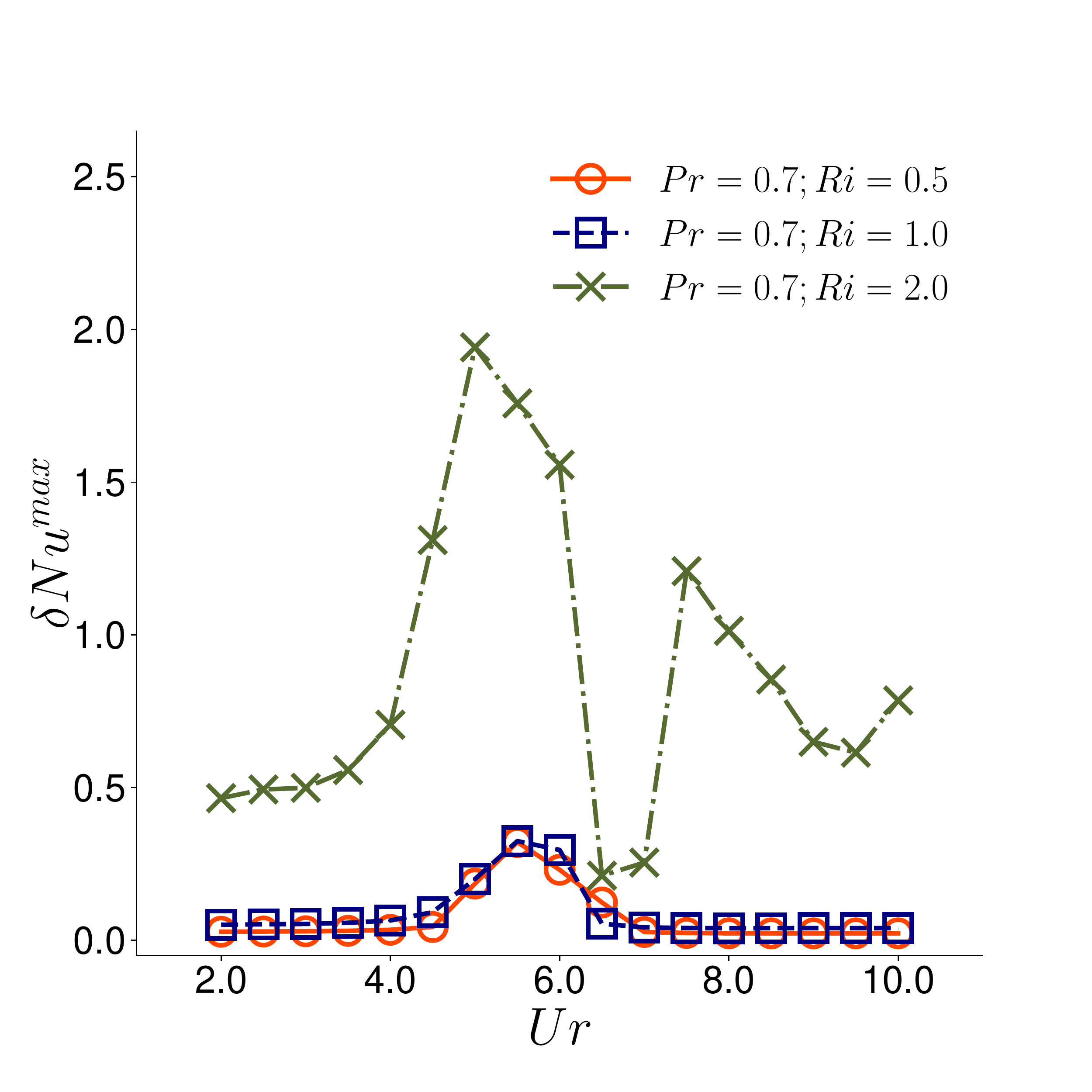}
	\caption{}
	\label{fig:Nu2}
\end{subfigure}
\begin{subfigure}{0.5\textwidth}
	\centering
	\includegraphics[trim=0.0cm 0.1cm 0.1cm 0.1cm,scale=0.25,clip]{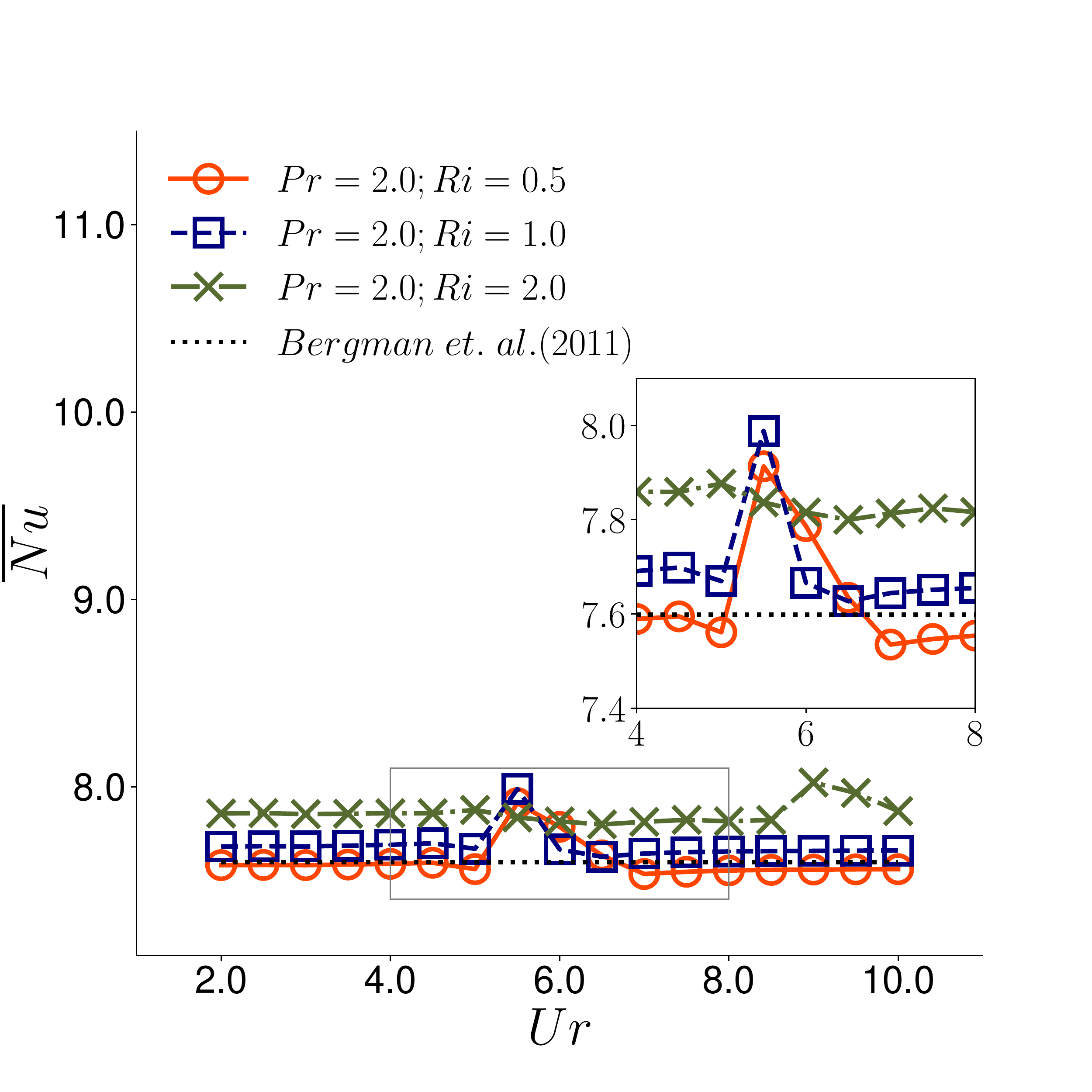}
	\caption{}
	\label{fig:Nu3}
\end{subfigure}%
\begin{subfigure}{0.5\textwidth}
	\centering
	\includegraphics[trim=0.0cm 0.1cm 0.1cm 0.1cm,scale=0.25,clip]{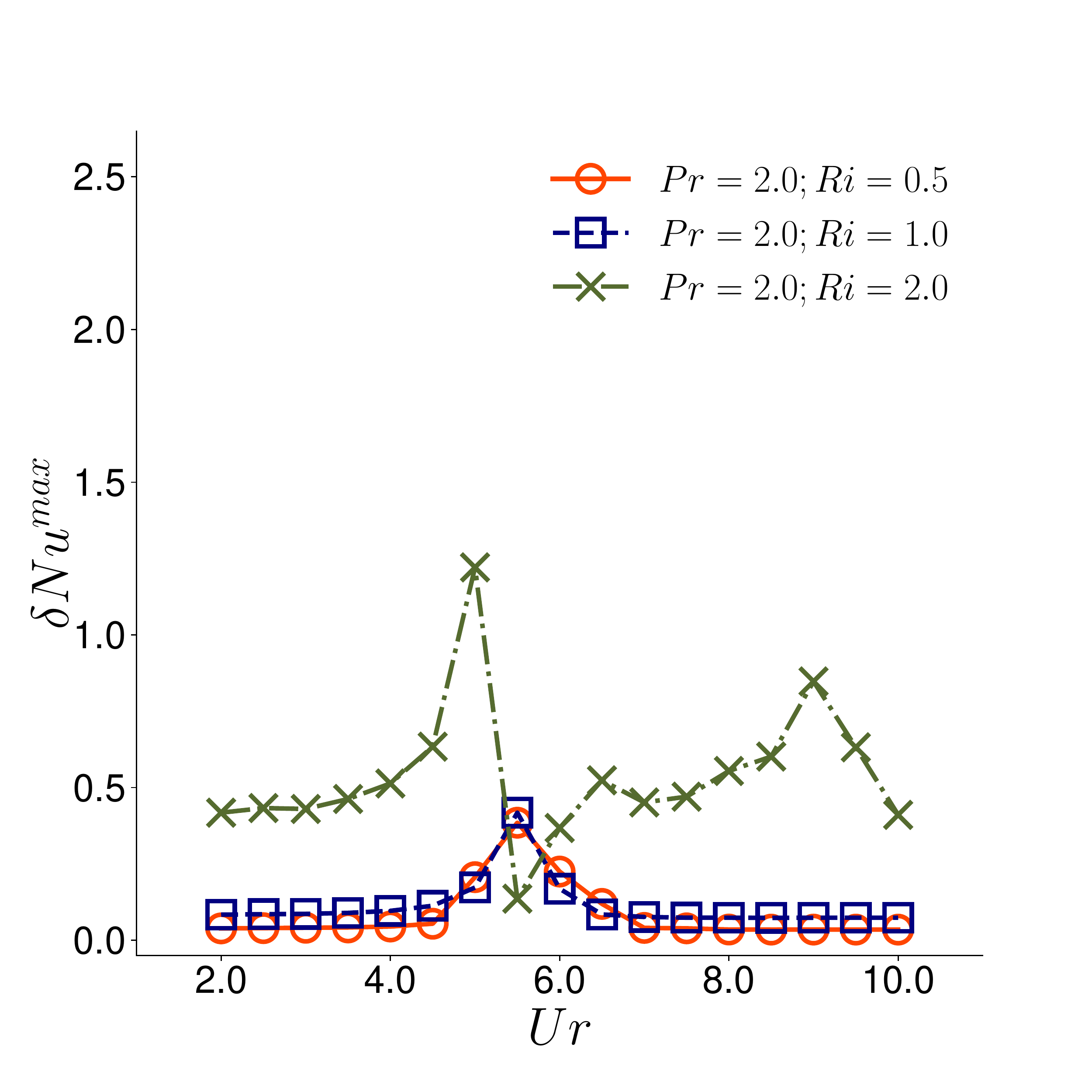}
	\caption{}
	\label{fig:Nu4}
\end{subfigure}
\begin{subfigure}{0.5\textwidth}
	\centering
	\includegraphics[trim=0.0cm 0.1cm 0.1cm 0.1cm,scale=0.25,clip]{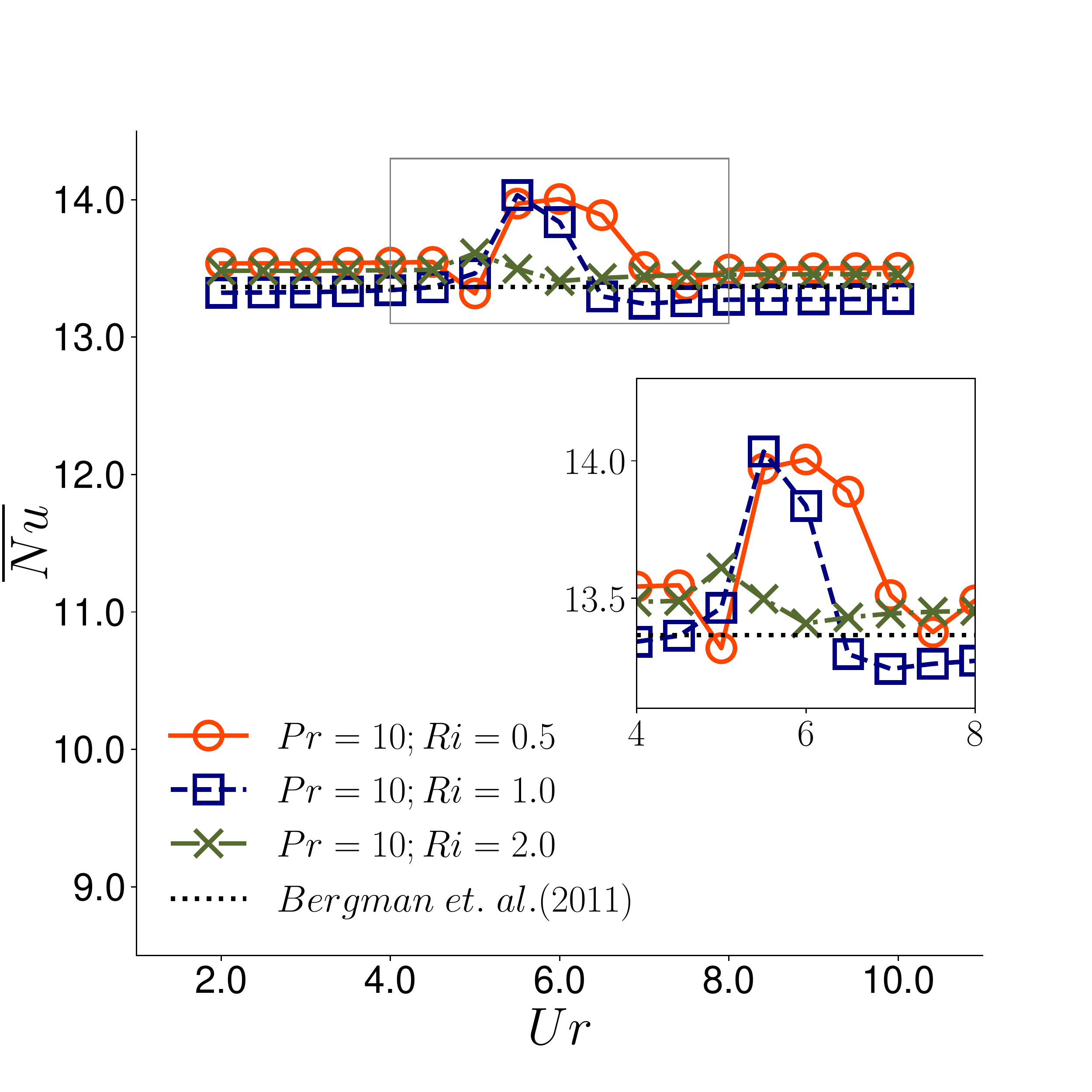}
	\caption{}
	\label{fig:Nu5}
\end{subfigure}%
\begin{subfigure}{0.5\textwidth}
	\centering
	\includegraphics[trim=0.0cm 0.1cm 0.1cm 0.1cm,scale=0.25,clip]{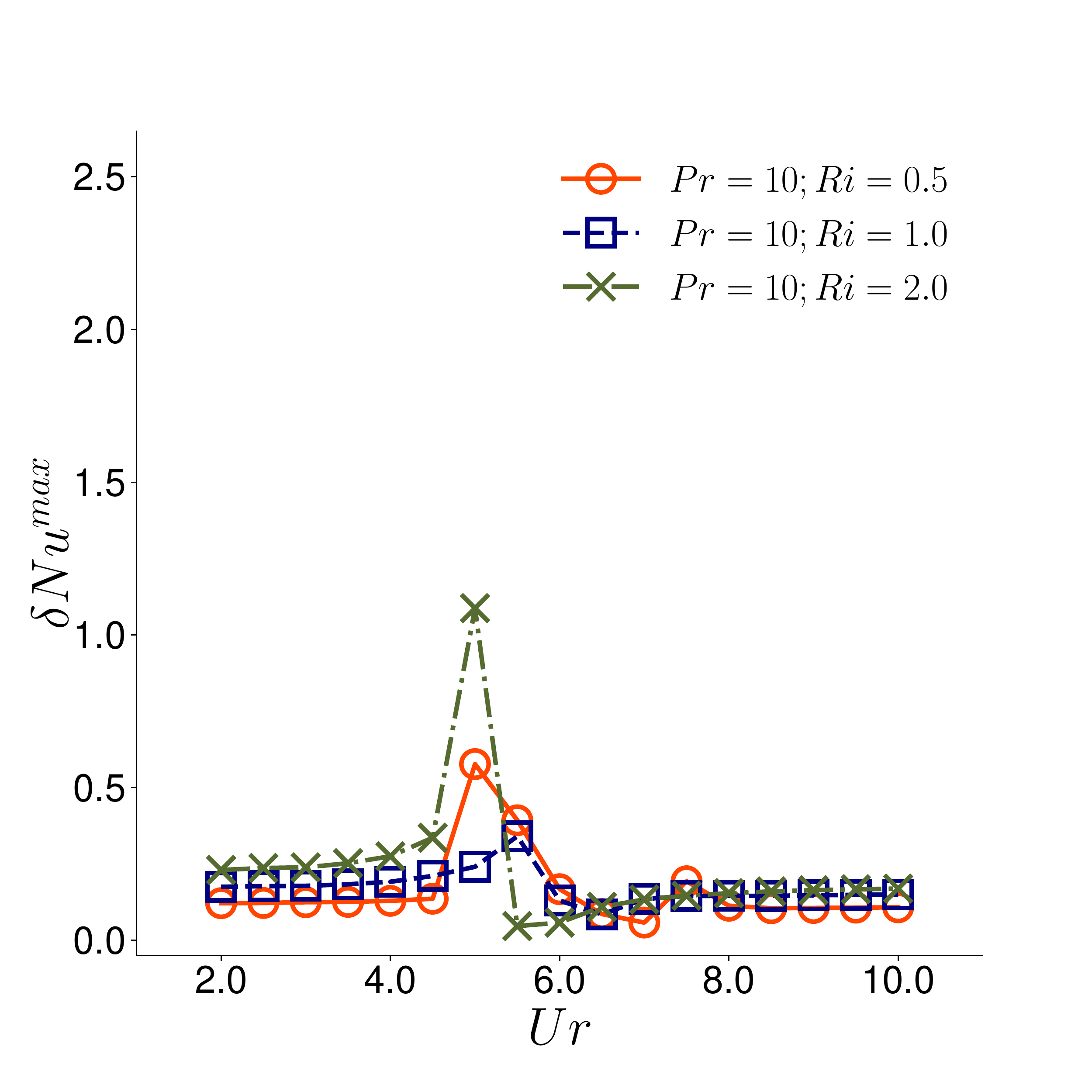}
	\caption{}
	\label{fig:Nu6}
\end{subfigure}
	\caption{Nusselt number of a freely-vibrating cylinder at $Re=100$, $m^*=10$, $\zeta = 0.01$, $Ur \in [2, 10]$, $Pr \in [0.7, 10]$ and $Ri \in [0.5, 2.0]$: (a, c, e) the mean Nusselt number and (b, d, f) the maximum fluctuation of Nusselt number}
	\label{fig:Nu}
\end{figure*}
Based on the previous analyses, it is found the heat energy field has a tremendous influence to the structural dynamics and hydrodynamics in wake, especially for high reduced velocity values, the typical post lock-in region in isothermal flow. This observation is critical. Normally, it was reported that the energy transfer between fluid and structure is inhibited during pre and post lock-in regions in isothermal flow~\cite{liu2018dynamics}. The appearance of secondary VIV lock-in region means that the fluid kinetic energy is further transferred into structures in mixed convection flow for high $Ur$ values. This expectation is affirmatively supported by the results in Figure~\ref{fig:Ce}, where the energy transfer between fluid and structure in transverse direction for $\tau \in [200, 400]$ (more than 30 cycles of vortex shedding) is plotted with respect to the reduced velocity values. It can be seen that a significant amount of fluid kinetic energy is transferred into the structure in the case of $Pr = 0.7$ and $Ri=2.0$ for high reduced velocity values ($Ur \in [7.0, 9.0]$) in Figure~\ref{fig:Ce1}, besides the primary VIV lock-in region. The kinetic energy transfer is excited at almost the same $Ur$ value for both mixed convection flow and isothermal flow, e.g., $Ur \approx 5.25$. In contrast, the amount of transferred fluid kinetic energy is significantly reduced in the cases of higher Richardson numbers. Especially, the fluid kinetic energy transfer is almost suppressed in the cases of high Prandtl and Richardson numbers, e.g., the green dotted line with \emph{cross} markers in Figure~\ref{fig:Ce3}. Consequently, the intensive buoyancy-driven flow in the cases of low Prandtl numbers widens the range of $Ur$ values for an effective transfer of kinetic energy between fluid and structure; whereas high Prandtl numbers could potentially inhibits the energy transfer and narrow the effective range of $Ur$ values, as plotted in Figure~\ref{fig:Ce3}. This finding is extremely meaningful in hydropower energy harvesting.               

On the other hand, the phase angle difference is another important indicator of energy transfer. It could be used to check if a signal provides positive feedback to one another. In this article, the instantaneous phase angle difference ($\phi$) between $A_y$ and $C_l$ is approximated by Hilbert-Huang Transform (HHT)~\cite{huang2014hilbert,liu2016interaction} and averaged over $\tau = [200,400]$. Liu \& Jaiman~\cite{liu2018dynamics} reported that the maximum energy transfer between fluid and structure occurs for $\phi \approx 90^{\circ}$, in which the cylinder could obtain the maximum acceleration from the fluid force. Based on the results in Figure~\ref{fig:phase}, overall, it is confirmed that maximum $C_e$ values indeed appears around the $Ur$ values where $\phi \approx 90^{\circ}$. Comparing Figure~\ref{fig:Ce} with Figure~\ref{fig:phase}, it is further realized that either in-phase or anti-phase phase angle difference is ineffective to transfer the fluid kinetic energy into structure, where very low $C_e$ values are associated. In addition, it is also found that the responses of $A_y$ and $C_l$ are generally anti-phase during pre lock-in regions for different $Pr$ and $Ri$ values. Nevertheless, the phase angle difference is much smaller during the pre lock-in in the cases of high Prandtl numbers in Figure~\ref{fig:phase3}. During the onset and end of energy transfer, the phase angle differences tend to switch their states, e.g., from in-phase to anti-phase or vice versa. The onsets of energy transfer almost occur at about the identical reduced velocity value $Ur \approx 5.25$ for different $Pr$ and $Ri$ values, albeit they ends at very different $Ur$ values for various Richardson numbers. This observation agrees well with the results of $C_e$ in Figure~\ref{fig:Ce}. As the process of energy transfer becomes suppressed further, the phase angle differences are stabilized, either in-phase or anti-phase.      

Different from isothermal flow, the heat energy is diffused and advected downstream from a vibrating and heated cylinders in this study. The perturbation of natural convection to hydrodynamics and structural dynamics is subtly linked with the efficiency of heat convection over the cylinder's surface, which is quantified by the value of mean Nusselt number. In Figure~\ref{fig:Nu}, overall, the values of mean Nusselt number ($\overline{Nu}$) increase proportionally with the values of Prandtl number over a wide range of $Ur$ values. This observation agrees well with the conclusion drawn in Section~\ref{sec:struct}, which says the temperature contours are concentrated around the cylinder's surface for high $Pr$ values. Consequently, a strong heat convection over cylinder's surface is observed in the cases of $Pr = 10$ in Figure~\ref{fig:Nu5}. In terms of Richardson number, the value of $\overline{Nu}$ becomes less sensitive to the changes in buoyancy-driven flow in the cases of $Ri \in [0.5, 2.0]$. The obtained numerical results of $\overline{Nu}$ during the off lock-in regions match well with the empirical formula derived for external flow over a stationary and heated cylinder~\cite{bergman2011fundamentals}, the dotted black horizontal line in Figure~\ref{fig:Nu}. In contrast, it is found the value of $\overline{Nu}$ becomes significantly excited by approximately 10\% during VIV lock-in for different combination of $Pr$ and $Ri$ values.  
Unlike the values of $\overline{Nu}$, the maximum fluctuation of Nusselt number $\delta Nu^{max}$ is very sensitive to the changes of Richardson numbers, but insensitive to the variation in Prandtl number. Figure~\ref{fig:Nu}(b, d, f) show that the responses of $\delta Nu^{max}$ with respect to $Ur$ values are almost identical for different Prandtl numbers and $Ri < 2.0$. The values of $\delta Nu^{max}$ are almost zero during off lock-in and merely excited by approximately 0.5 during VIV lock-in. The only exception is the cases of $Ri=2.0$, where the influence of natural convection is non-trivial, e.g., the green dotted lines with \emph{cross} markers in Figure~\ref{fig:Nu} (b, d, f). Again, the aforementioned secondary VIV lock-in regions are affirmatively noticed in the cases of $Pr = [0.7, 2.0]$ and $Ri = 2.0$. An intense fluctuation of Nusselt number is associated with VIV lock-in regions. Specifically, it is also found the maximum fluctuation of Nusselt number could surprisingly reach as high as 2.0 in the case of $Ri = 2.0$ during VIV lock-in in Figure~\ref{fig:Nu2}. Generally, the maximum fluctuation of Nusselt number is significantly large in the cases of $Ri = 2.0$ during off lock-in too. Like the lift forces, these intensive fluctuation of $Nu$ during off lock-in regions do not effectively excite the structural vibration. Based on the analysis of fluid kinetic energy transfer in Figure~\ref{fig:Ce} and Figure~\ref{fig:phase}, it is understood that the fluid kinetic energy transfer is almost inhibited during off lock-in regions. Hence this tendency in $Nu$ values is well appreciated.

\section{Conclusions} \label{sec:conclusion}
In this article, the hydrodynamic and thermal characteristics of a freely-vibrating circular cylinder in mixed convection flow were numerically investigated for $Re = 100$, $Pr \in [0.7, 10.0]$ and $Ri \in [0.5, 2.0]$. In those cases, both forced convection and natural convection are important, but the forced convection is dominant, which is of our particular interest. It was found that the values of mean streamwise displacement was relatively insensitive to Prandtl and Richardson numbers, but increased proportionally with the values of reduced velocity and Richardson number. Whereas both the streamwise and transverse maximum fluctuations of structure were tremendously amplified during VIV lock-in. Overall, the hydrodynamics and structural dynamics tended to approach those in isothermal flow for high Prandtl numbers, e.g., $Pr = 10$ in this study, except for $Ri = 2.0$, where the interference of natural convection is non-trivial. Strong buoyancy-driven flow had strong influences on all aspects of the results, e.g., hydrodynamics, structural dynamics and heat transfer, specially for high reduced velocity values. The region of VIV lock-in was narrowed for higher values of Richardson number. In particular, VIV lock-in was almost suppressed in cases of $Pr > 10$ and $Ri = 2.0$. Overall, it was found that the influence of Richardson number on structural and hydrodynamic responses was much severe than Prandtl number. This influence primarily occurred for high reduced velocity values, which was typically within the post lock-in region in isothermal flow. Especially, a strong secondary VIV lock-in region was formed for $Ri = 2.0$ and large $Ur$ values. It further formed a huge VIV lock-in region over a wide range of $Ur$ values by coalescing the primary and secondary VIV lock-in regions for $Pr = 2.0$ and $Ri = 2.0$. The wide VIV lock-in region was associated with tremendous fluid kinetic energy transfer between fluid and structure, which is extremely meaningful for hydropower harvesting. However, the fluid kinetic energy transfer could be significantly inhibited in the cases of $Pr > 2.0$ and $Ri = 2.0$ too. It was further confirmed that the maximum fluid kinetic energy transfer occurred at approximately $\phi = 90^{\circ}$ for different combination of $Pr$ and $Ri$ values. In contrast, the mean Nusselt number was found generally insensitive to Richardson numbers and reduced velocity, albeit high values of $\overline{Nu}$ were observed for VIV lock-in. In contrast, Prandtl number primarily controls the heat transfer over cylinder's surface. The maximum fluctuation of Nusselt number is apparently amplified in the cases of $Ri=2.0$ and VIV lock-in, but it was still incapable of exciting the structural dynamics during off lock-in, since the fluid kinetic energy transfer is significantly inhibited.      



\nocite{*}
\bibliography{refs}

\providecommand{\noopsort}[1]{}\providecommand{\singleletter}[1]{#1}%
\begin{thebibliography}{55}%
\makeatletter
\providecommand \@ifxundefined [1]{%
 \@ifx{#1\undefined}
}%
\providecommand \@ifnum [1]{%
 \ifnum #1\expandafter \@firstoftwo
 \else \expandafter \@secondoftwo
 \fi
}%
\providecommand \@ifx [1]{%
 \ifx #1\expandafter \@firstoftwo
 \else \expandafter \@secondoftwo
 \fi
}%
\providecommand \natexlab [1]{#1}%
\providecommand \enquote  [1]{``#1''}%
\providecommand \bibnamefont  [1]{#1}%
\providecommand \bibfnamefont [1]{#1}%
\providecommand \citenamefont [1]{#1}%
\providecommand \href@noop [0]{\@secondoftwo}%
\providecommand \href [0]{\begingroup \@sanitize@url \@href}%
\providecommand \@href[1]{\@@startlink{#1}\@@href}%
\providecommand \@@href[1]{\endgroup#1\@@endlink}%
\providecommand \@sanitize@url [0]{\catcode `\\12\catcode `\$12\catcode
  `\&12\catcode `\#12\catcode `\^12\catcode `\_12\catcode `\%12\relax}%
\providecommand \@@startlink[1]{}%
\providecommand \@@endlink[0]{}%
\providecommand \url  [0]{\begingroup\@sanitize@url \@url }%
\providecommand \@url [1]{\endgroup\@href {#1}{\urlprefix }}%
\providecommand \urlprefix  [0]{URL }%
\providecommand \Eprint [0]{\href }%
\providecommand \doibase [0]{http://dx.doi.org/}%
\providecommand \selectlanguage [0]{\@gobble}%
\providecommand \bibinfo  [0]{\@secondoftwo}%
\providecommand \bibfield  [0]{\@secondoftwo}%
\providecommand \translation [1]{[#1]}%
\providecommand \BibitemOpen [0]{}%
\providecommand \bibitemStop [0]{}%
\providecommand \bibitemNoStop [0]{.\EOS\space}%
\providecommand \EOS [0]{\spacefactor3000\relax}%
\providecommand \BibitemShut  [1]{\csname bibitem#1\endcsname}%
\let\auto@bib@innerbib\@empty
\bibitem [{\citenamefont {De~Groot}\ and\ \citenamefont
  {De~Groot}(1951)}]{de1951thermodynamics}%
  \BibitemOpen
  \bibfield  {author} {\bibinfo {author} {\bibfnamefont {S.~R.}\ \bibnamefont
  {De~Groot}}\ and\ \bibinfo {author} {\bibfnamefont {S.~R.}\ \bibnamefont
  {De~Groot}},\ }\href@noop {} {\emph {\bibinfo {title} {Thermodynamics of
  irreversible processes}}},\ Vol.\ \bibinfo {volume} {336}\ (\bibinfo
  {publisher} {North-Holland Amsterdam},\ \bibinfo {year} {1951})\BibitemShut
  {NoStop}%
\bibitem [{\citenamefont {Hirschfelder}\ \emph {et~al.}(1964)\citenamefont
  {Hirschfelder}, \citenamefont {Curtiss}, \citenamefont {Bird},\ and\
  \citenamefont {Mayer}}]{hirschfelder1964molecular}%
  \BibitemOpen
  \bibfield  {author} {\bibinfo {author} {\bibfnamefont {J.~O.}\ \bibnamefont
  {Hirschfelder}}, \bibinfo {author} {\bibfnamefont {C.~F.}\ \bibnamefont
  {Curtiss}}, \bibinfo {author} {\bibfnamefont {R.~B.}\ \bibnamefont {Bird}}, \
  and\ \bibinfo {author} {\bibfnamefont {M.~G.}\ \bibnamefont {Mayer}},\
  }\href@noop {} {\emph {\bibinfo {title} {Molecular theory of gases and
  liquids}}},\ Vol.\ \bibinfo {volume} {165}\ (\bibinfo  {publisher} {Wiley New
  York},\ \bibinfo {year} {1964})\BibitemShut {NoStop}%
\bibitem [{\citenamefont {Chapman}\ and\ \citenamefont
  {Cowling}(1970)}]{chapman1970mathematical}%
  \BibitemOpen
  \bibfield  {author} {\bibinfo {author} {\bibfnamefont {S.}~\bibnamefont
  {Chapman}}\ and\ \bibinfo {author} {\bibfnamefont {T.}~\bibnamefont
  {Cowling}},\ }\bibfield  {title} {\enquote {\bibinfo {title} {The
  mathematical theory of non-uniform gases, cambridge univ},}\ }\href@noop {}
  {\bibfield  {journal} {\bibinfo  {journal} {Press, Cambridge, England}\ }
  (\bibinfo {year} {1970})}\BibitemShut {NoStop}%
\bibitem [{\citenamefont {Carslaw}\ and\ \citenamefont
  {Jaeger}(1959)}]{carslaw1959conduction}%
  \BibitemOpen
  \bibfield  {author} {\bibinfo {author} {\bibfnamefont {H.}~\bibnamefont
  {Carslaw}}\ and\ \bibinfo {author} {\bibfnamefont {J.}~\bibnamefont
  {Jaeger}},\ }\bibfield  {title} {\enquote {\bibinfo {title} {Conduction of
  heat in solids (london: Oxford university)},}\ }\href@noop {} {\  (\bibinfo
  {year} {1959})}\BibitemShut {NoStop}%
\bibitem [{\citenamefont {Ozisik}(2002)}]{ozisik2002boundary}%
  \BibitemOpen
  \bibfield  {author} {\bibinfo {author} {\bibfnamefont {M.~N.}\ \bibnamefont
  {Ozisik}},\ }\bibfield  {title} {\enquote {\bibinfo {title} {Boundary value
  problems of heat conduction, 1968},}\ }\href@noop {} {\bibfield  {journal}
  {\bibinfo  {journal} {International Textbook, Scranton}\ } (\bibinfo {year}
  {2002})}\BibitemShut {NoStop}%
\bibitem [{\citenamefont {Iba{\~n}ez}(2002)}]{ibanez2002advanced}%
  \BibitemOpen
  \bibfield  {author} {\bibinfo {author} {\bibfnamefont {M.~T.}\ \bibnamefont
  {Iba{\~n}ez}},\ }\href@noop {} {\emph {\bibinfo {title} {Advanced boundary
  elements for heat transfer}}},\ Vol.~\bibinfo {volume} {42}\ (\bibinfo
  {publisher} {Wit Pr/Computational Mechanics},\ \bibinfo {year}
  {2002})\BibitemShut {NoStop}%
\bibitem [{\citenamefont {{\"O}zi{\c{s}}ik}\ \emph {et~al.}(2017)\citenamefont
  {{\"O}zi{\c{s}}ik}, \citenamefont {Orlande}, \citenamefont {Cola{\c{c}}o},\
  and\ \citenamefont {Cotta}}]{ozicsik2017finite}%
  \BibitemOpen
  \bibfield  {author} {\bibinfo {author} {\bibfnamefont {M.~N.}\ \bibnamefont
  {{\"O}zi{\c{s}}ik}}, \bibinfo {author} {\bibfnamefont {H.~R.}\ \bibnamefont
  {Orlande}}, \bibinfo {author} {\bibfnamefont {M.~J.}\ \bibnamefont
  {Cola{\c{c}}o}}, \ and\ \bibinfo {author} {\bibfnamefont {R.~M.}\
  \bibnamefont {Cotta}},\ }\href@noop {} {\emph {\bibinfo {title} {Finite
  difference methods in heat transfer}}}\ (\bibinfo  {publisher} {CRC press},\
  \bibinfo {year} {2017})\BibitemShut {NoStop}%
\bibitem [{\citenamefont {Patankar}(2018)}]{patankar2018numerical}%
  \BibitemOpen
  \bibfield  {author} {\bibinfo {author} {\bibfnamefont {S.}~\bibnamefont
  {Patankar}},\ }\href@noop {} {\emph {\bibinfo {title} {Numerical heat
  transfer and fluid flow}}}\ (\bibinfo  {publisher} {Taylor \& Francis},\
  \bibinfo {year} {2018})\BibitemShut {NoStop}%
\bibitem [{\citenamefont {Lewis}, \citenamefont {Nithiarasu},\ and\
  \citenamefont {Seetharamu}(2004)}]{lewis2004fundamentals}%
  \BibitemOpen
  \bibfield  {author} {\bibinfo {author} {\bibfnamefont {R.~W.}\ \bibnamefont
  {Lewis}}, \bibinfo {author} {\bibfnamefont {P.}~\bibnamefont {Nithiarasu}}, \
  and\ \bibinfo {author} {\bibfnamefont {K.~N.}\ \bibnamefont {Seetharamu}},\
  }\href@noop {} {\emph {\bibinfo {title} {Fundamentals of the finite element
  method for heat and fluid flow}}}\ (\bibinfo  {publisher} {John Wiley \&
  Sons},\ \bibinfo {year} {2004})\BibitemShut {NoStop}%
\bibitem [{\citenamefont {Reddy}\ and\ \citenamefont
  {Gartling}(2010)}]{reddy2010finite}%
  \BibitemOpen
  \bibfield  {author} {\bibinfo {author} {\bibfnamefont {J.~N.}\ \bibnamefont
  {Reddy}}\ and\ \bibinfo {author} {\bibfnamefont {D.~K.}\ \bibnamefont
  {Gartling}},\ }\href@noop {} {\emph {\bibinfo {title} {The finite element
  method in heat transfer and fluid dynamics}}}\ (\bibinfo  {publisher} {CRC
  press},\ \bibinfo {year} {2010})\BibitemShut {NoStop}%
\bibitem [{\citenamefont {Biswas}\ \emph {et~al.}(1990)\citenamefont {Biswas},
  \citenamefont {Laschefski}, \citenamefont {Mitra},\ and\ \citenamefont
  {Fiebig}}]{biswas1990numerical}%
  \BibitemOpen
  \bibfield  {author} {\bibinfo {author} {\bibfnamefont {G.}~\bibnamefont
  {Biswas}}, \bibinfo {author} {\bibfnamefont {H.}~\bibnamefont {Laschefski}},
  \bibinfo {author} {\bibfnamefont {N.}~\bibnamefont {Mitra}}, \ and\ \bibinfo
  {author} {\bibfnamefont {M.}~\bibnamefont {Fiebig}},\ }\bibfield  {title}
  {\enquote {\bibinfo {title} {Numerical investigation of mixed convection heat
  transfer in a horizontal channel with a built-in square cylinder},}\
  }\href@noop {} {\bibfield  {journal} {\bibinfo  {journal} {Numerical heat
  transfer}\ }\textbf {\bibinfo {volume} {18}},\ \bibinfo {pages} {173--188}
  (\bibinfo {year} {1990})}\BibitemShut {NoStop}%
\bibitem [{\citenamefont {Sanitjai}\ and\ \citenamefont
  {Goldstein}(2004)}]{sanitjai2004heat}%
  \BibitemOpen
  \bibfield  {author} {\bibinfo {author} {\bibfnamefont {S.}~\bibnamefont
  {Sanitjai}}\ and\ \bibinfo {author} {\bibfnamefont {R.~J.}\ \bibnamefont
  {Goldstein}},\ }\bibfield  {title} {\enquote {\bibinfo {title} {Heat transfer
  from a circular cylinder to mixtures of water and ethylene glycol},}\
  }\href@noop {} {\bibfield  {journal} {\bibinfo  {journal} {International
  journal of heat and mass transfer}\ }\textbf {\bibinfo {volume} {47}},\
  \bibinfo {pages} {4785--4794} (\bibinfo {year} {2004})}\BibitemShut {NoStop}%
\bibitem [{\citenamefont {Juncu}(2007)}]{Juncu2007}%
  \BibitemOpen
  \bibfield  {author} {\bibinfo {author} {\bibfnamefont {G.}~\bibnamefont
  {Juncu}},\ }\bibfield  {title} {\enquote {\bibinfo {title} {A numerical study
  of momentum and forced convection heat transfer around two tandem circular
  cylinders at low reynolds numbers. part ii: Forced convection heat
  transfer},}\ }\href@noop {} {\bibfield  {journal} {\bibinfo  {journal}
  {International Journal of Heat and Mass Transfer}\ }\textbf {\bibinfo
  {volume} {50}},\ \bibinfo {pages} {3799--3808} (\bibinfo {year}
  {2007})}\BibitemShut {NoStop}%
\bibitem [{\citenamefont {Biswas}\ and\ \citenamefont
  {Sarkar}(2009)}]{biswas2009effect}%
  \BibitemOpen
  \bibfield  {author} {\bibinfo {author} {\bibfnamefont {G.}~\bibnamefont
  {Biswas}}\ and\ \bibinfo {author} {\bibfnamefont {S.}~\bibnamefont
  {Sarkar}},\ }\bibfield  {title} {\enquote {\bibinfo {title} {Effect of
  thermal buoyancy on vortex shedding past a circular cylinder in cross-flow at
  low reynolds numbers},}\ }\href@noop {} {\bibfield  {journal} {\bibinfo
  {journal} {International Journal of Heat and Mass Transfer}\ }\textbf
  {\bibinfo {volume} {52}},\ \bibinfo {pages} {1897--1912} (\bibinfo {year}
  {2009})}\BibitemShut {NoStop}%
\bibitem [{\citenamefont {Sarpkaya}(1979)}]{Sarpkaya1979JoAM}%
  \BibitemOpen
  \bibfield  {author} {\bibinfo {author} {\bibfnamefont {T.}~\bibnamefont
  {Sarpkaya}},\ }\bibfield  {title} {\enquote {\bibinfo {title} {Vortex-induced
  oscillations: a selective review},}\ }\href@noop {} {\bibfield  {journal}
  {\bibinfo  {journal} {Journal of Applied Mechanics}\ }\textbf {\bibinfo
  {volume} {46}},\ \bibinfo {pages} {241--258} (\bibinfo {year}
  {1979})}\BibitemShut {NoStop}%
\bibitem [{\citenamefont {Blackburn}, \citenamefont {Govardhan},\ and\
  \citenamefont {Williamson}(2001)}]{Blackburn2001JoFaS}%
  \BibitemOpen
  \bibfield  {author} {\bibinfo {author} {\bibfnamefont {H.~M.}\ \bibnamefont
  {Blackburn}}, \bibinfo {author} {\bibfnamefont {R.}~\bibnamefont
  {Govardhan}}, \ and\ \bibinfo {author} {\bibfnamefont {C.}~\bibnamefont
  {Williamson}},\ }\bibfield  {title} {\enquote {\bibinfo {title} {A
  complementary numerical and physical investigation of vortex-induced
  vibration},}\ }\href@noop {} {\bibfield  {journal} {\bibinfo  {journal}
  {Journal of Fluids and Structures}\ }\textbf {\bibinfo {volume} {15}},\
  \bibinfo {pages} {481--488} (\bibinfo {year} {2001})}\BibitemShut {NoStop}%
\bibitem [{\citenamefont {Williamson}\ and\ \citenamefont
  {Govardhan}(2004)}]{Williamson2004ARFM}%
  \BibitemOpen
  \bibfield  {author} {\bibinfo {author} {\bibfnamefont {C.}~\bibnamefont
  {Williamson}}\ and\ \bibinfo {author} {\bibfnamefont {R.}~\bibnamefont
  {Govardhan}},\ }\bibfield  {title} {\enquote {\bibinfo {title}
  {Vortex-induced vibrations},}\ }\href@noop {} {\bibfield  {journal} {\bibinfo
   {journal} {Annu. Rev. Fluid Mech.}\ }\textbf {\bibinfo {volume} {36}},\
  \bibinfo {pages} {413--455} (\bibinfo {year} {2004})}\BibitemShut {NoStop}%
\bibitem [{\citenamefont {Sarpkaya}(2004)}]{Sarpkaya2004JoFaS}%
  \BibitemOpen
  \bibfield  {author} {\bibinfo {author} {\bibfnamefont {T.}~\bibnamefont
  {Sarpkaya}},\ }\bibfield  {title} {\enquote {\bibinfo {title} {A critical
  review of the intrinsic nature of vortex-induced vibrations},}\ }\href@noop
  {} {\bibfield  {journal} {\bibinfo  {journal} {Journal of Fluids and
  Structures}\ }\textbf {\bibinfo {volume} {19}},\ \bibinfo {pages} {389--447}
  (\bibinfo {year} {2004})}\BibitemShut {NoStop}%
\bibitem [{\citenamefont {Prasanth}\ and\ \citenamefont
  {Mittal}(2008)}]{Prasanth2008JoFM}%
  \BibitemOpen
  \bibfield  {author} {\bibinfo {author} {\bibfnamefont {T.}~\bibnamefont
  {Prasanth}}\ and\ \bibinfo {author} {\bibfnamefont {S.}~\bibnamefont
  {Mittal}},\ }\bibfield  {title} {\enquote {\bibinfo {title} {Vortex-induced
  vibrations of a circular cylinder at low reynolds numbers},}\ }\href@noop {}
  {\bibfield  {journal} {\bibinfo  {journal} {Journal of Fluid Mechanics}\
  }\textbf {\bibinfo {volume} {594}},\ \bibinfo {pages} {463--491} (\bibinfo
  {year} {2008})}\BibitemShut {NoStop}%
\bibitem [{\citenamefont {Bearman}(2011)}]{Bearman2011JoFaS}%
  \BibitemOpen
  \bibfield  {author} {\bibinfo {author} {\bibfnamefont {P.}~\bibnamefont
  {Bearman}},\ }\bibfield  {title} {\enquote {\bibinfo {title} {Circular
  cylinder wakes and vortex-induced vibrations},}\ }\href@noop {} {\bibfield
  {journal} {\bibinfo  {journal} {Journal of Fluids and Structures}\ }\textbf
  {\bibinfo {volume} {27}},\ \bibinfo {pages} {648--658} (\bibinfo {year}
  {2011})}\BibitemShut {NoStop}%
\bibitem [{\citenamefont {Zhu}, \citenamefont {Zhang},\ and\ \citenamefont
  {Liu}(2019)}]{zhu2019wake}%
  \BibitemOpen
  \bibfield  {author} {\bibinfo {author} {\bibfnamefont {H.}~\bibnamefont
  {Zhu}}, \bibinfo {author} {\bibfnamefont {C.}~\bibnamefont {Zhang}}, \ and\
  \bibinfo {author} {\bibfnamefont {W.}~\bibnamefont {Liu}},\ }\bibfield
  {title} {\enquote {\bibinfo {title} {Wake-induced vibration of a circular
  cylinder at a low reynolds number of 100},}\ }\href@noop {} {\bibfield
  {journal} {\bibinfo  {journal} {Physics of Fluids}\ }\textbf {\bibinfo
  {volume} {31}},\ \bibinfo {pages} {073606} (\bibinfo {year}
  {2019})}\BibitemShut {NoStop}%
\bibitem [{\citenamefont {Williamson}(1985)}]{Williamson1985JoFM}%
  \BibitemOpen
  \bibfield  {author} {\bibinfo {author} {\bibfnamefont {C.~H.~K.}\
  \bibnamefont {Williamson}},\ }\bibfield  {title} {\enquote {\bibinfo {title}
  {Evolution of a single wake behind a pair of bluff bodies},}\ }\href
  {\doibase http://dx.doi.org/10.1017/S002211208500307X} {\bibfield  {journal}
  {\bibinfo  {journal} {Journal of Fluid Mechanics}\ }\textbf {\bibinfo
  {volume} {159}},\ \bibinfo {pages} {1--18} (\bibinfo {year}
  {1985})}\BibitemShut {NoStop}%
\bibitem [{\citenamefont {Carini}, \citenamefont {Giannetti},\ and\
  \citenamefont {Auteri}(2014)}]{Carini2014JoFM}%
  \BibitemOpen
  \bibfield  {author} {\bibinfo {author} {\bibfnamefont {M.}~\bibnamefont
  {Carini}}, \bibinfo {author} {\bibfnamefont {F.}~\bibnamefont {Giannetti}}, \
  and\ \bibinfo {author} {\bibfnamefont {F.}~\bibnamefont {Auteri}},\
  }\bibfield  {title} {\enquote {\bibinfo {title} {On the origin of the
  flip-flop instability of two side-by-side cylinder wakes},}\ }\href {\doibase
  10.1017/jfm.2014.9} {\bibfield  {journal} {\bibinfo  {journal} {Journal of
  Fluid Mechanics}\ }\textbf {\bibinfo {volume} {742}},\ \bibinfo {pages}
  {552--576} (\bibinfo {year} {2014})}\BibitemShut {NoStop}%
\bibitem [{\citenamefont {Liu}\ and\ \citenamefont
  {Jaiman}(2016{\natexlab{a}})}]{liu2016interaction}%
  \BibitemOpen
  \bibfield  {author} {\bibinfo {author} {\bibfnamefont {B.}~\bibnamefont
  {Liu}}\ and\ \bibinfo {author} {\bibfnamefont {R.~K.}\ \bibnamefont
  {Jaiman}},\ }\bibfield  {title} {\enquote {\bibinfo {title} {Interaction
  dynamics of gap flow with vortex-induced vibration in side-by-side cylinder
  arrangement},}\ }\href@noop {} {\bibfield  {journal} {\bibinfo  {journal}
  {Physics of Fluids}\ }\textbf {\bibinfo {volume} {28}},\ \bibinfo {pages}
  {127103} (\bibinfo {year} {2016}{\natexlab{a}})}\BibitemShut {NoStop}%
\bibitem [{\citenamefont {Liu}\ and\ \citenamefont
  {Jaiman}(2016{\natexlab{b}})}]{liu2016effect}%
  \BibitemOpen
  \bibfield  {author} {\bibinfo {author} {\bibfnamefont {B.}~\bibnamefont
  {Liu}}\ and\ \bibinfo {author} {\bibfnamefont {R.~K.}\ \bibnamefont
  {Jaiman}},\ }\bibfield  {title} {\enquote {\bibinfo {title} {The effect of
  gap flow on vortex-induced vibration of side-by-side cylinder arrangement},}\
  }in\ \href@noop {} {\emph {\bibinfo {booktitle} {International Conference on
  Offshore Mechanics and Arctic Engineering}}},\ Vol.\ \bibinfo {volume}
  {49934}\ (\bibinfo {organization} {American Society of Mechanical
  Engineers},\ \bibinfo {year} {2016})\ p.\ \bibinfo {pages}
  {V002T08A013}\BibitemShut {NoStop}%
\bibitem [{\citenamefont {Borazjani}\ and\ \citenamefont
  {Sotiropoulos}(2009)}]{Borazjani2009Jofm}%
  \BibitemOpen
  \bibfield  {author} {\bibinfo {author} {\bibfnamefont {I.}~\bibnamefont
  {Borazjani}}\ and\ \bibinfo {author} {\bibfnamefont {F.}~\bibnamefont
  {Sotiropoulos}},\ }\bibfield  {title} {\enquote {\bibinfo {title}
  {Vortex-induced vibrations of two cylinders in tandem arrangement in the
  proximity--wake interference region},}\ }\href@noop {} {\bibfield  {journal}
  {\bibinfo  {journal} {Journal of fluid mechanics}\ }\textbf {\bibinfo
  {volume} {621}},\ \bibinfo {pages} {321--364} (\bibinfo {year} {2009})},\
  \bibinfo {note} {three dimensionality suppression in the gap between two
  tandem cylinders}\BibitemShut {NoStop}%
\bibitem [{\citenamefont {Assi}, \citenamefont {Bearman},\ and\ \citenamefont
  {Meneghini}(2010)}]{Assi2010JoFM}%
  \BibitemOpen
  \bibfield  {author} {\bibinfo {author} {\bibfnamefont {G.}~\bibnamefont
  {Assi}}, \bibinfo {author} {\bibfnamefont {P.}~\bibnamefont {Bearman}}, \
  and\ \bibinfo {author} {\bibfnamefont {J.~R.}\ \bibnamefont {Meneghini}},\
  }\bibfield  {title} {\enquote {\bibinfo {title} {On the wake-induced
  vibration of tandem circular cylinders: the vortex interaction excitation
  mechanism},}\ }\href@noop {} {\bibfield  {journal} {\bibinfo  {journal}
  {Journal of Fluid Mechanics}\ }\textbf {\bibinfo {volume} {661}},\ \bibinfo
  {pages} {365--401} (\bibinfo {year} {2010})}\BibitemShut {NoStop}%
\bibitem [{\citenamefont {Li}\ \emph {et~al.}(2016)\citenamefont {Li},
  \citenamefont {Yao}, \citenamefont {Yang}, \citenamefont {Jaiman},\ and\
  \citenamefont {Khoo}}]{Li2016JoFaS}%
  \BibitemOpen
  \bibfield  {author} {\bibinfo {author} {\bibfnamefont {Z.}~\bibnamefont
  {Li}}, \bibinfo {author} {\bibfnamefont {W.}~\bibnamefont {Yao}}, \bibinfo
  {author} {\bibfnamefont {K.}~\bibnamefont {Yang}}, \bibinfo {author}
  {\bibfnamefont {R.~K.}\ \bibnamefont {Jaiman}}, \ and\ \bibinfo {author}
  {\bibfnamefont {B.~C.}\ \bibnamefont {Khoo}},\ }\bibfield  {title} {\enquote
  {\bibinfo {title} {On the vortex-induced oscillations of a freely vibrating
  cylinder in the vicinity of a stationary plane wall},}\ }\href@noop {}
  {\bibfield  {journal} {\bibinfo  {journal} {Journal of Fluids and
  Structures}\ }\textbf {\bibinfo {volume} {65}},\ \bibinfo {pages} {495--526}
  (\bibinfo {year} {2016})}\BibitemShut {NoStop}%
\bibitem [{\citenamefont {Liu}\ and\ \citenamefont
  {Magee}(2020)}]{liu2020numerical}%
  \BibitemOpen
  \bibfield  {author} {\bibinfo {author} {\bibfnamefont {B.}~\bibnamefont
  {Liu}}\ and\ \bibinfo {author} {\bibfnamefont {A.}~\bibnamefont {Magee}},\
  }\bibfield  {title} {\enquote {\bibinfo {title} {Numerical stability and
  three dimensionality of a streamline hyperbolic critical point in wake at low
  reynolds number},}\ }\href@noop {} {\bibfield  {journal} {\bibinfo  {journal}
  {arXiv preprint arXiv:2006.05306}\ } (\bibinfo {year} {2020})}\BibitemShut
  {NoStop}%
\bibitem [{\citenamefont {Ju}\ \emph {et~al.}(2020)\citenamefont {Ju},
  \citenamefont {An}, \citenamefont {Cheng},\ and\ \citenamefont
  {Tong}}]{Ju2020JoFM}%
  \BibitemOpen
  \bibfield  {author} {\bibinfo {author} {\bibfnamefont {X.}~\bibnamefont
  {Ju}}, \bibinfo {author} {\bibfnamefont {H.}~\bibnamefont {An}}, \bibinfo
  {author} {\bibfnamefont {L.}~\bibnamefont {Cheng}}, \ and\ \bibinfo {author}
  {\bibfnamefont {F.}~\bibnamefont {Tong}},\ }\bibfield  {title} {\enquote
  {\bibinfo {title} {Modes of synchronisation around a near-wall oscillating
  cylinder in streamwise directions},}\ }\href@noop {} {\bibfield  {journal}
  {\bibinfo  {journal} {Journal of Fluid Mechanics}\ }\textbf {\bibinfo
  {volume} {893}} (\bibinfo {year} {2020})}\BibitemShut {NoStop}%
\bibitem [{\citenamefont {Joshi}, \citenamefont {Liu},\ and\ \citenamefont
  {Jaiman}(2016)}]{joshi2016flow}%
  \BibitemOpen
  \bibfield  {author} {\bibinfo {author} {\bibfnamefont {V.}~\bibnamefont
  {Joshi}}, \bibinfo {author} {\bibfnamefont {B.}~\bibnamefont {Liu}}, \ and\
  \bibinfo {author} {\bibfnamefont {R.~K.}\ \bibnamefont {Jaiman}},\ }\bibfield
   {title} {\enquote {\bibinfo {title} {Flow-induced vibrations of riser array
  system},}\ }in\ \href@noop {} {\emph {\bibinfo {booktitle} {International
  Conference on Offshore Mechanics and Arctic Engineering}}},\ Vol.\ \bibinfo
  {volume} {49934}\ (\bibinfo {organization} {American Society of Mechanical
  Engineers},\ \bibinfo {year} {2016})\ p.\ \bibinfo {pages}
  {V002T08A012}\BibitemShut {NoStop}%
\bibitem [{\citenamefont {Tang}\ \emph {et~al.}(2020)\citenamefont {Tang},
  \citenamefont {Yu}, \citenamefont {Shan}, \citenamefont {Li},\ and\
  \citenamefont {Yu}}]{Tang2020PoFa}%
  \BibitemOpen
  \bibfield  {author} {\bibinfo {author} {\bibfnamefont {T.}~\bibnamefont
  {Tang}}, \bibinfo {author} {\bibfnamefont {P.}~\bibnamefont {Yu}}, \bibinfo
  {author} {\bibfnamefont {X.}~\bibnamefont {Shan}}, \bibinfo {author}
  {\bibfnamefont {J.}~\bibnamefont {Li}}, \ and\ \bibinfo {author}
  {\bibfnamefont {S.}~\bibnamefont {Yu}},\ }\bibfield  {title} {\enquote
  {\bibinfo {title} {On the transition behavior of laminar flow through and
  around a multi-cylinder array},}\ }\href@noop {} {\bibfield  {journal}
  {\bibinfo  {journal} {Physics of Fluids}\ }\textbf {\bibinfo {volume} {32}},\
  \bibinfo {pages} {013601} (\bibinfo {year} {2020})}\BibitemShut {NoStop}%
\bibitem [{\citenamefont {Venkatasubbaiah}\ and\ \citenamefont
  {Sengupta}(2009)}]{venkatasubbaiah2009mixed}%
  \BibitemOpen
  \bibfield  {author} {\bibinfo {author} {\bibfnamefont {K.}~\bibnamefont
  {Venkatasubbaiah}}\ and\ \bibinfo {author} {\bibfnamefont {T.}~\bibnamefont
  {Sengupta}},\ }\bibfield  {title} {\enquote {\bibinfo {title} {Mixed
  convection flow past a vertical plate: Stability analysis and its direct
  simulation},}\ }\href@noop {} {\bibfield  {journal} {\bibinfo  {journal}
  {International Journal of Thermal Sciences}\ }\textbf {\bibinfo {volume}
  {48}},\ \bibinfo {pages} {461--474} (\bibinfo {year} {2009})}\BibitemShut
  {NoStop}%
\bibitem [{\citenamefont {Sengupta}\ and\ \citenamefont
  {Poinsot}(2010)}]{sengupta2010instabilities}%
  \BibitemOpen
  \bibfield  {author} {\bibinfo {author} {\bibfnamefont {T.}~\bibnamefont
  {Sengupta}}\ and\ \bibinfo {author} {\bibfnamefont {T.}~\bibnamefont
  {Poinsot}},\ }\href@noop {} {\emph {\bibinfo {title} {Instabilities of flows:
  With and without heat transfer and chemical reaction}}},\ Vol.\ \bibinfo
  {volume} {517}\ (\bibinfo  {publisher} {Springer Science \& Business Media},\
  \bibinfo {year} {2010})\BibitemShut {NoStop}%
\bibitem [{\citenamefont {Sengupta}\ \emph {et~al.}(2011)\citenamefont
  {Sengupta}, \citenamefont {Unnikrishnan}, \citenamefont {Bhaumik},
  \citenamefont {Singh},\ and\ \citenamefont {Usman}}]{sengupta2011linear}%
  \BibitemOpen
  \bibfield  {author} {\bibinfo {author} {\bibfnamefont {T.}~\bibnamefont
  {Sengupta}}, \bibinfo {author} {\bibfnamefont {S.}~\bibnamefont
  {Unnikrishnan}}, \bibinfo {author} {\bibfnamefont {S.}~\bibnamefont
  {Bhaumik}}, \bibinfo {author} {\bibfnamefont {P.}~\bibnamefont {Singh}}, \
  and\ \bibinfo {author} {\bibfnamefont {S.}~\bibnamefont {Usman}},\ }\bibfield
   {title} {\enquote {\bibinfo {title} {Linear spatial stability analysis of
  mixed convection boundary layer over a heated plate},}\ }\href@noop {}
  {\bibfield  {journal} {\bibinfo  {journal} {Progress in Applied Mathematics}\
  }\textbf {\bibinfo {volume} {1}},\ \bibinfo {pages} {71--89} (\bibinfo {year}
  {2011})}\BibitemShut {NoStop}%
\bibitem [{\citenamefont {Sengupta}(2012)}]{sengupta2012instabilities}%
  \BibitemOpen
  \bibfield  {author} {\bibinfo {author} {\bibfnamefont {T.~K.}\ \bibnamefont
  {Sengupta}},\ }\href@noop {} {\emph {\bibinfo {title} {Instabilities of flows
  and transition to turbulence}}}\ (\bibinfo  {publisher} {CRC Press},\
  \bibinfo {year} {2012})\BibitemShut {NoStop}%
\bibitem [{\citenamefont {Khan}\ \emph {et~al.}(2020)\citenamefont {Khan},
  \citenamefont {Anwer}, \citenamefont {Khan},\ and\ \citenamefont
  {Hasan}}]{Khan2020IJoTS}%
  \BibitemOpen
  \bibfield  {author} {\bibinfo {author} {\bibfnamefont {M.~A.}\ \bibnamefont
  {Khan}}, \bibinfo {author} {\bibfnamefont {S.~F.}\ \bibnamefont {Anwer}},
  \bibinfo {author} {\bibfnamefont {S.~A.}\ \bibnamefont {Khan}}, \ and\
  \bibinfo {author} {\bibfnamefont {N.}~\bibnamefont {Hasan}},\ }\bibfield
  {title} {\enquote {\bibinfo {title} {Hydrodynamic and heat transfer
  characteristics of vortex-induced vibration of square cylinder with various
  flow approach angle},}\ }\href@noop {} {\bibfield  {journal} {\bibinfo
  {journal} {International Journal of Thermal Sciences}\ }\textbf {\bibinfo
  {volume} {156}},\ \bibinfo {pages} {106454} (\bibinfo {year}
  {2020})}\BibitemShut {NoStop}%
\bibitem [{\citenamefont {Yang}\ \emph {et~al.}(2020)\citenamefont {Yang},
  \citenamefont {Ding}, \citenamefont {Zhang}, \citenamefont {Yang},\ and\
  \citenamefont {He}}]{Yang2020IJoHaMT}%
  \BibitemOpen
  \bibfield  {author} {\bibinfo {author} {\bibfnamefont {Z.}~\bibnamefont
  {Yang}}, \bibinfo {author} {\bibfnamefont {L.}~\bibnamefont {Ding}}, \bibinfo
  {author} {\bibfnamefont {L.}~\bibnamefont {Zhang}}, \bibinfo {author}
  {\bibfnamefont {L.}~\bibnamefont {Yang}}, \ and\ \bibinfo {author}
  {\bibfnamefont {H.}~\bibnamefont {He}},\ }\bibfield  {title} {\enquote
  {\bibinfo {title} {Two degrees of freedom flow-induced vibration and heat
  transfer of an isothermal cylinder},}\ }\href@noop {} {\bibfield  {journal}
  {\bibinfo  {journal} {International Journal of Heat and Mass Transfer}\
  }\textbf {\bibinfo {volume} {154}},\ \bibinfo {pages} {119766} (\bibinfo
  {year} {2020})}\BibitemShut {NoStop}%
\bibitem [{\citenamefont {Liu}\ and\ \citenamefont
  {Jaiman}(2018)}]{liu2018dynamics}%
  \BibitemOpen
  \bibfield  {author} {\bibinfo {author} {\bibfnamefont {B.}~\bibnamefont
  {Liu}}\ and\ \bibinfo {author} {\bibfnamefont {R.}~\bibnamefont {Jaiman}},\
  }\bibfield  {title} {\enquote {\bibinfo {title} {Dynamics and stability of
  gap-flow interference in a vibrating side-by-side arrangement of two circular
  cylinders},}\ }\href@noop {} {\bibfield  {journal} {\bibinfo  {journal}
  {Journal of Fluid Mechanics}\ }\textbf {\bibinfo {volume} {855}},\ \bibinfo
  {pages} {804--838} (\bibinfo {year} {2018})}\BibitemShut {NoStop}%
\bibitem [{\citenamefont {Brooks}\ and\ \citenamefont
  {Hughes}(1982)}]{brooks1982streamline}%
  \BibitemOpen
  \bibfield  {author} {\bibinfo {author} {\bibfnamefont {A.~N.}\ \bibnamefont
  {Brooks}}\ and\ \bibinfo {author} {\bibfnamefont {T.~J.}\ \bibnamefont
  {Hughes}},\ }\bibfield  {title} {\enquote {\bibinfo {title} {Streamline
  upwind/petrov-galerkin formulations for convection dominated flows with
  particular emphasis on the incompressible navier-stokes equations},}\
  }\href@noop {} {\bibfield  {journal} {\bibinfo  {journal} {Computer methods
  in applied mechanics and engineering}\ }\textbf {\bibinfo {volume} {32}},\
  \bibinfo {pages} {199--259} (\bibinfo {year} {1982})}\BibitemShut {NoStop}%
\bibitem [{\citenamefont {Hughes}, \citenamefont {Franca},\ and\ \citenamefont
  {Balestra}(1986)}]{hughes1986new}%
  \BibitemOpen
  \bibfield  {author} {\bibinfo {author} {\bibfnamefont {T.~J.}\ \bibnamefont
  {Hughes}}, \bibinfo {author} {\bibfnamefont {L.~P.}\ \bibnamefont {Franca}},
  \ and\ \bibinfo {author} {\bibfnamefont {M.}~\bibnamefont {Balestra}},\
  }\bibfield  {title} {\enquote {\bibinfo {title} {A new finite element
  formulation for computational fluid dynamics: V. circumventing the
  babu{\v{s}}ka-brezzi condition: A stable petrov-galerkin formulation of the
  stokes problem accommodating equal-order interpolations},}\ }\href@noop {}
  {\bibfield  {journal} {\bibinfo  {journal} {Computer Methods in Applied
  Mechanics and Engineering}\ }\textbf {\bibinfo {volume} {59}},\ \bibinfo
  {pages} {85--99} (\bibinfo {year} {1986})}\BibitemShut {NoStop}%
\bibitem [{\citenamefont {Heywood}, \citenamefont {Rannacher},\ and\
  \citenamefont {Turek}(1996)}]{heywood1996artificial}%
  \BibitemOpen
  \bibfield  {author} {\bibinfo {author} {\bibfnamefont {J.~G.}\ \bibnamefont
  {Heywood}}, \bibinfo {author} {\bibfnamefont {R.}~\bibnamefont {Rannacher}},
  \ and\ \bibinfo {author} {\bibfnamefont {S.}~\bibnamefont {Turek}},\
  }\bibfield  {title} {\enquote {\bibinfo {title} {Artificial boundaries and
  flux and pressure conditions for the incompressible navier--stokes
  equations},}\ }\href@noop {} {\bibfield  {journal} {\bibinfo  {journal}
  {International Journal for numerical methods in fluids}\ }\textbf {\bibinfo
  {volume} {22}},\ \bibinfo {pages} {325--352} (\bibinfo {year}
  {1996})}\BibitemShut {NoStop}%
\bibitem [{\citenamefont {Stein}, \citenamefont {Tezduyar},\ and\ \citenamefont
  {Benney}(2003)}]{stein2003mesh}%
  \BibitemOpen
  \bibfield  {author} {\bibinfo {author} {\bibfnamefont {K.}~\bibnamefont
  {Stein}}, \bibinfo {author} {\bibfnamefont {T.}~\bibnamefont {Tezduyar}}, \
  and\ \bibinfo {author} {\bibfnamefont {R.}~\bibnamefont {Benney}},\
  }\bibfield  {title} {\enquote {\bibinfo {title} {Mesh moving techniques for
  fluid-structure interactions with large displacements},}\ }\href@noop {}
  {\bibfield  {journal} {\bibinfo  {journal} {J. Appl. Mech.}\ }\textbf
  {\bibinfo {volume} {70}},\ \bibinfo {pages} {58--63} (\bibinfo {year}
  {2003})}\BibitemShut {NoStop}%
\bibitem [{\citenamefont {Dettmer}\ and\ \citenamefont
  {Peri{\'c}}(2013)}]{Dettmer2013IJfNMiE}%
  \BibitemOpen
  \bibfield  {author} {\bibinfo {author} {\bibfnamefont {W.~G.}\ \bibnamefont
  {Dettmer}}\ and\ \bibinfo {author} {\bibfnamefont {D.}~\bibnamefont
  {Peri{\'c}}},\ }\bibfield  {title} {\enquote {\bibinfo {title} {A new
  staggered scheme for fluid--structure interaction},}\ }\href@noop {}
  {\bibfield  {journal} {\bibinfo  {journal} {International Journal for
  Numerical Methods in Engineering}\ }\textbf {\bibinfo {volume} {93}},\
  \bibinfo {pages} {1--22} (\bibinfo {year} {2013})}\BibitemShut {NoStop}%
\bibitem [{\citenamefont {Chung}\ and\ \citenamefont
  {Hulbert}(1993)}]{Chung1993Joam}%
  \BibitemOpen
  \bibfield  {author} {\bibinfo {author} {\bibfnamefont {J.}~\bibnamefont
  {Chung}}\ and\ \bibinfo {author} {\bibfnamefont {G.}~\bibnamefont
  {Hulbert}},\ }\bibfield  {title} {\enquote {\bibinfo {title} {A time
  integration algorithm for structural dynamics with improved numerical
  dissipation: the generalized-$\alpha$ method},}\ }\href@noop {} {\bibfield
  {journal} {\bibinfo  {journal} {Journal of applied mechanics}\ }\textbf
  {\bibinfo {volume} {60}},\ \bibinfo {pages} {371--375} (\bibinfo {year}
  {1993})}\BibitemShut {NoStop}%
\bibitem [{\citenamefont {Jansen}, \citenamefont {Whiting},\ and\ \citenamefont
  {Hulbert}(2000)}]{Jansen2000Cmiamae}%
  \BibitemOpen
  \bibfield  {author} {\bibinfo {author} {\bibfnamefont {K.~E.}\ \bibnamefont
  {Jansen}}, \bibinfo {author} {\bibfnamefont {C.~H.}\ \bibnamefont {Whiting}},
  \ and\ \bibinfo {author} {\bibfnamefont {G.~M.}\ \bibnamefont {Hulbert}},\
  }\bibfield  {title} {\enquote {\bibinfo {title} {A generalized-$\alpha$
  method for integrating the filtered navier--stokes equations with a
  stabilized finite element method},}\ }\href@noop {} {\bibfield  {journal}
  {\bibinfo  {journal} {Computer methods in applied mechanics and engineering}\
  }\textbf {\bibinfo {volume} {190}},\ \bibinfo {pages} {305--319} (\bibinfo
  {year} {2000})}\BibitemShut {NoStop}%
\bibitem [{\citenamefont {Sarkar}, \citenamefont {Dalal},\ and\ \citenamefont
  {Biswas}(2011)}]{sarkar2011unsteady}%
  \BibitemOpen
  \bibfield  {author} {\bibinfo {author} {\bibfnamefont {S.}~\bibnamefont
  {Sarkar}}, \bibinfo {author} {\bibfnamefont {A.}~\bibnamefont {Dalal}}, \
  and\ \bibinfo {author} {\bibfnamefont {G.}~\bibnamefont {Biswas}},\
  }\bibfield  {title} {\enquote {\bibinfo {title} {Unsteady wake dynamics and
  heat transfer in forced and mixed convection past a circular cylinder in
  cross flow for high prandtl numbers},}\ }\href@noop {} {\bibfield  {journal}
  {\bibinfo  {journal} {International Journal of Heat and Mass Transfer}\
  }\textbf {\bibinfo {volume} {54}},\ \bibinfo {pages} {3536--3551} (\bibinfo
  {year} {2011})}\BibitemShut {NoStop}%
\bibitem [{\citenamefont {Churchill}\ and\ \citenamefont
  {Bernstein}(1977)}]{churchill1977correlating}%
  \BibitemOpen
  \bibfield  {author} {\bibinfo {author} {\bibfnamefont {S.}~\bibnamefont
  {Churchill}}\ and\ \bibinfo {author} {\bibfnamefont {M.}~\bibnamefont
  {Bernstein}},\ }\bibfield  {title} {\enquote {\bibinfo {title} {A correlating
  equation for forced convection from gases and liquids to a circular cylinder
  in crossflow},}\ }\href@noop {} {\  (\bibinfo {year} {1977})}\BibitemShut
  {NoStop}%
\bibitem [{\citenamefont {Lu}\ \emph {et~al.}(2011)\citenamefont {Lu},
  \citenamefont {Qin}, \citenamefont {Teng},\ and\ \citenamefont
  {Li}}]{lu2011numerical}%
  \BibitemOpen
  \bibfield  {author} {\bibinfo {author} {\bibfnamefont {L.}~\bibnamefont
  {Lu}}, \bibinfo {author} {\bibfnamefont {J.-M.}\ \bibnamefont {Qin}},
  \bibinfo {author} {\bibfnamefont {B.}~\bibnamefont {Teng}}, \ and\ \bibinfo
  {author} {\bibfnamefont {Y.-C.}\ \bibnamefont {Li}},\ }\bibfield  {title}
  {\enquote {\bibinfo {title} {Numerical investigations of lift suppression by
  feedback rotary oscillation of circular cylinder at low reynolds number},}\
  }\href@noop {} {\bibfield  {journal} {\bibinfo  {journal} {Physics of
  Fluids}\ }\textbf {\bibinfo {volume} {23}},\ \bibinfo {pages} {033601}
  (\bibinfo {year} {2011})}\BibitemShut {NoStop}%
\bibitem [{\citenamefont {Liu}\ and\ \citenamefont
  {Tan}(2020)}]{liu2020nitsche}%
  \BibitemOpen
  \bibfield  {author} {\bibinfo {author} {\bibfnamefont {B.}~\bibnamefont
  {Liu}}\ and\ \bibinfo {author} {\bibfnamefont {D.}~\bibnamefont {Tan}},\
  }\bibfield  {title} {\enquote {\bibinfo {title} {A nitsche stabilized finite
  element method for embedded interfaces: Application to fluid-structure
  interaction and rigid-body contact},}\ }\href@noop {} {\bibfield  {journal}
  {\bibinfo  {journal} {Journal of Computational Physics}\ ,\ \bibinfo {pages}
  {109461}} (\bibinfo {year} {2020})}\BibitemShut {NoStop}%
\bibitem [{\citenamefont {Mittal}\ and\ \citenamefont
  {Singh}(2005)}]{mittal2005vortex}%
  \BibitemOpen
  \bibfield  {author} {\bibinfo {author} {\bibfnamefont {S.}~\bibnamefont
  {Mittal}}\ and\ \bibinfo {author} {\bibfnamefont {S.}~\bibnamefont {Singh}},\
  }\bibfield  {title} {\enquote {\bibinfo {title} {Vortex-induced vibrations at
  subcritical re},}\ }\href@noop {} {\bibfield  {journal} {\bibinfo  {journal}
  {Journal of Fluid Mechanics}\ }\textbf {\bibinfo {volume} {534}},\ \bibinfo
  {pages} {185} (\bibinfo {year} {2005})}\BibitemShut {NoStop}%
\bibitem [{\citenamefont {Dolci}\ and\ \citenamefont
  {Carmo}(2019)}]{dolci2019bifurcation}%
  \BibitemOpen
  \bibfield  {author} {\bibinfo {author} {\bibfnamefont {D.~I.}\ \bibnamefont
  {Dolci}}\ and\ \bibinfo {author} {\bibfnamefont {B.~S.}\ \bibnamefont
  {Carmo}},\ }\bibfield  {title} {\enquote {\bibinfo {title} {Bifurcation
  analysis of the primary instability in the flow around a flexibly mounted
  circular cylinder},}\ }\href@noop {} {\bibfield  {journal} {\bibinfo
  {journal} {Journal of Fluid Mechanics}\ }\textbf {\bibinfo {volume} {880}}
  (\bibinfo {year} {2019})}\BibitemShut {NoStop}%
\bibitem [{\citenamefont {Bao}\ \emph {et~al.}(2012)\citenamefont {Bao},
  \citenamefont {Huang}, \citenamefont {Zhou}, \citenamefont {Tu},\ and\
  \citenamefont {Han}}]{bao2012two}%
  \BibitemOpen
  \bibfield  {author} {\bibinfo {author} {\bibfnamefont {Y.}~\bibnamefont
  {Bao}}, \bibinfo {author} {\bibfnamefont {C.}~\bibnamefont {Huang}}, \bibinfo
  {author} {\bibfnamefont {D.}~\bibnamefont {Zhou}}, \bibinfo {author}
  {\bibfnamefont {J.}~\bibnamefont {Tu}}, \ and\ \bibinfo {author}
  {\bibfnamefont {Z.}~\bibnamefont {Han}},\ }\bibfield  {title} {\enquote
  {\bibinfo {title} {Two-degree-of-freedom flow-induced vibrations on isolated
  and tandem cylinders with varying natural frequency ratios},}\ }\href@noop {}
  {\bibfield  {journal} {\bibinfo  {journal} {Journal of Fluids and
  Structures}\ }\textbf {\bibinfo {volume} {35}},\ \bibinfo {pages} {50--75}
  (\bibinfo {year} {2012})}\BibitemShut {NoStop}%
\bibitem [{\citenamefont {Huang}(2014)}]{huang2014hilbert}%
  \BibitemOpen
  \bibfield  {author} {\bibinfo {author} {\bibfnamefont {N.~E.}\ \bibnamefont
  {Huang}},\ }\href@noop {} {\emph {\bibinfo {title} {Hilbert-Huang transform
  and its applications}}},\ Vol.~\bibinfo {volume} {16}\ (\bibinfo  {publisher}
  {World Scientific},\ \bibinfo {year} {2014})\BibitemShut {NoStop}%
\bibitem [{\citenamefont {Bergman}\ \emph {et~al.}(2011)\citenamefont
  {Bergman}, \citenamefont {Incropera}, \citenamefont {DeWitt},\ and\
  \citenamefont {Lavine}}]{bergman2011fundamentals}%
  \BibitemOpen
  \bibfield  {author} {\bibinfo {author} {\bibfnamefont {T.~L.}\ \bibnamefont
  {Bergman}}, \bibinfo {author} {\bibfnamefont {F.~P.}\ \bibnamefont
  {Incropera}}, \bibinfo {author} {\bibfnamefont {D.~P.}\ \bibnamefont
  {DeWitt}}, \ and\ \bibinfo {author} {\bibfnamefont {A.~S.}\ \bibnamefont
  {Lavine}},\ }\href@noop {} {\emph {\bibinfo {title} {Fundamentals of heat and
  mass transfer}}}\ (\bibinfo  {publisher} {John Wiley \& Sons},\ \bibinfo
  {year} {2011})\BibitemShut {NoStop}%
\end{thebibliography}%

\end{document}